\documentclass[10pt,a4paper]{article}

\usepackage[utf8]{inputenc}
\usepackage[T1]{fontenc}
\usepackage{amssymb}
\usepackage{amsmath}
\usepackage{graphicx}
\usepackage{epsfig,latexsym}
\usepackage{slashed}
\usepackage{geometry}
\usepackage{simplewick}
\usepackage{nicefrac}
\usepackage{hyperref}
\usepackage[dvipsnames]{xcolor}
\usepackage[normalem]{ulem}
\usepackage{url}
\usepackage[mathscr]{eucal}
\usepackage[affil-sl]{authblk}

\def\App#1{Appendix~\ref{#1}}

\newcommand{\gettitle}{Novel tools and observables for jet physics in heavy-ion collisions}
\hypersetup{
	colorlinks,
	linkcolor={blue},
	citecolor={blue},
	urlcolor={blue},
	pdftitle={\gettitle},
	pdfkeywords={QCD jets} {Jet quenching} {Jet substructure}
	bookmarksopen=true,
	bookmarksopenlevel=2,
	bookmarksnumbered=true
}

\def\be{\begin{eqnarray*}}
\def\ee{\end{eqnarray*}}
\def\beq{\begin{eqnarray}}
\def\eeq{\end{eqnarray}}
\def\bem{\begin{multline}}
\def\eem{\end{multline}}

\def\med{\text{med}}

\def\tot{\text{tot}}

\def\dd{\text{d}}

\def\tdecoh{t_\text{d}}
\def\tform{t_\text{f}}

\newcommand{\onehalf}{{\nicefrac{1}{2}}}

\newcommand{\onethird}{{\nicefrac{1}{3}}}

\newcommand{\kT}{k_{\text{\tiny T}}}
\newcommand{\pT}{p_{\text{\tiny T}}}

\newcommand{\zg}{\ensuremath{z_\text{g}}}

\begin{document}

\title{Novel tools and observables for jet physics in heavy-ion collisions}

\author[1]{Harry Arthur Andrews}
\author[2,3]{Liliana Apolinario}
\author[4]{Redmer Alexander Bertens}
\author[5,6]{Christian Bierlich}
\author[7,8]{Matteo Cacciari}
\author[9]{Yi Chen}
\author[10]{Yang-Ting Chien}
\author[11,4]{Leticia Cunqueiro Mendez}
\author[12]{Michal Deak}
\author[9]{David d'Enterria}
\author[13]{Fabio Dominguez}
\author[14]{Philip Coleman Harris}
\author[12]{Krzysztof Kutak}
\author[14]{Yen-Jie Lee}
\author[15,16]{Yacine Mehtar-Tani}
\author[17]{James Mulligan}
\author[18]{Matthew Nguyen}
\author[19]{Chang Ning-Bo}
\author[20]{Dennis Perepelitsa}
\author[21]{Gavin Salam\footnote{On leave from CNRS, UMR 7589, LPTHE, F-75005, Paris, France and from Rudolf Peierls Centre for Theoretical Physics, 1 Keble Road, Oxford OX1 3NP, UK.}}
\author[22]{Martin Spousta}
\author[2,3,21]{Jos\'e Guilherme Milhano}
\author[21]{Konrad Tywoniuk}
\author[23]{Marco Van Leeuwen}
\author[24,25]{Marta Verweij}
\author[13]{Victor Vila}
\author[21]{Urs A. Wiedemann}
\author[2,21]{Korinna C. Zapp}

\affil[1]{University of Birmingham, Birmingham B15 2TT, United Kingdom}
\affil[2]{LIP, Av. Prof. Gama Pinto, 2, P-1649-003 Lisboa , Portugal}
\affil[3]{Instituto Superior T{\' e}cnico (IST), Universidade de Lisboa, Av. Rovisco Pais 1, 1049-001, Lisbon, Portugal}
\affil[4]{University of Tennessee, Knoxville, TN, USA-37996}
\affil[5]{Dept. of Astronomy and Theoretical Physics, S{\" o}lvegatan 14A, S-223 62 Lund, Sweden}
\affil[6]{Niels Bohr Institute, Blegdamsvej 17, 2100 Copenhagen, Denmark}
\affil[7]{Universit\'e Paris Diderot, F-75013 Paris, France}
\affil[8]{Laboratoire de Physique Th\'eorique et Hautes Energies (LPTHE), UMR 7589 CNRS \& Sorbonne Universit\'e, 4 Place Jussieu, F-75252, Paris, France}
\affil[9]{EP Department, CERN, 1211 Geneva, Switzerland}
\affil[10]{Center for Theoretical Physics, Massachusetts Institute of Technology, 77 Massachusetts Ave, Cambridge, MA 02139, USA}
\affil[11]{Oak Ridge National Laboratory, Oak Ridge, Tennessee, USA}
\affil[12]{Instytut Fizyki J{\c a}drowej (PAN), ul. Radzikowskiego 152, 31-342 Krak{\' o}w, Poland}
\affil[13]{Instituto Galego de F{\' i}sica de Altas Enerx{\' i}as IGFAE, Universidade de Santiago de Compostela, Santiago de Compostela, 15782 Galicia, Spain}
\affil[14]{Massachusetts Institute of Technology, 77 Massachusetts Ave, Cambridge, MA 02139, USA}
\affil[15]{Institute for Nuclear Theory, University of Washington, Box 351550, Seattle, WA 98195-1550, USA}
\affil[16]{Physics Department, Brookhaven National Laboratory, Upton, NY 11973, USA}
\affil[17]{Yale University, New Haven, Connecticut, USA}
\affil[18]{Laboratoire Leprince-Ringuet, Ecole polytechnique, CNRS/IN2P3, Universit\'e Paris-Saclay, Palaiseau, France}
\affil[19]{Institute of Theoretical Physics, Xinyang Normal University, Xinyang, Henan 464000, China}
\affil[20]{University of Colorado Boulder, Boulder, CO 80309, USA}
\affil[21]{Theoretical Physics Department, CERN, 1211 Geneva 23, Switzerland}
\affil[22]{Charles University, V Holesovickach 2, 180 00 Prague, Czech Republic}
\affil[23]{Institute for Subatomic Physics, Utrecht University/Nikhef, Utrecht, Netherlands}
\affil[24]{Departement of Physics and Astronomy, Vanderbilt University, Nashville, TN 37235, USA}
\affil[25]{RIKEN BNL Research Center, Brookhaven National Laboratory, Upton, NY 11973, USA}

\maketitle
\newpage

\begin{abstract}
Studies of fully-reconstructed jets in heavy-ion collisions aim at extracting thermodynamical and transport properties of hot and dense QCD matter. Recently, a plethora of new jet substructure observables have been theoretically and experimentally developed that provide novel precise insights on the modifications of the parton radiation pattern induced by a QCD medium. This report, summarizing the main lines of discussion at the 5th Heavy Ion Jet Workshop and CERN TH institute {\sl ``Novel tools and observables for jet physics in heavy-ion collisions''} in 2017, presents a first attempt at outlining a strategy for isolating and identifying the relevant physical processes that are responsible for the observed medium-induced jet modifications. These studies combine theory insights, based on the Lund parton splitting map, with sophisticated jet reconstruction techniques, including grooming and background subtraction algorithms.
\end{abstract}

\begin{flushright}
CERN-TH-2018-186, LU-TP 18-14, IFJPAN-IV-2018-8, MCNET-18-19
\end{flushright}

\tableofcontents


\section{Introduction}
\label{sec:intro}

Energetic partons, produced in high-energy hadron-hadron collisions, initiate a cascade of lower-energy quarks and gluons that eventually hadronize into collimated sprays of colorless hadrons called jets. Monte Carlo (MC) event generators of $e^+e^-$ and pp collisions, such as PYTHIA \cite{Sjostrand:2007gs}, describe reasonably well both the perturbative cascade, dominated by soft gluon emissions and collinear parton splittings, as well as the final hadronization via non-perturbative models at the end of the parton shower below some cutoff scale of the order of $1$ GeV. In heavy ion collisions at LHC and at RHIC, essentially all hadronic high-$\pT$ observables deviate from baseline measurements in proton-proton collisions. The totality of these findings is generically referred to as ``jet-quenching''. It was established first at RHIC on the level of single inclusive hadron spectra and high-$\pT$ hadron correlations. With the higher center of mass energies reached in nucleus-nucleus collisions at CERN's LHC, the focus has now moved to characterizing jet quenching in multi-hadron final states identified by modern jet finding algorithms.

There is overwhelming evidence that these jet quenching phenomena arise predominantly from final state interactions of the high-$\pT$ partons in scattering processes within the dense QCD matter produced in the collision region. Initial state effects such as nuclear modifications of parton distribution functions are well constrained and play a non-negligible but generally sub-dominant role. As a consequence, dynamical models of jet quenching focus mainly on the medium-induced modifications of final state parton showers. 

For the central aims of the relativistic heavy-ion programs at LHC and at RHIC, jet quenching is of interest for at least two reasons. First, experimental access to information about the dense QCD matter produced in heavy ion collisions can be obtained from the scattering of calibrated hard probes inside the medium. 
In this respect, jet physics has been extensively constrained during decades of high-energy collider experiments and is a workhorse for studying perturbative and non-perturbative aspects of QCD \cite{Ellis:1991qj}.
Second, since high-$\pT$ partons are far out-of-equilibrium, characterizing their medium-induced softening and isotropization may give access to the QCD equilibration processes and thereby help to understand whether, how and on which time scale heavy ion collisions evolve towards equilibrium. 

Jet quenching models, i.e., dynamical models of jet-medium interactions and the ensuing modifications of final state parton showers, are essential to 
infer, from the measurements of quenched jets, information about dense QCD matter and the equilibration processes that lead to it. Obviously, firm 
conclusions about QCD matter properties are only those that are independent of model-specific details, that are consistent with QCD theory and that are---within the range of validity of the dynamical model---consistent with the totality of experimental data. This necessitates the benchmarking of jet quenching models to establish where various models differ, and how experimental data can best discriminate between them.

With the aim of clarifying differences and similarities between various theoretical implementations of jet quenching on the level of single-hadron spectra, such a benchmarking exercise was organized by the TECHQM collaboration \cite{TECHQM}. For the success of this exercise, it was crucial to identify a simplified 
benchmark calculation that would allow for a comprehensive comparison easily performed in all model set-ups. For the case of single inclusive
high-$\pT$ hadron spectra, this joint effort was the so-called ``brick problem'' \cite{Armesto:2011ht,Burke:2013yra}.

In this report, we consider how to extend such studies to  medium-modified multiparticle final states. Our approach summarises the consensus view arrived at in the CERN TH Institute \textsl{``Novel tools and observables for jet physics in heavy-ion collisions''} \cite{THinst}, and uses the kinematic Lund plane to study how hadronic fragments are distributed in different quenching models. 
This representation provides a common language to discuss features of final state showering on an operational level that provides a basis for comparing qualitative and quantitative features of theoretical models and physical observables. We also discuss the closely related jet substructure observables with the help of modern grooming (and tagging) techniques.
The focus of the ensuing discussion is two-fold. On the one hand, we set out to understand how and for which dynamical reasons the fragment distributions of different jet quenching models show marked differences. Second, we wish to demonstrate how modern jet substructure techniques can be utilized to focus on particular regions in the Lund plane, thus providing a possibility of constructing jet observables that discriminate optimally between different models. 

The main body of numerical results shown in this report are based 
on  the Monte Carlo (MC) event generators PYTHIA 8 \cite{Sjostrand:2007gs}  (Monash 2013 tune) for simulating the proton-proton baseline, and on the 
codes QPYTHIA \cite{Armesto:2009fj} and JEWEL (v.2.2.0) \cite{Zapp:2011ya,Zapp:2012ak} for simulating in-medium jet evolution at center-of-mass energy per nucleus-nucleus collisions $\sqrt{s_{NN}}=$ 5.02 TeV.\footnote{The generated samples used in the analyses can be recovered at \url{https://twiki.cern.ch/twiki/bin/viewauth/JetQuenchingTools/PU14Samples}.}
For jet reconstruction we made extensive use of FastJet \cite{Cacciari:2005hq,Cacciari:2011ma} and for the purposes of additional pile-up mitigation in heavy-ion context, we mainly used constituent subtraction (CS) \cite{Berta:2014eza}, see \autoref{app:background} for more details. A suite of useful analysis tools were prepared in GitHub repository \cite{cerninstitute-github} and a TWiki page was also set up \cite{cerninstitute-twiki}.

It would have been of obvious interest to include in this comparison on equal par other existing jet quenching Monte Carlos, such as MARTINI \cite{Schenke:2009gb,Young:2011ug}, LBT \cite{Wang:2013cia,He:2015pra}, HYDJET++ \cite{Lokhtin:2008xi}, YaJEM \cite{Renk:2010zx} and MATTER \cite{Majumder:2013re}. 
Nevertheless, a fully comprehensive comparison is beyond the scope of the current endeavor.
The numerical studies included in this report are therefore illustrative for the general strategy of jet model comparison 
but they are not exhaustive given that they are based only on a small number of simulations done with a subset of all existing tools. In the coming years,
we  foresee the further development of existing jet quenching models, and the advent of new ones (such as those envisaged in the framework of 
JETSCAPE \cite{Cao:2017zih}). The main aim of this write-up is to formulate a simple, generally applicable strategy for characterizing differences
between jet quenching models and devising observables that allow one to best discriminate among them. The report is organized as follows.

 In \autoref{sec:phasespace}, we introduce for the first time, in the context of heavy-ion studies, the concept of a splitting map, based on the kinematical Lund diagram \cite{Andersson:1988gp}.\footnote{During the preparation of this report, a similar approach was discussed in \cite{Chien:2018dfn}.}
 This map provides a representation of the radiation pattern implemented in Monte Carlo showering algorithms and allows to directly compare their features in different kinematical regimes. It also provides a direct, visual impression of what phase space region is being most significantly modified by medium effects. In detail, 
\begin{description}

\item[\autoref{sec:phasespace-theory}] gives a brief introduction to theoretical concepts that are useful for understanding the Lund diagram on the level of a single splitting, both for vacuum showers and showers in the medium.

\item[\autoref{sec:iterative-declustering}] describes in detail the procedure to fill the splitting map, by describing the steps related to jet reclustering and calculation of the variables that go into the map. 

\item[\autoref{sec:phasespace-mc}] presents a study of the splitting maps of in-medium MC parton showers, QPYTHIA and JEWEL. 
We refer the interested reader to Appendix~\ref{app:models} for further details on the MC's utilized in the studies presented below. 
Finally, in \autoref{sec:uncorrelatedbackground}, we study the resilience of the observed   features at generator level to uncorrelated background by embedding the jet samples into a realistic heavy-ion environment.

\end{description}
While the splitting map contains the maximal amount of information, since it convolves the kinematics of every splitting, it is also amenable to more selective examinations, for instance, through the implementation of jet ``tagging'' and ``grooming'' procedures. These tools are extensively used in the pp collider community \cite{Bendavid:2018nar} for a wide range of purposes, spanning jet substructure studies and leveraging this control for studies of observables beyond pure QCD, see e.g. \cite{Larkoski:2017jix,Asquith:2018igt,Marzani:2019hun} for the most recent reviews. They have also been previously applied in Monte Carlo studies for heavy-ion collisions \cite{Zhang:2015trf,Apolinario:2017qay,Milhano:2017nzm}. 
In studying concrete observables, we have mainly focussed on applying the so-called Soft Drop (SD) grooming procedure with $\beta = 0$ (equivalent to the so-called modified mass-drop tagging) \cite{Dasgupta:2013ihk} and $\beta \neq 0$ \cite{Larkoski:2014wba}, to be detailed below, that aims at identifying the first hard jet branching. 
Hence, in the second part of the report, \autoref{sec:jetsubstructure}, we perform a set of MC studies, on generator level and including embedding into a realistic heavy-ion background, using state-of-the-art grooming techniques. These observables are  not limited to substructure but are also used in order to extract more differential aspects from inclusive jet observables. In detail, 
\begin{description}
\item[\autoref{sec:groomedobservables}] presents the result for the groomed momentum-fraction $z_g$, subjet angle $\Delta R_{12}$ and the groomed mass  $M_g$ for QPYTHIA and JEWEL using three grooming settings. In this section, the studies were performed without embedding in a realistic background. We shed more light on the robustness of these results by checking the size of hadronization effects in the $z_g$ distribution for three different SD settings \autoref{sec:hadronization}. 
\item[\autoref{sec:dissecting}] suggests a complementary look on substructure by submitting the jet sample that goes in to constructing a fully inclusive observable to an additional grooming step. Concretely, we propose to bin the inclusive jet sample in terms of the angular separation of the hardest subjets, $\Delta R_{12}$, using SD grooming. We demonstrate this procedure on QPYTHIA and JEWEL samples for the nuclear modification factor $R_{AA}$ and the photon-jet momentum imbalance in terms of the variable $x_{J\gamma}$. For these observables we also took into account embedding into a heavy-ion background \cite{deBarros:2012ws}.
\end{description}
Finally, we wrap up with an outlook in \autoref{sec:outlook}.


\section{Mapping the splittings of in-medium jets}
\label{sec:phasespace}

\subsection{Theoretical considerations}
\label{sec:phasespace-theory}

Jets are multi-particle objects and are experimentally accessed by assembling measured tracks or calorimeter energy depositions, or a combination of both, according to a jet algorithm that, ideally, is infrared and collinear (IRC) safe \cite{Buttar:2008jx}. In the context of perturbative QCD, multiple splittings inside the jet cone have to be taken into account due to the mass singularity of QCD. 
In the medium, these splittings happen concurrently with, and are affected by, final-state interactions with the surrounding medium. It is therefore worth considering how medium scales, related to the size\footnote{For simplicity, in this section we treat the medium as static, where all jets traverse the same length $L$.} and local medium properties, will have an impact on the variety of jet observables. 

\begin{figure}
\centering
\includegraphics[width=0.49\textwidth]{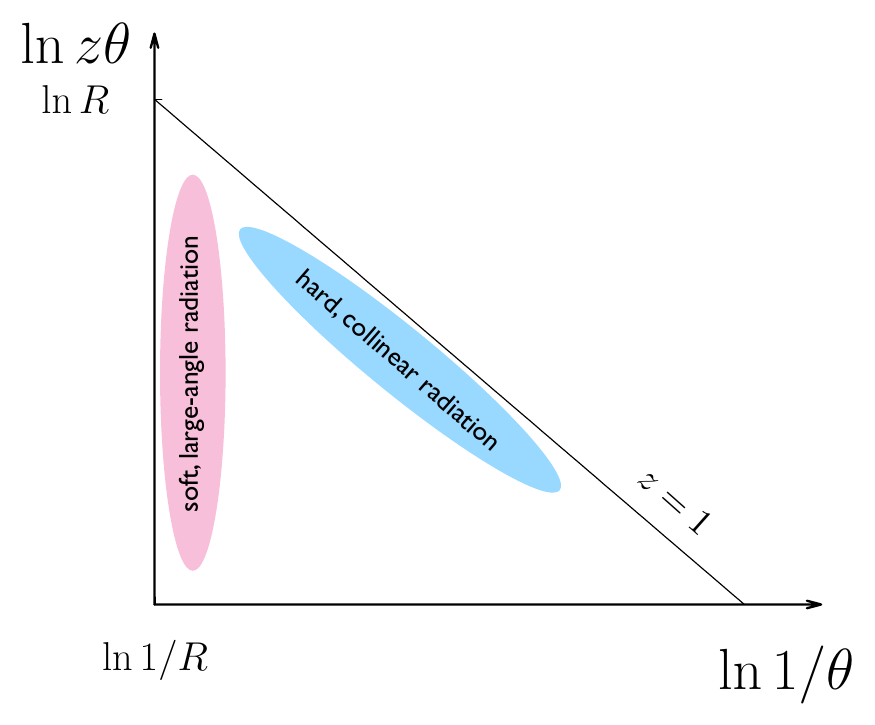}
\includegraphics[width=0.49\textwidth]{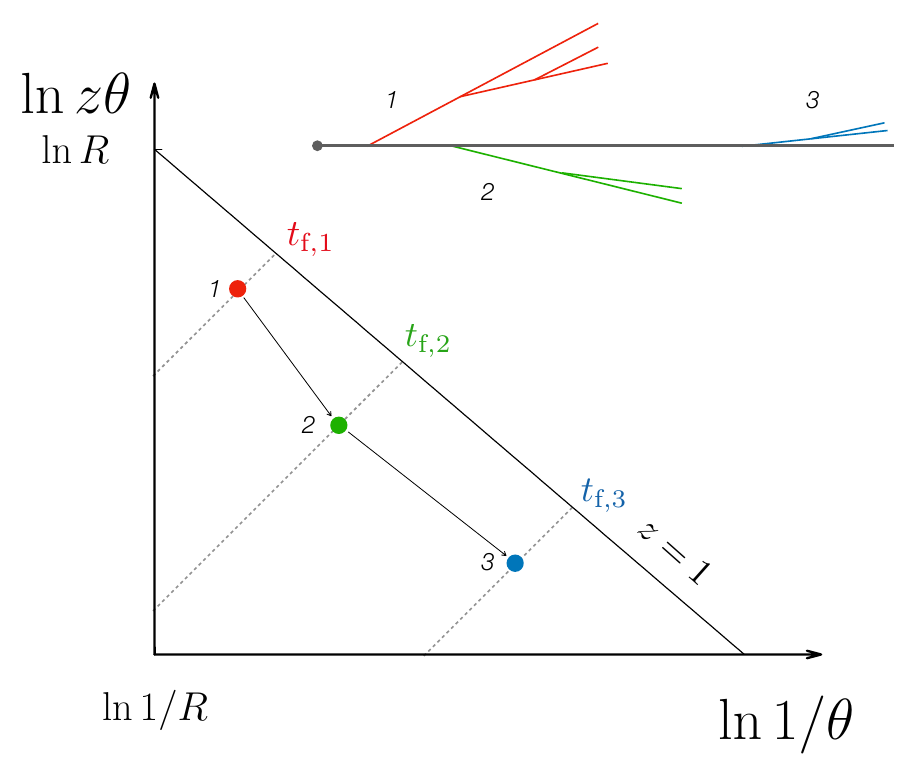}
\caption{Left: The kinematical Lund plane spanned by $\ln 1/\theta$ and $\ln z\theta$ for jets with opening angle $R$, see text for details. Right: clustering history with the formation time of {\sl primary} emissions in the kinematical Lund plane.}
\label{fig:PS0}
\end{figure}

In this context, it is very useful to introduce the kinematical Lund diagram \cite{Andersson:1988gp} for an arbitrary $1\to 2$ splitting process. Working in the small-angle limit, and denoting by $z$ the energy sharing fraction and $\theta$ the dipole splitting angle, we find that the dipole invariant mass reads
\beq
\label{eq:DipoleMass}
M^2 = z(1-z) \pT^2 \theta^2 \,,
\eeq
where $\pT$ is the total momentum of the dipole, corresponding roughly to the transverse momentum in the detector. The characteristic time-scale of the splitting is usually referred to as the \textsl{formation time}, and is related to the finite energy resolution, $\tform \sim \Delta E^{-1}$. It is explicitly given by
\beq
\label{eq:FormationTime}
\tform = \frac{2 z(1-z)\pT }{\kT^2} = \frac{2 \pT}{M^2} \,,
\eeq
where $\kT = z(1-z) \pT \theta$ is the (relative) transverse momentum of the dipole in the small angle limit. 
This formula can easily be understood as the time-scale for decaying in the rest frame of the parent times its boost factor $\sim (1/M) \times (\pT/M)$. 

The Lund diagram exists in various forms, the common feature being that the variables spanning the plane exploit the logarithmic phase space due to the soft, $1/z$, and collinear, $1/\theta$, divergences of typical QCD splittings (except $g \to q\bar q $).
Here, the phase-space for emission from each particle is represented in the Lund map, as a triangle in a $\ln 1/\theta$ and $\ln \kT/\pT$ plane, where $\theta$ and $\kT$ are respectively the angle and transverse momentum of an emission with respect to its emitter.
In the soft and collinear limit, usually referred to as the double-logarithmic regime, the differential probability $\dd P$ of one splitting is given by \cite{Dokshitzer:1991wu,Ellis:1991qj}
\beq
\label{eq:vacuum-phase-space}
\dd P = 2\frac{\alpha_s C_i}{\pi}\, \dd \ln z\theta \, \dd \ln\frac{1}{\theta} \,,
\eeq
in terms of its kinematical variables, and approximating $\kT \approx z \pT \theta$, and in arbitrary color representation ($C_i=C_F$ for quark and $C_i = N_c$ for gluon splitting, respectively). Due to the self-similar nature of the phase space \eqref{eq:vacuum-phase-space}, the effect of {\sl multiple} splittings corresponds to higher-order corrections.\footnote{For a given phase space point with $\{z,\theta\}$, the feed-down contribution from one additional splitting yields a contribution $\sim \mathcal{O}\big(\alpha_s^2 \ln1/z\, \ln R/\theta\big)$ which contributes at higher-logarithmic order.} Hence, the area spanned below the line $z=1$, see \autoref{fig:PS0} (left), is uniformly populated by emissions with the weight $2\alpha_s C_i/\pi$ where emissions can take place up to the jet opening angle $R$. The density at fixed $\kT$ is mainly modulated by running coupling effects, down to the QCD scale, $\kT \sim \Lambda_\text{\tiny  QCD}$, where non-perturbative effects will dominate. 
In \autoref{fig:PS0} (left) we have also explicitly denoted the regimes of soft, large-angle  and hard, collinear radiation. 

For a full-fledged jet, the diagram is built up by mapping every branching to a point on the Lund plane. In \autoref{fig:PS0} (right), we illustrate how to fill this plane for multiple, {\sl primary} emissions off the main branch---three in the illustrated case. Resolving subsequent splittings along the primary branches will, in turn, generate new, orthogonal Lund planes, and so on. Simplifying the graphical representation, these planes are ultimately collapsed onto the original one. Due to the strong ordering of emissions in the double-logarithmic regime, especially in angles, the emissions are roughly evenly distributed in the splitting variables.

Each point in the diagram is uniquely associated with a specific formation time $\tform$ that is a proxy for the physical time $t$ when a splitting can be resolved. Solving $t = \tform$, results in a line parametrized by
\beq
\label{eq:FormationTimeLund}
\ln z \theta = \ln \frac{1}{\theta} + \ln \frac{1}{\pT t} \,,
\eeq
with a positive, unit slope in the plane. Hence, each splitting in \autoref{fig:PS0} (right) happens approximately at $t_{\text{f},i}$ ($i=\{1,2,3\}$). This representation is particularly useful when considering branchings in the presence of a spatially extended density of scattering centers.

\begin{figure}
\centering
\includegraphics[width=0.6\textwidth]{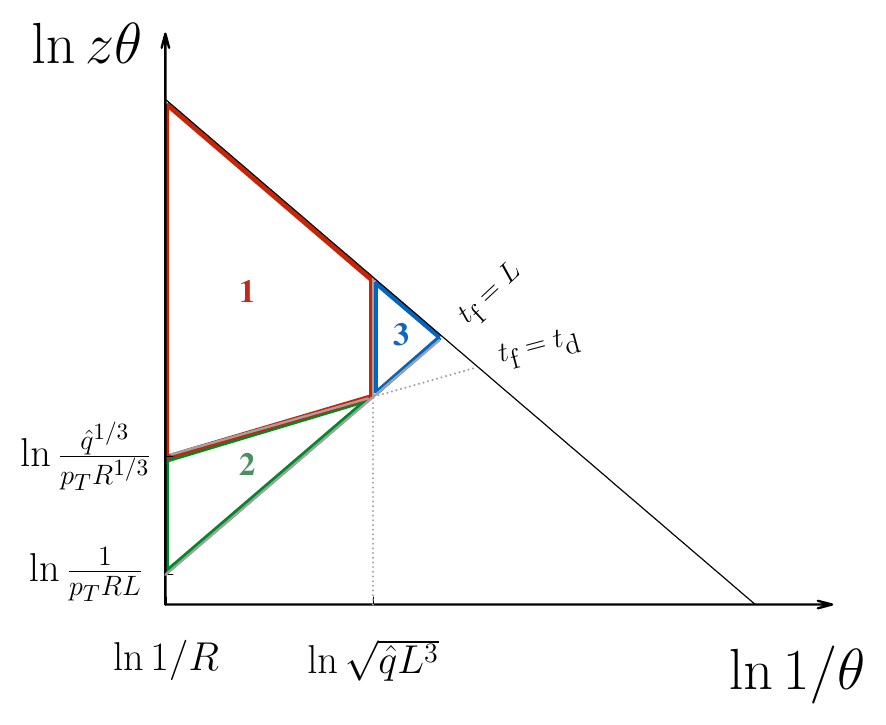}%
\caption{Lund diagram of parton splittings with the inclusion of relevant medium scales related to creation and decoherence of partons in the medium. The chosen parameters are $\hat q = $ 2 GeV$^2$/fm, L = 2 fm and $p_{\scriptscriptstyle T}$ = 300 GeV.}
\label{fig:PS1}
\end{figure}

Turning now to medium effects, it is most natural to consider which of the splittings happen inside a medium of length $L$.  The line corresponding to $\tform = L$ is found by substituting $t = L$ in \eqref{eq:FormationTimeLund}, and is also represented in \autoref{fig:PS1}, where we have chosen a particular value for $L$. Hence, the area above the line marked $\tform = L$ corresponds to emissions that occur inside the medium. Emissions with $\tform > L$ occupy the region below the line.

Providing a comprehensive overview of models of medium interactions proposed in the literature is beyond the scope of this report. Instead, we consider for the moment a well-known picture that shares commonalities between a wide class of approaches by assuming that all propagating particles experience diffusive momentum broadening. The amount of accumulated momentum is characterized by the diffusion relation $\langle \kT^2 \rangle = \hat q t$, where $t$ corresponds to the time of in-medium propagation and where the jet transport coefficient $\hat q$ acts as a diffusion constant in transverse space.\footnote{Here we neglect the influence of rare, hard kicks in the medium that go beyond this definition. Their discussion follows closely what we describe below, with the resolution scale $\lambda_\perp \sim q_\perp^{-1}$, where $q_\perp$ is the transverse momentum kick from the medium. We refer, e.g. to \cite{Kurkela:2014tla} for a comprehensive discussion.} 
For a given splitting, with a given transverse momentum $\kT$ and formation time $\tform$, the accumulated transverse momentum $\sim \hat q \tform$ could either be a small correction or a dominating contribution. The limiting line, $\kT^2 = \hat q \tform$, is parametrized by,
\beq
\ln z\theta = \frac{1}{3} \ln \frac{1}{\theta} + \ln \frac{\hat q ^\onethird}{\pT} \,.
\eeq
Note that the slope is a factor $1/3$ smaller than in \eqref{eq:FormationTimeLund}.
More generally, at any instant $t$ we can compare the intrinsic transverse momentum $\kT^2 \sim (\theta t)^{-2}$ to the accumulated one $\sim \hat q t$. This allows us to identify a characteristic time-scale when the two are of the same order that is usually referred to as the {\sl decoherence} time
\beq
\tdecoh \sim (\hat q \theta^2)^{-1/3} \,.
\eeq
Hence, the line $\tform = \tdecoh$, indicated in \autoref{fig:PS1}, divides the region above, where emissions are not resolved by medium interactions, from the region below, where the dipole splitting kinematics is dominated by diffusion.
Also note that this broadening-dominated regime ceases to exist for decoherence times longer than the medium length, $\tdecoh \gtrsim L$. This corresponds to small dipole configurations that have a vanishing probability of ever being resolved in the medium. The condition $\tform = \tdecoh = L$ corresponds therefore to the minimal decoherence angle $\theta_c \sim (\hat q L^3)^{-\onehalf}$, see \autoref{fig:PS1}.

The kinematical Lund plane for {\sl one} splitting inside the medium is therefore divided into three main regimes, in addition to the possibility of fragmenting outside of the medium, $\tform > L$.
Emissions that fall within the area marked by ``1'' (red lines) in \autoref{fig:PS1}, i.e. with $\tform< \tdecoh < L$, correspond to pure vacuum splittings inside the medium. On the other hand, in-medium splittings with $\tdecoh > L$, falling within the area marked by ``3'' (blue lines) in \autoref{fig:PS1}, are never resolved by medium interactions.
Finally, whenever $\tdecoh \lesssim \tform$, see the area marked by ``2'' (green lines) in \autoref{fig:PS1}, the splitting kinematics is dominated by medium effects and the probability of splitting is no longer described by \autoref{eq:vacuum-phase-space}, i.e. the Lund plane is not expected to be uniformly filled. These splittings will also undergo further momentum broadening in the medium. For a comprehensive discussion of this regime, we refer to \cite{Kurkela:2014tla,Blaizot:2014rla}.

Generalizing this schematic picture to multiple splittings in the medium is largely an open theoretical question, with some recent progress \cite{Mehtar-Tani:2017web,Caucal:2018dla}. However, our discussion has so far not taken into consideration important effects related to in-medium transverse momentum broadening or momentum conservation in medium interactions, which so far only are modeled in state-of-the-art Monte Carlo parton shower generators, e.g. such as JEWEL \cite{Zapp:2011ya,Zapp:2012ak}. A direct comparison with fully-fledged dynamical models, employed in \autoref{sec:phasespace-mc}, is therefore not directly applicable. We believe, nevertheless, that this discussion serves as a useful sketch for separating out regimes dominated, within the scope of the current discussion, by widely different physical mechanisms. In the studies presented in this report, we have also not considered in detail the jet flavor (quark or gluon) dependence.

\subsection{Filling the map from reclustered jets}
\label{sec:iterative-declustering}

As already pointer out above, the generalization of this picture to multiple emissions is more delicate. In vacuum, subsequent emissions are self-similar (apart from the running of $\alpha_s$) which allows to iterate the splitting process with the jet opening angle R replaced by the splitting angle of the parent dipole (angular ordering) \cite{Ellis:1991qj,Dokshitzer:1991wu}. 

In order to connect theory with experimental observables, one relies on an operational definition of what a jet is \cite{Salam:2009jx}. Such procedures cluster the final-state stable particles using sequential recombination algorithms, e.g. as implemented in FastJet \cite{Cacciari:2005hq,Cacciari:2011ma}. Final state particles $i$ and $j$ are assigned a mutual distance $d_{ij}$ and a distance to the beam $d_{i\rm B}$. The pair with the smallest distance is recombined first, and the algorithm repeats until the distance to the beam is the smallest quantity. In this case, the algorithm terminates labelling $i$ a jet. The distance metric is generally defined as
\begin{align}
\label{eq:JetDistanceMetric}
d_{ij} &= \min\left(p_{{\rm \tiny T},i}^{2\alpha},p_{{\rm \tiny T},j}^{2\alpha} \right) \frac{\Delta R_{ij}^2}{R^2} \,, \\
d_{i \rm B} &= p_{{\rm \tiny T},i}^{2\alpha} \,,
\end{align}
where $\Delta R_{ij}^2 = (\Delta \phi_{ij})^2 + (\Delta y_{ij})^2$ and the choice of the $\alpha$ (integer) exponent defines the algorithm; $\Delta \phi_{ij}$ ($\Delta y_{ij}$) being the separation of the particles in azimuthal angle (rapidity). In our studies, we have used the anti-$k_{\rm \tiny T}$ algorithm ($\alpha = -1$) \cite{Cacciari:2008gp}, the Cambridge/Aachen (C/A) algorithm ($\alpha = 0$) \cite{Dokshitzer:1997in,Wobisch:1998wt}, and the $k_{\rm \tiny T}$ algorithm ($\alpha = 1$) \cite{Catani:1993hr,Ellis:1993tq}.

Given a reconstructed jet, obtained from a full heavy-ion event, with a list of constituents belonging to it, one can repeat the recombination using one of the algorithms described above. In this context, this is referred to as a reclustering of the jet, providing a complete hierarchical tree (aka ``history'') of the jet evolution. Substructure techniques, to be used extensively throughout this report, define observables based on the information organized in such a tree. 
Traditionally, the C/A algorithm is preferred since it resolves the jet as a function of its angular scale. Although the $\kT$ algorithm also could be used in these applications, it is known to be sensitive to local clusters of soft particles (so-called `junk jets') \cite{Dokshitzer:1997in}. Finally, the anti-$\kT$ algorithm will recombine particles around the hardest branch, rather than recombining individual subjets first (see, e.g., \autoref{fig:PS0} (right)) \cite{Salam:2009jx}, thus spoiling the close correspondence to a typical QCD branching history. 
We have however studied aspects using all three recombination schemes, and will discuss them briefly below. This was partly motivated by considering medium-modified branching, and associated thermalization, with no preferred ordering variable, analogous to the angular ordering in the vacuum. Nevertheless, we observe that the C/A algorithm seems to be most resilient and predictable, even in the presence of medium effects, and will be the default scheme applied to most of the studies performed below.

In order to extract relevant information from a sample of real (simulated) jets, we apply the following procedure. For a given jet in the sample,
\begin{description}

\item[1)] build a history of splittings by reclustering a jet with a given reclustering algorithm,

\item[2)] at each branching, extract the variables $z$ and $\theta$. Here, we define $z \equiv z_\text{rel} = p_{t,j_2}/(p_{t,j_1}+p_{t,j_2})$ and use $\ln z \Delta R_{j_1 j_2}/R$ as the quantity on the $y$-axis, where $j_i$ ($i=1,2$) refer to two branches of the tree. This definition of the variable $z$ has the property that it always reflects the momentum sharing within the local branching (see also the description at \cite{LundFromMC}).

\item[3)] enter the corresponding $z,\theta$ point in the Lund diagram.

\end{description}
The C/A reclustering, where the distance metric is only determined by the angular separation, see Eq.~(\ref{eq:JetDistanceMetric}), should correspond most closely to an angular-ordered sequence of splittings based on our arguments above. That means that the last step of jet reclustering merges two substructures separated at large angles. As mentioned above, alternative reclustering strategies can be used, albeit with caution. In the case of the $k_{\rm \tiny T}$-algorithm, the softest particles are clustered first. As a consequence, the last reclustering step will merge hard splittings. The anti-$k_{\rm \tiny T}$ clusters hard particles first, thus splittings at the last reclustering steps will be generally soft. 

\begin{figure}[t!]
\centering
\includegraphics[width=0.33\textwidth]{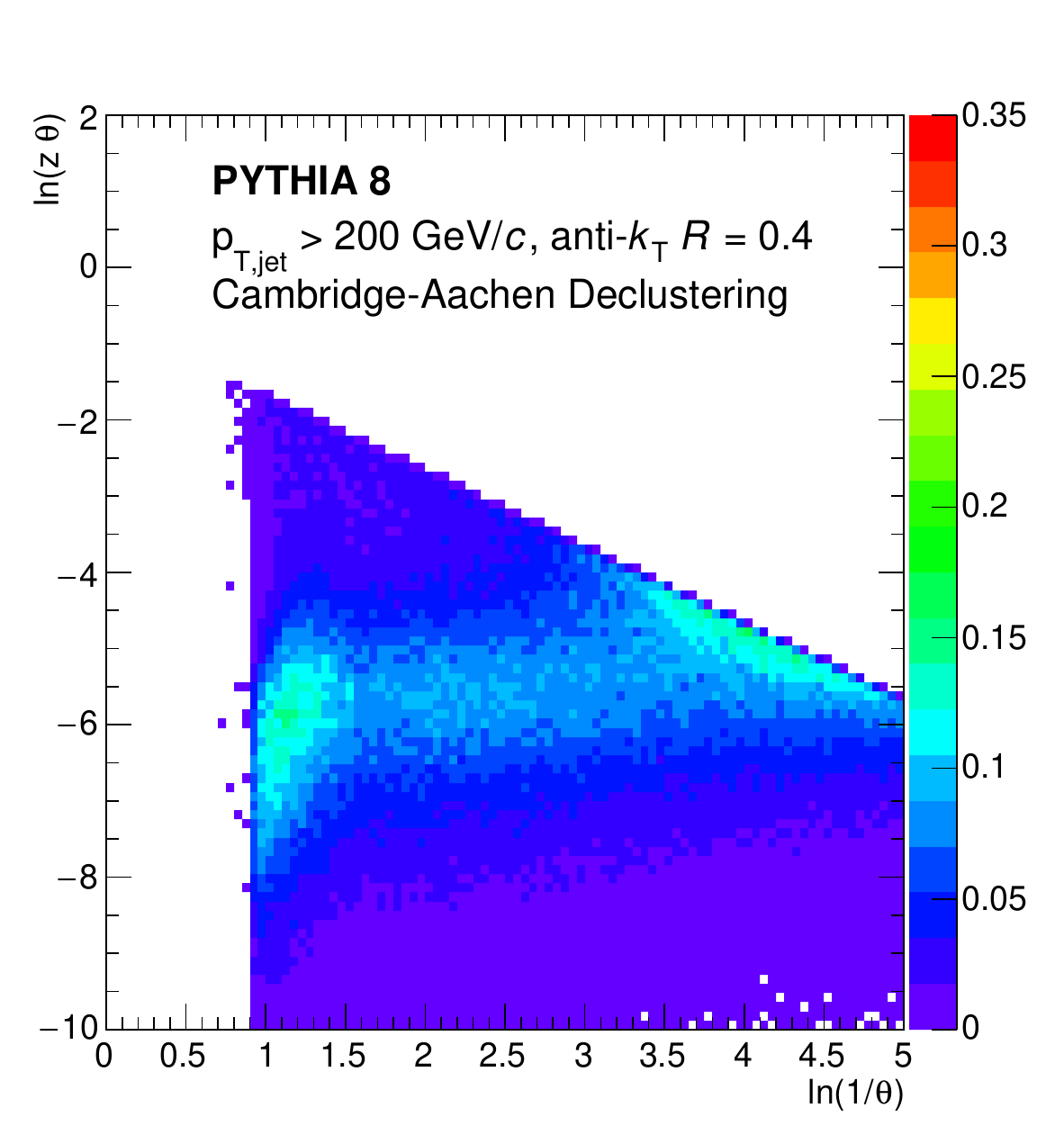}%
\includegraphics[width=0.33\textwidth]{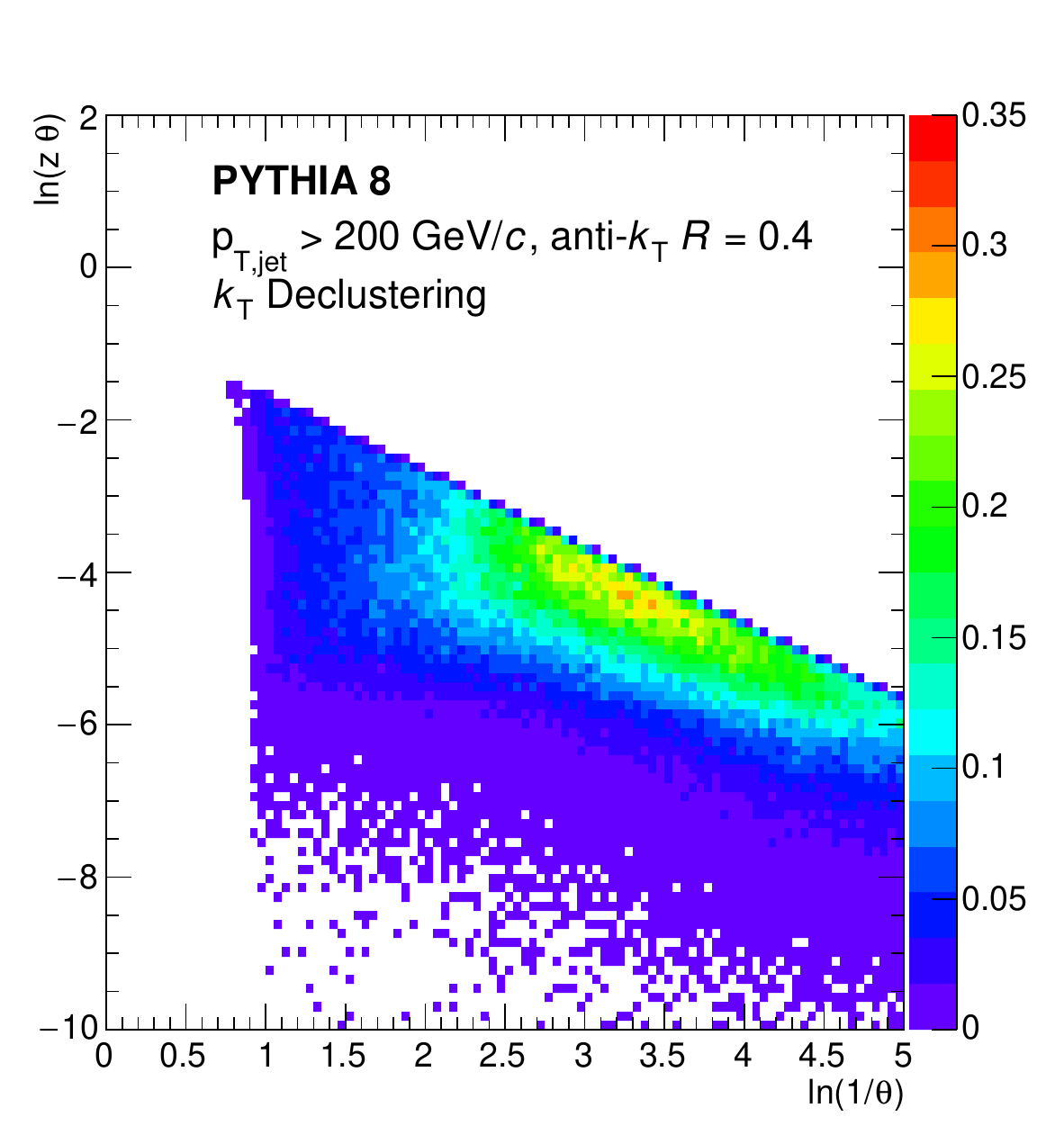}%
\includegraphics[width=0.33\textwidth]{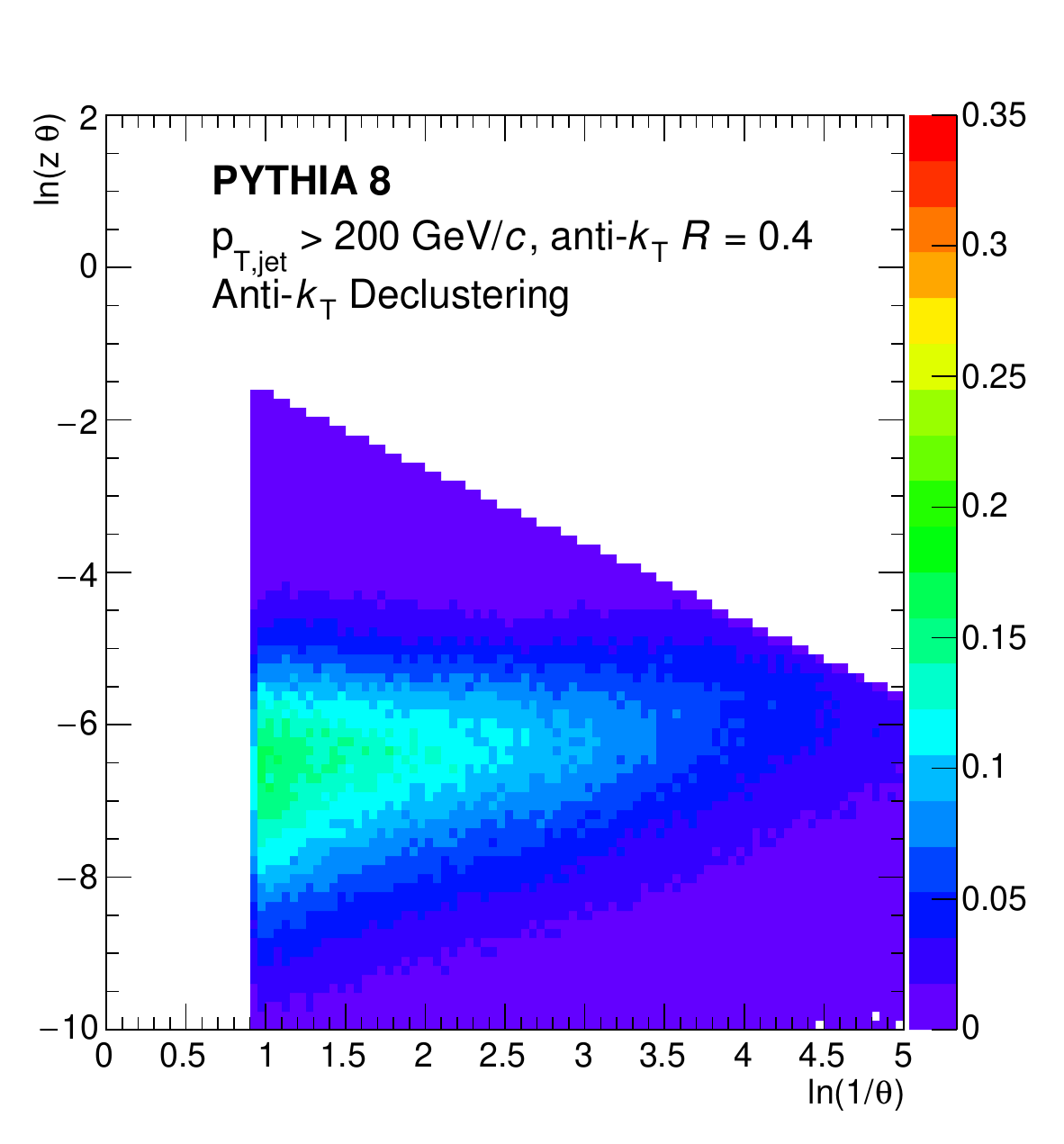}%
\caption{Lund diagrams reconstructed from a sample anti-$\kT$ $R = 0.4$ jets generated by PYTHIA8. Three reclustering strategies were considered: C/A (left), $\kT$(middle), and anti-$\kT$ (right).}
\label{fig:PS2Vac}
\end{figure}

Using this procedure for the three different reclustering algorithms, we analyzed a sample of jets generated by PYTHIA8 in \autoref{fig:PS2Vac}.\footnote{The figures produced in this report were built using G. Salam's code at \cite{LundFromMC}.
} The jet sample corresponds to reconstructed anti-$k_{\rm \tiny T} = 0.4$ jets with $\pT > 200 $ GeV. The expected, simple features are nicely realized for the C/A reclustering, see \autoref{fig:PS2Vac} (left).\footnote{Some features near the edges of the figure are artefacts related to the use of anti-$\kT$ jets. We appreciate the discussion with the authors of \cite{Dreyer:2018nbf} about these aspects.} In particular, we see a slow enhancement of radiation with increasing $\kT$ that can mainly be attributed to running-coupling effects. The additional features, e.g. at large angles, can be attributed to effects from the underlying event that was not subtracted in this sample. Indeed, the maps generated by the (anti-)$k_{\rm \tiny T}$ reclustering are not uniform and possess and enhanced sensitivity to collinear, \autoref{fig:PS2Vac} (center), and soft, large-angle configurations, \autoref{fig:PS2Vac} (right), as naively expected. See also \autoref{sec:reclusteringalgo} for a further discussion in the context of grooming studies.

As pointed out before, medium-induced radiation does not per se follow the same (angular) ordering as described above. In fact, the resummation of soft radiation leads to quite different characteristics. We will however continue to apply the procedure outlined above to identify regions of particular medium modification in the following Section. 

\subsection{Radiation phase space and sensitivity to jet quenching}
\label{sec:phasespace-mc}

\begin{figure}[h]
\centering
\includegraphics[width=0.3\textwidth]{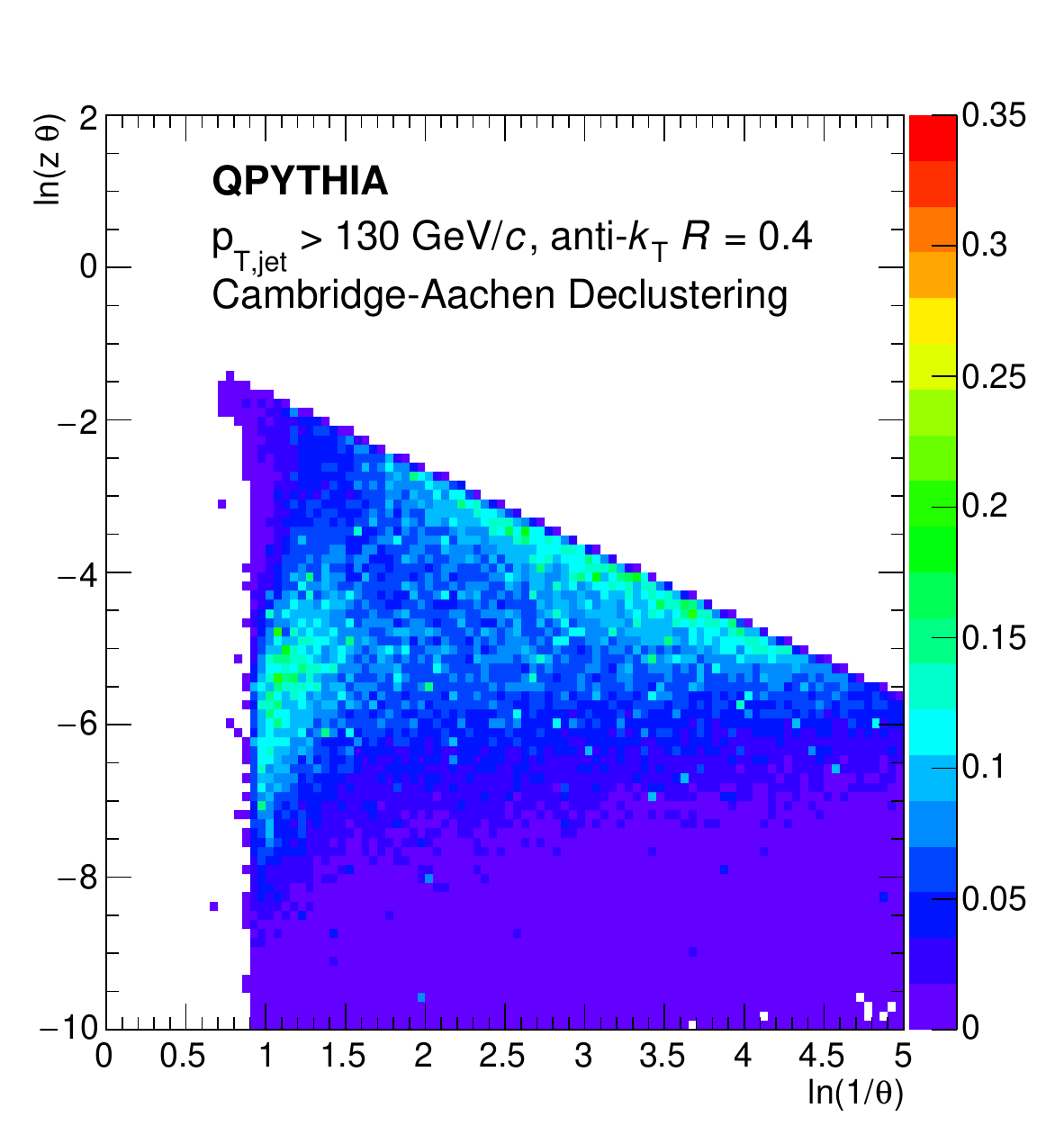}
\includegraphics[width=0.3\textwidth]{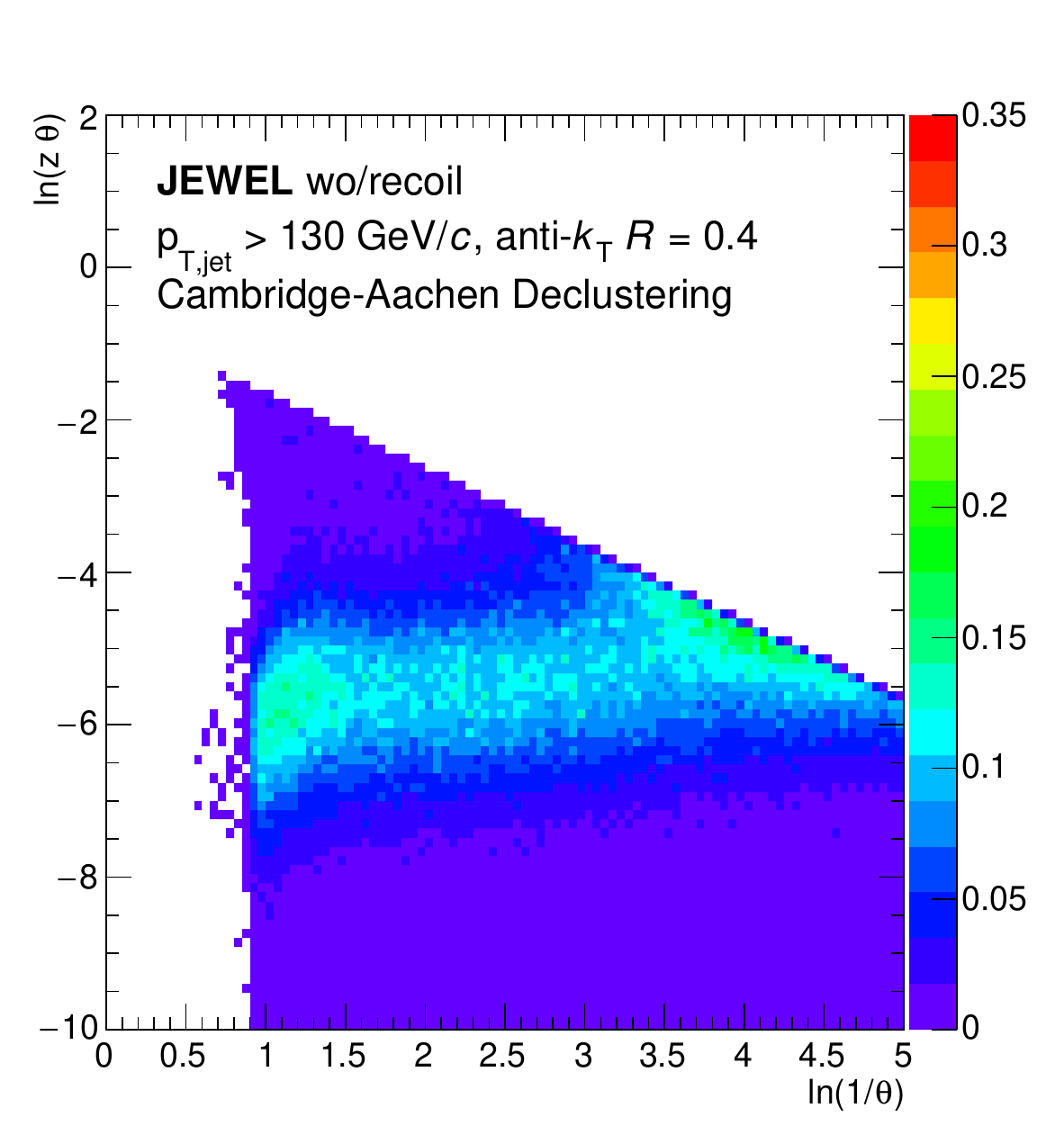}
\includegraphics[width=0.3\textwidth]{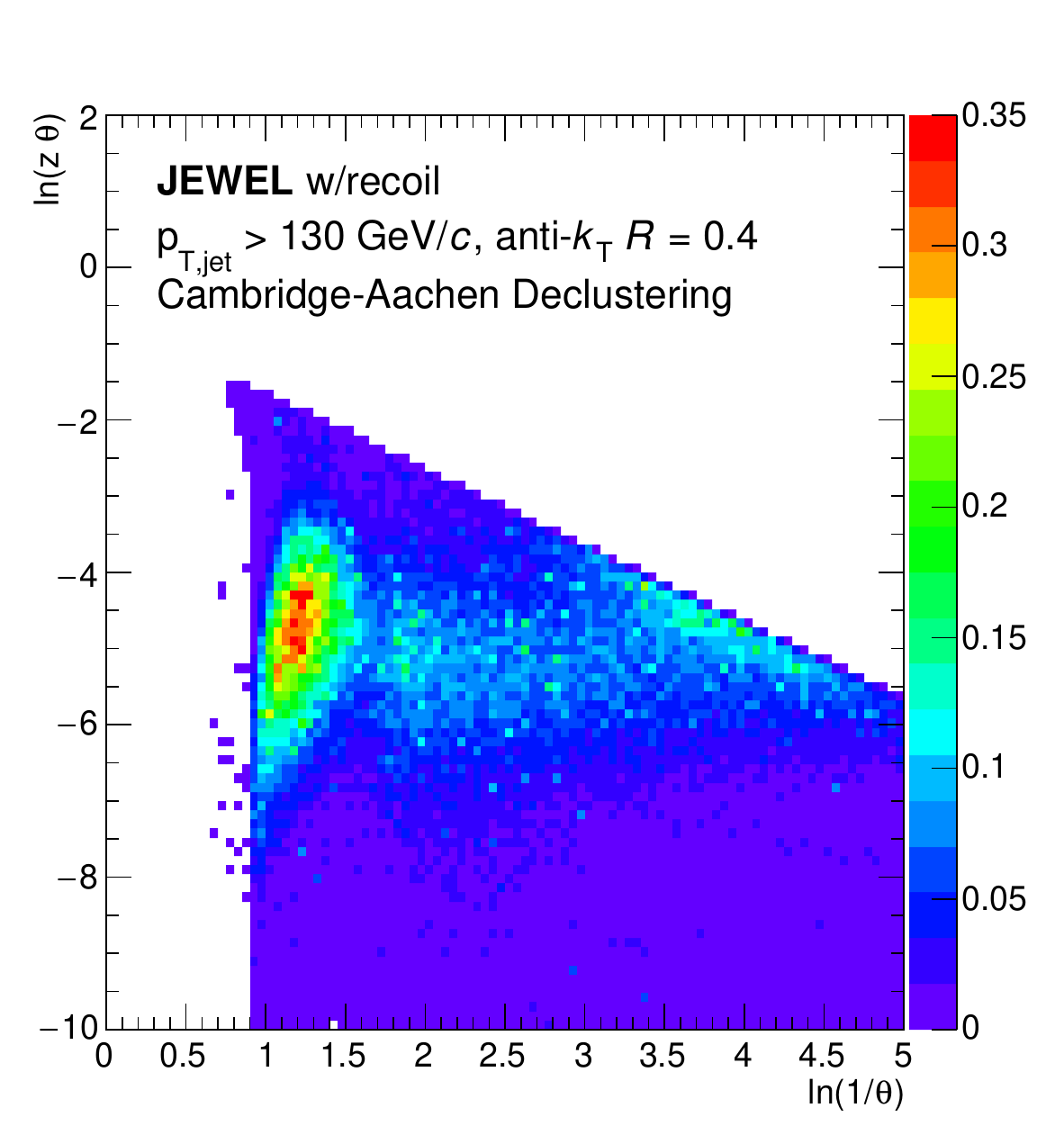}
\includegraphics[width=0.3\textwidth]{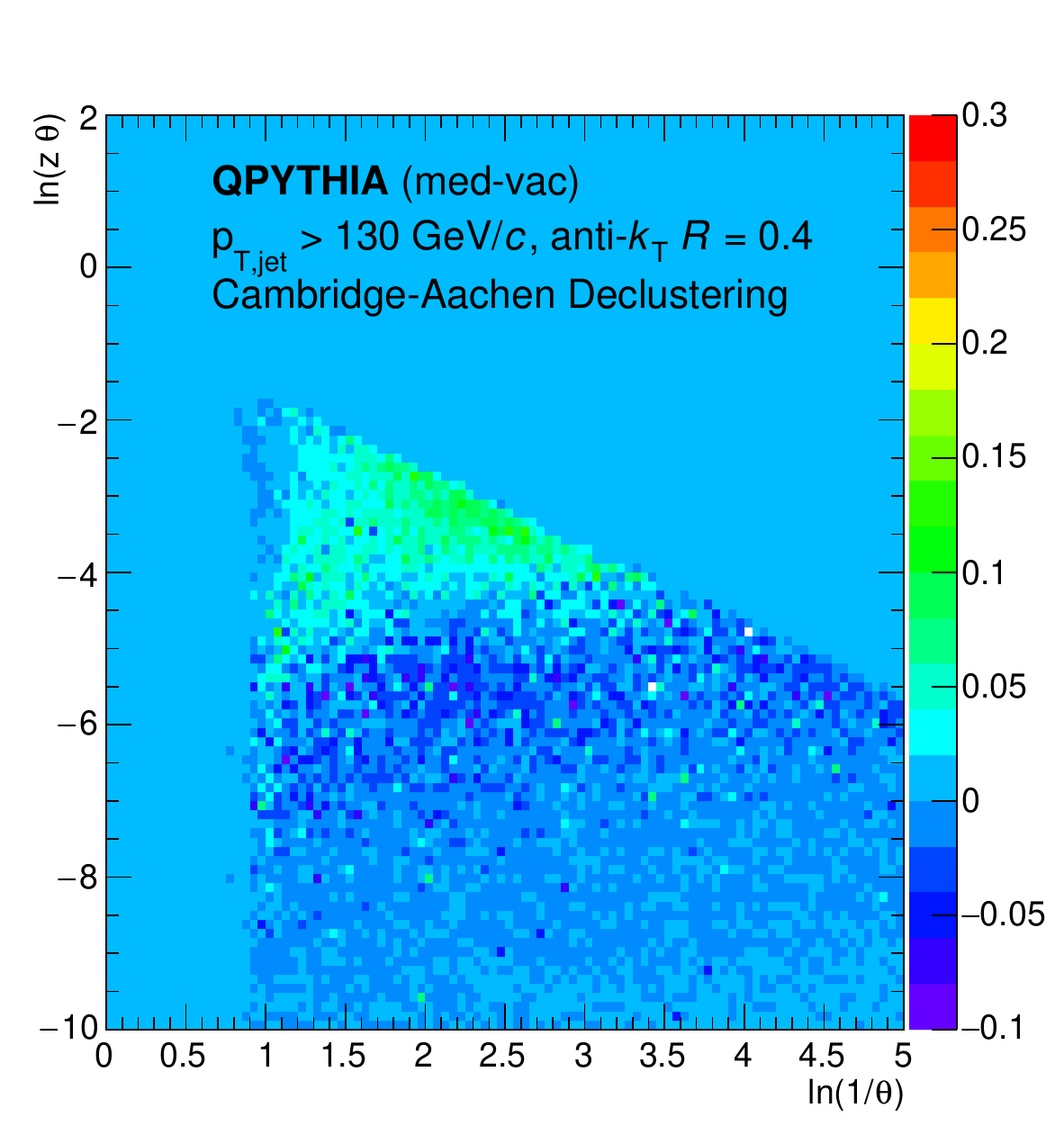}
\includegraphics[width=0.3\textwidth]{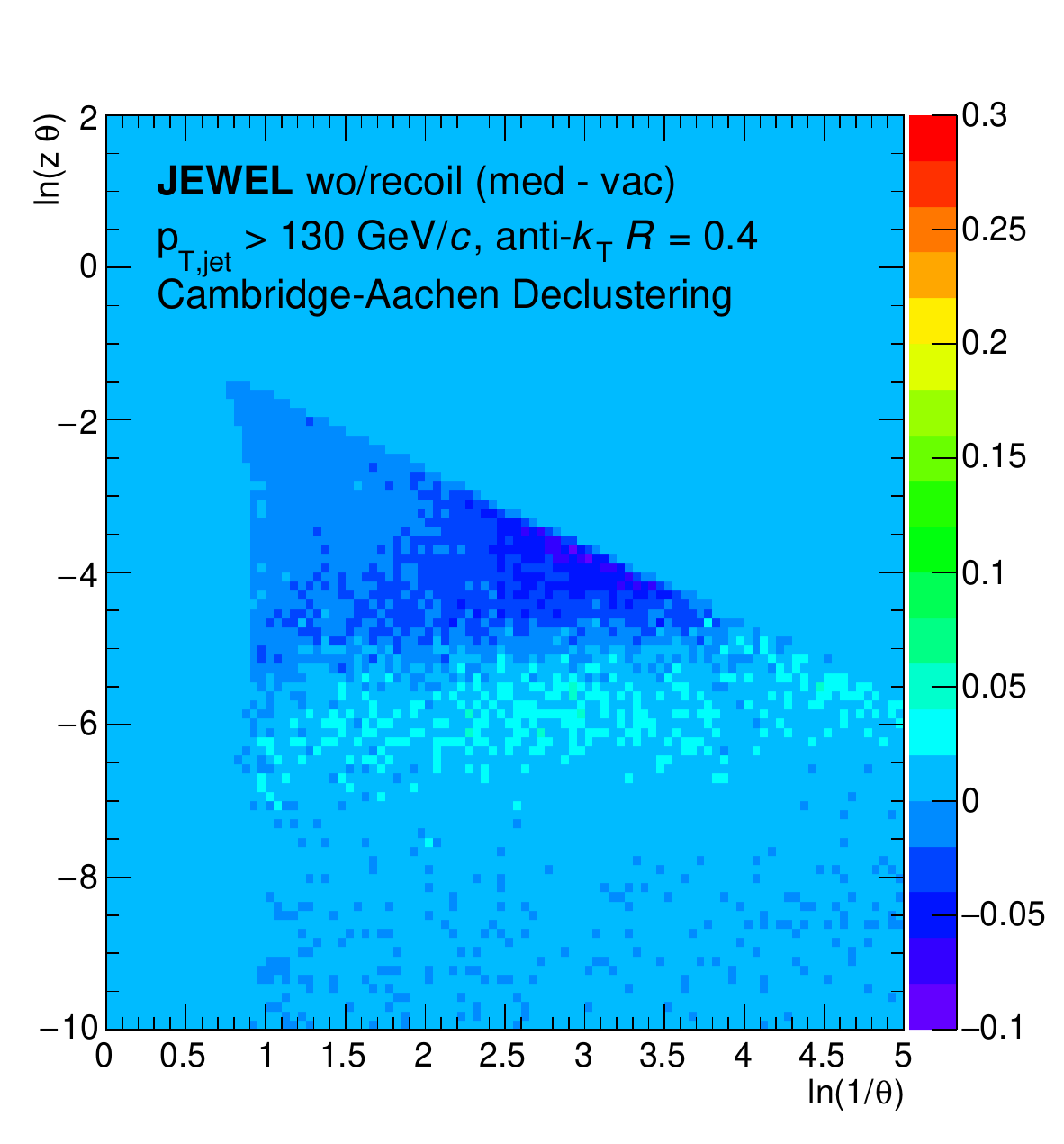}
\includegraphics[width=0.3\textwidth]{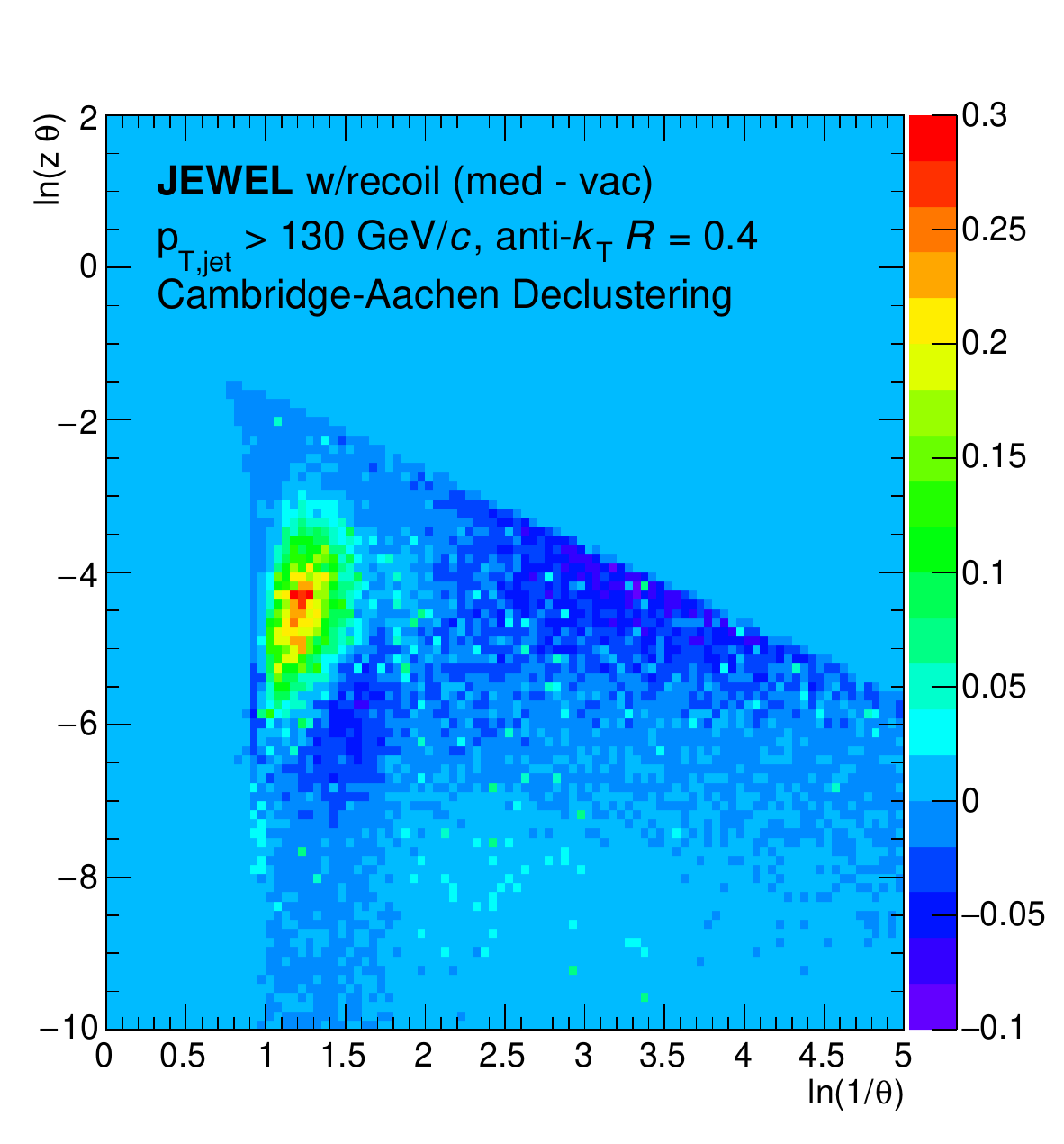}
\caption{Lund diagram reconstructed from jets generated by QPYTHIA (left column), JEWEL without recoils (middle column) and JEWEL with recoils (right column).
The lower panels correspond to the difference of the radiation pattern with and without jet quenching effects. Note that the scale of the $z$-axes varies between the panels.}
\label{fig:PS2}
\end{figure}

As a demonstration of the general ideas outlined above, we fill the Lund diagram using two QCD-based models for jet quenching, namely QPYTHIA \cite{Armesto:2009fj} and JEWEL \cite{Zapp:2011ya,Zapp:2012ak}. Both models implement the possibility for medium-induced bremsstrahlung. However, only JEWEL (i) evaluates dynamically the kinematics of multiple scattering, (ii) implements additional momentum broadening of all particles and (iii) provides the possibility to track recoiling medium constituents that have interacted with the jet and, finally, includes them in the hadronization step.\footnote{Note, however, that in JEWEL medium particles that interact with the jet do not interact further with the medium.} The jet-induced medium response constitutes a correlated ``background'' component that can contribute to the modifications of the measured jet substructure. Recoil effects are expected to contribute in the soft-large angle sector of the phase space, similarly to the uncorrelated underlying event, discussed further in \autoref{sec:uncorrelatedbackground}. One can also neglect tracking the recoil particles altogether. For further details about the employed models, see \App{app:models}.

We present first the results of generator level studies, i.e. without embedding the models into a realistic heavy-ion background.
For the same jet criteria as in \autoref{fig:PS2Vac}, in \autoref{fig:PS2} (upper row) we plot the Lund plots generated by QPYTHIA, JEWEL without recoils and JEWEL with recoils, respectively. 
In this particular study, we employ the C/A reclustering. 
The lower plots show the differences to the corresponding vacuum diagrams. 
It is also important to keep in mind that there is a significant migration between $\pT$ bins in heavy-ion collisions, widely understood as the effect of jet energy-loss. Considering multiple gluon emissions, the $\pT$ shift can be as big as $\Delta \pT \sim \sqrt{\omega_{s} \pT/n}$ \cite{Baier:2001yt} where $\omega_s \sim 2-5$ GeV is the  energy of typical medium-induced emissions. This could result in a significant contribution, in a fixed $\pT$ bin, from jets that were minimally modified.

The results from QPYTHIA exhibit a modest excess $\sim 10\%$ of hard quanta relative to vacuum, see \autoref{fig:PS2} (lower, left). In the model, the number of splittings is increased relative to vacuum leading to a significant intra-jet momentum broadening at scales corresponding to very short formation times. In the case of JEWEL, the difference plot exhibits only a mild increase of splittings at moderate $\kT$ and a small suppression $\sim 6\%$ of hard quanta, see \autoref{fig:PS2} (lower, center). This suppression is consistent with a lack of strong intra-jet broadening and a more collimated fragmentation. 
This shows that the realistic modifications to the Lund diagram are highly non-trivial and calls for a better theoretical understanding, see \autoref{sec:phasespace-theory} for a discussion.
When the medium recoils are included, a significant excess of large-angle quanta is reconstructed, see \autoref{fig:PS2} (lower, right). 
We note that in our declustering approach the angles are always measured relative to the hardest parent or subjet, in which case the angular distribution can be broader than the angular distribution measured relative to the jet axis that is used to compute jet profiles, see for instance \cite{KunnawalkamElayavalli:2017hxo}. We also note that the concept of formation time, as defined in \eqref{eq:FormationTime}, does not apply to these particles since they are collimated with the jet through elastic scattering and not per se part of the branching process.

It is worth pointing out that the medium-induced signal populates different regions of phase space in the two jet quenching models. 
While these features ultimately will be reflected in the relevant observables, the mapping onto the kinematical Lund plane seems to be a powerful tool to identify the impact of various medium modifications. Performing additional grooming, that is picking out branchings with specific properties, allows to enhance the sensitivity to the signal depending on the grooming parameters, see \autoref{sec:jetsubstructure}. Furthermore, changing the reclustering algorithm could also boost the signal, cf. \autoref{fig:PS2Vac}. We have observed that, in the case of JEWEL, the suppression of hard splittings is enhanced by $\sim 14\%$ with $k_{\rm \tiny T}$ reclustering. In the case of QPYTHIA, the excess of hard splittings is enhanced by $\sim 20\%$ with $k_{\rm \tiny T}$ reclustering.

The impact of the recoils as modeled by JEWEL has been extensively documented \cite{KunnawalkamElayavalli:2017hxo,Milhano:2017nzm}. Its contribution is needed to describe most of the jet shapes measured so far at the LHC. In particular, if the medium response can smear the subleading subjet momentum above the given grooming cut, the subjet momentum balance or $z_{g}$ can become more asymmetric relative to vacuum.   As a correlated background, the medium response cannot be experimentally subtracted to isolate purely radiative modifications to the jet shower. However, a cross-correlation of jet substructure observables might help to suppress its influence \cite{Milhano:2017nzm}.

It is worth noting that, albeit in a complicated form, the splitting map contains all of the information about a given medium shower. Certainly, such a procedure can be directly applied to experimental data, apart from the aspect of uncorrelated background that we outline in the next Section. Hence, in the remaining part of the report, the observables we choose to analyze will reflect particular features that already appear in the splitting map.

\subsubsection{Sensitivity to uncorrelated background}
\label{sec:uncorrelatedbackground}

\begin{figure}[th]
\centering
\includegraphics[width=0.33\textwidth]{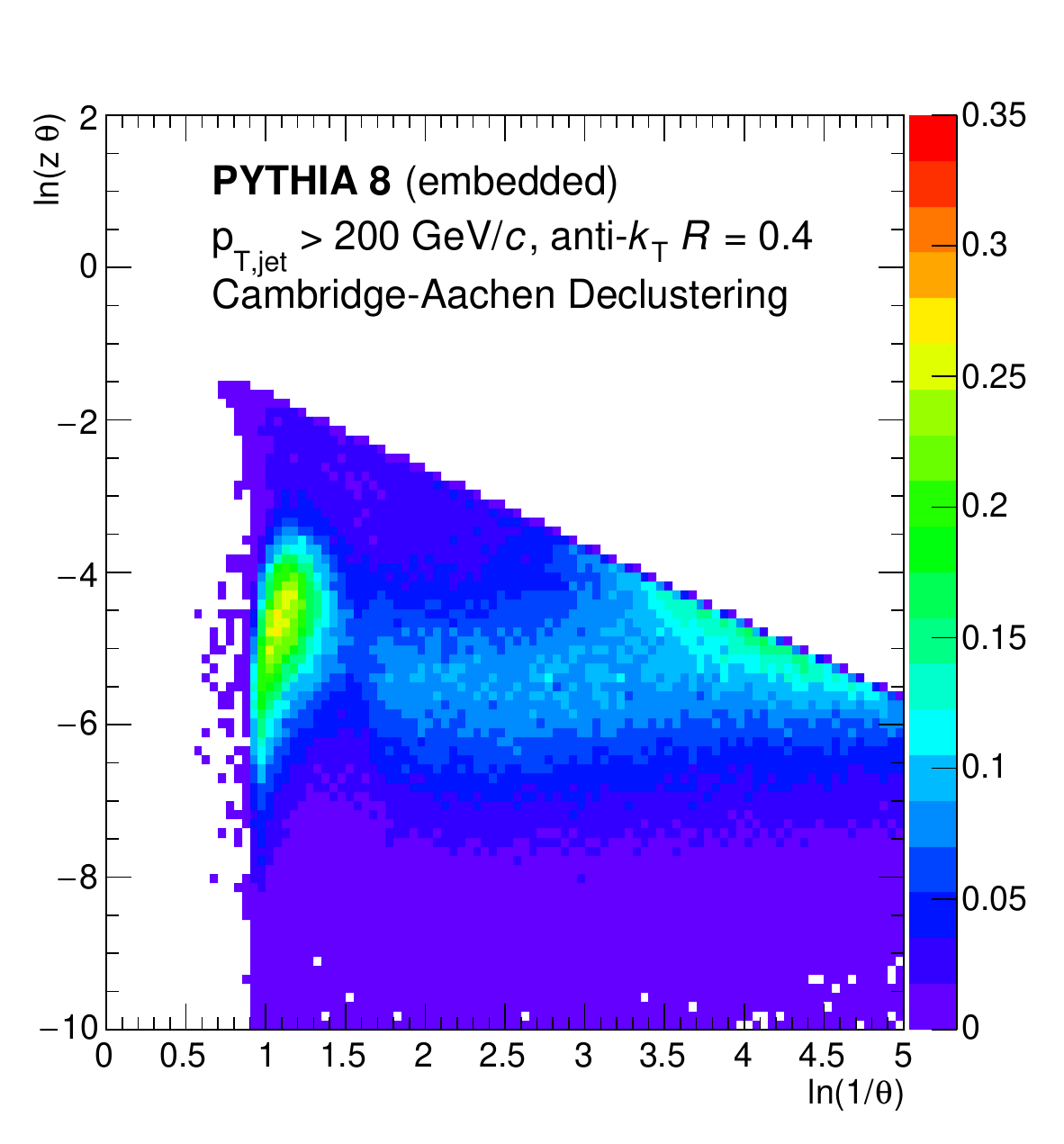}%
\includegraphics[width=0.33\textwidth]{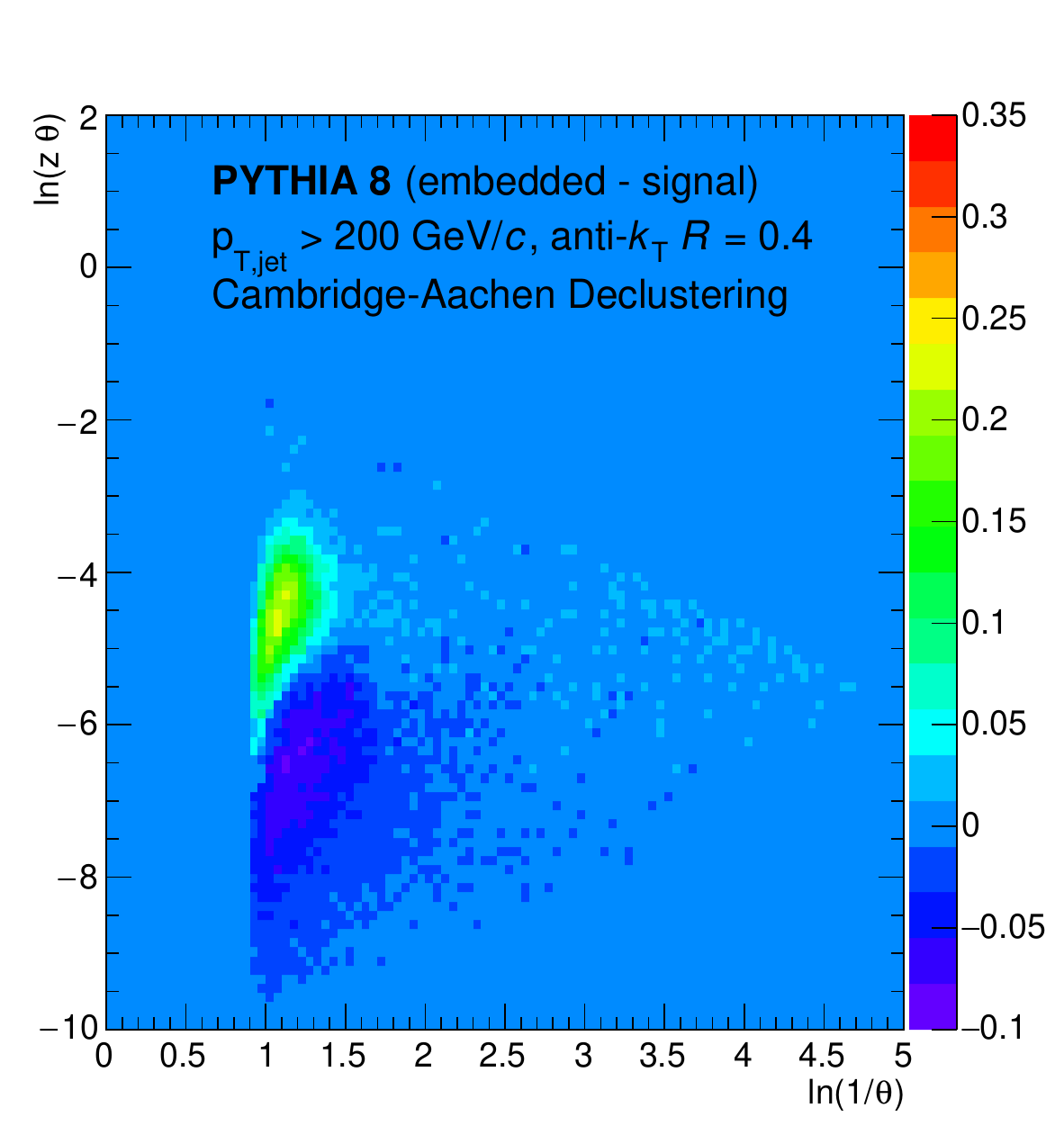}%
\caption{Impact of the uncorrelated background in the splitting map of the PYTHIA shower.}
\label{fig:UncorrelatedBkg}
\end{figure}

In all the studies performed so far in this Section, we have not included the effect of embedding the jets into a realistic heavy-ion background. In the studies presented in the following Sections, we have modeled the underlying event as in \cite{deBarros:2012ws}. 
This model randomly generates massless particles with uniform distribution at central rapidities. The transverse momentum distribution of the soft background is a Boltzmann distribution, with $\langle \pT \rangle$ = 1.2 GeV at the LHC. This is equivalent to an average momentum density of $\langle \rho \rangle =250$ GeV and a total multiplicity of $\sim 7000$ particles, which corresponds roughly to the most central events in the CMS detector, see also \cite{cerninstitute-twiki}. More details on the performance of jet reconstrution is such large background is provided in Appendix \ref{app:background}.

Hence, ending this section, we point out the fragility of using the Lund kinematical diagram in a realistic, noisy environment.
The heavy ion background that is uncorrelated to the jet will typically populate the phase space in the form of soft splittings at large angle $\theta \sim R$, where the area is maximal. Depending on the considered jet $p_{T}$, these fake splittings can contribute significantly to the distribution of groomed observables, cf. \autoref{sec:jetsubstructure}, by enhancing the number of asymmetric splittings and inducing a strong modification relative to vacuum jets. 
\autoref{fig:UncorrelatedBkg} shows the Lund diagram filled iteratively with PYTHIA jets embedded into a thermal background (left) and the difference plot to PYTHIA (right) where the latter clearly exhibits the enhancement of uncorrelated splittings at large angles after average background subtraction using constituent subtraction method \cite{Berta:2014eza}, see also \autoref{app:background} for more details. 
This provides a hint that uncorrelated background becomes a significant contribution to the observable the smaller the $\pT$ and significantly contaminates the angular resolution in the unfolding.

\begin{figure}
\centering
\includegraphics[width=0.32\textwidth]{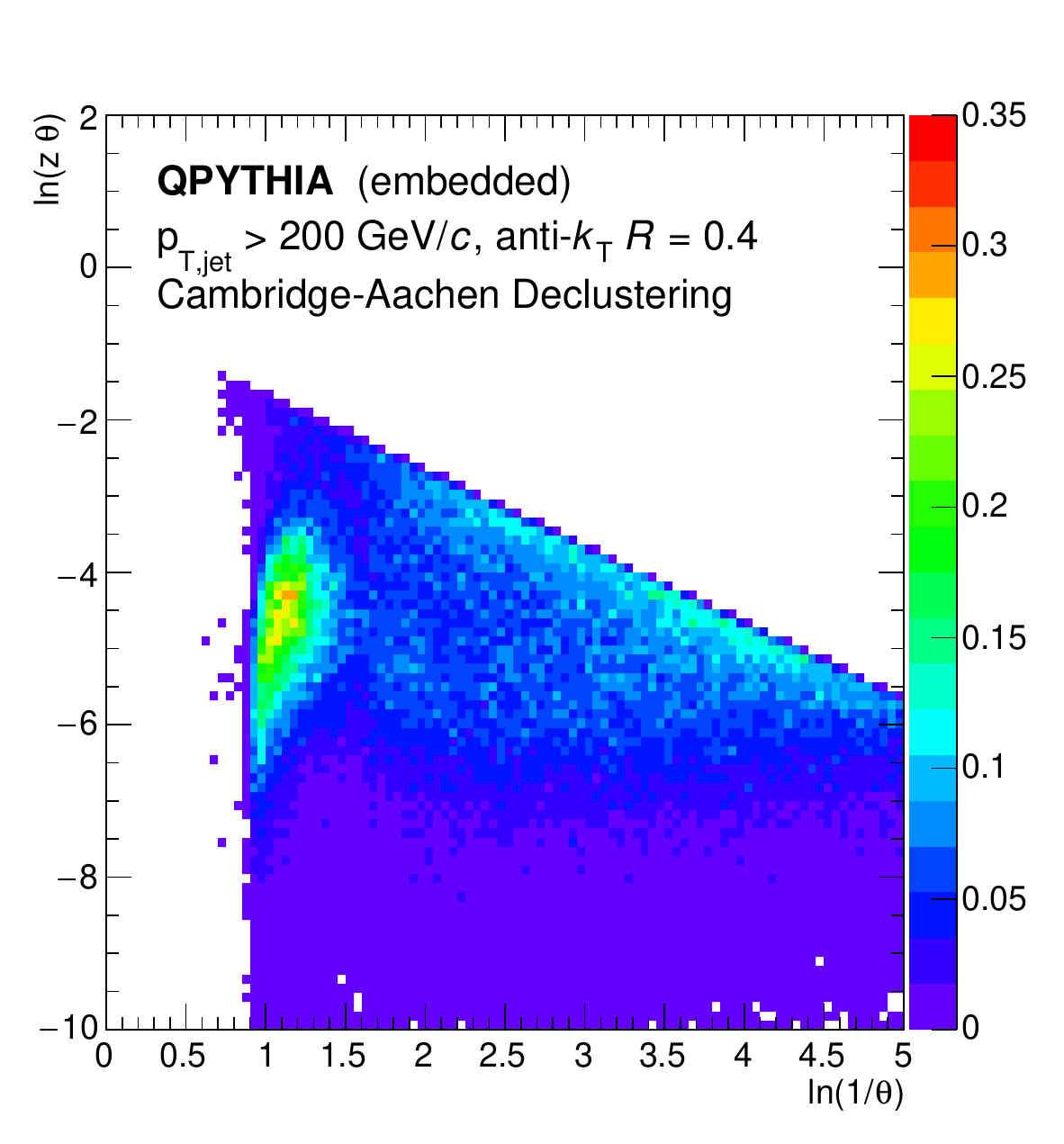}
\includegraphics[width=0.32\textwidth]{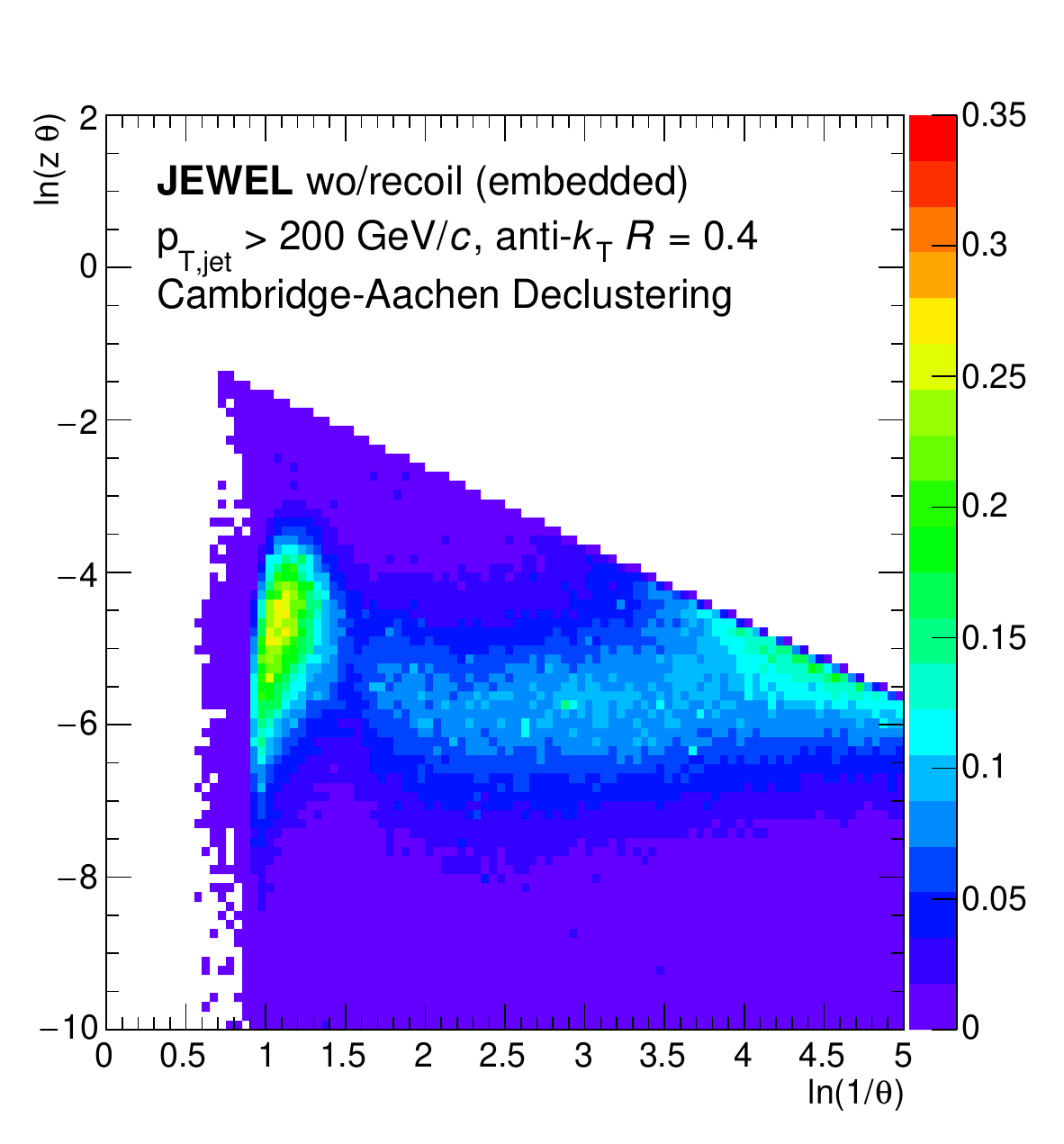}
\includegraphics[width=0.32\textwidth]{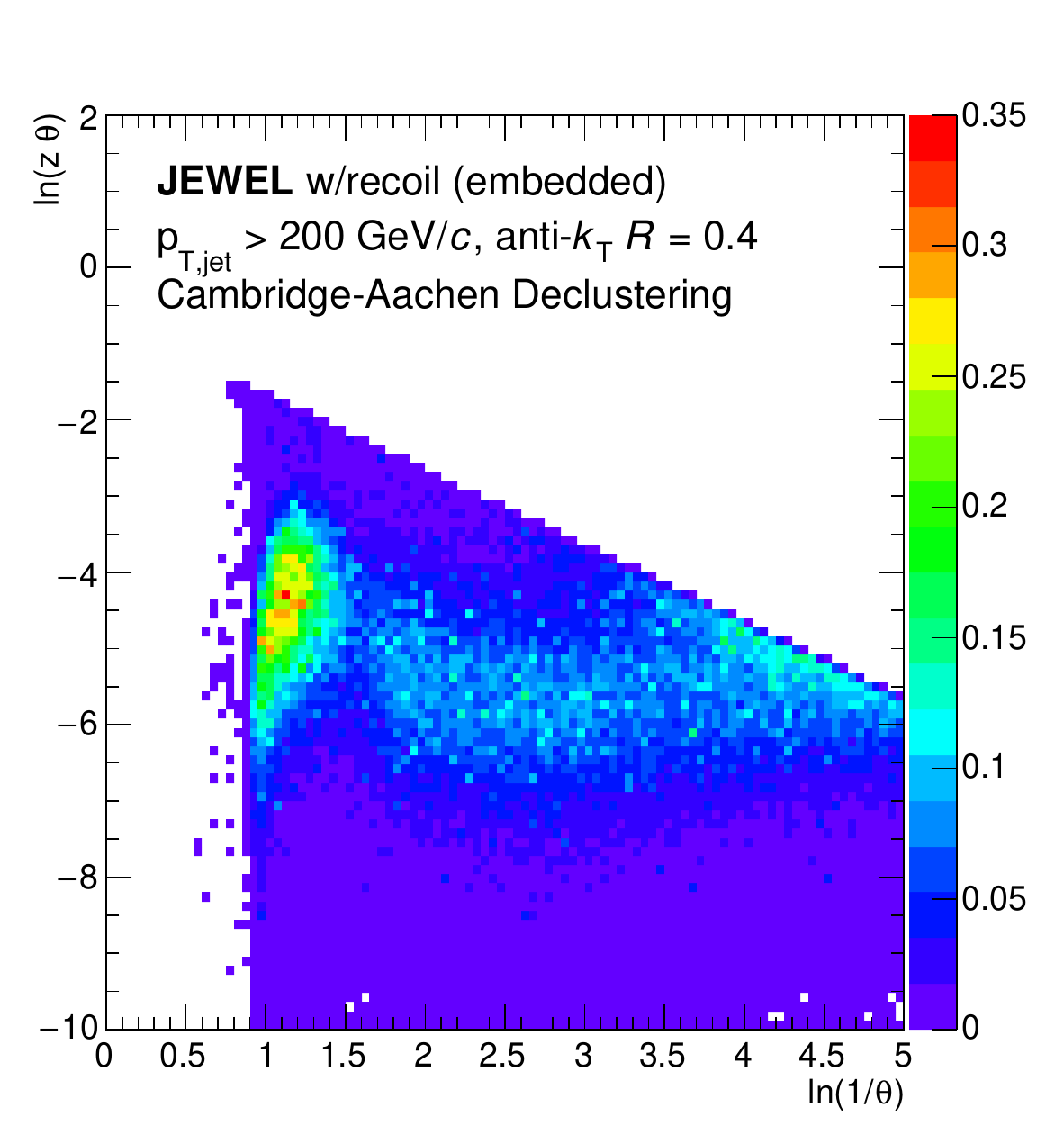}\\
\includegraphics[width=0.32\textwidth]{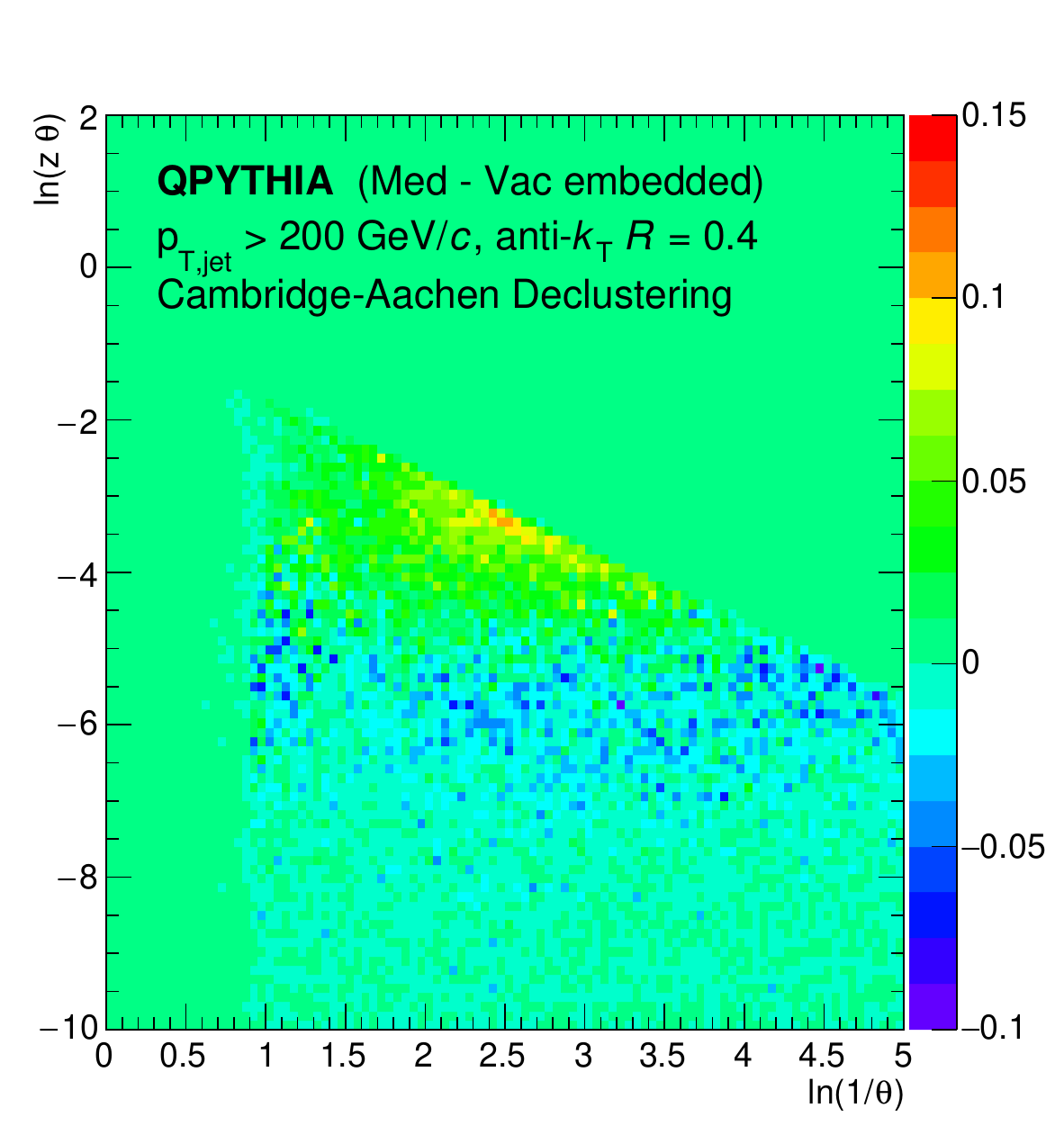}
\includegraphics[width=0.32\textwidth]{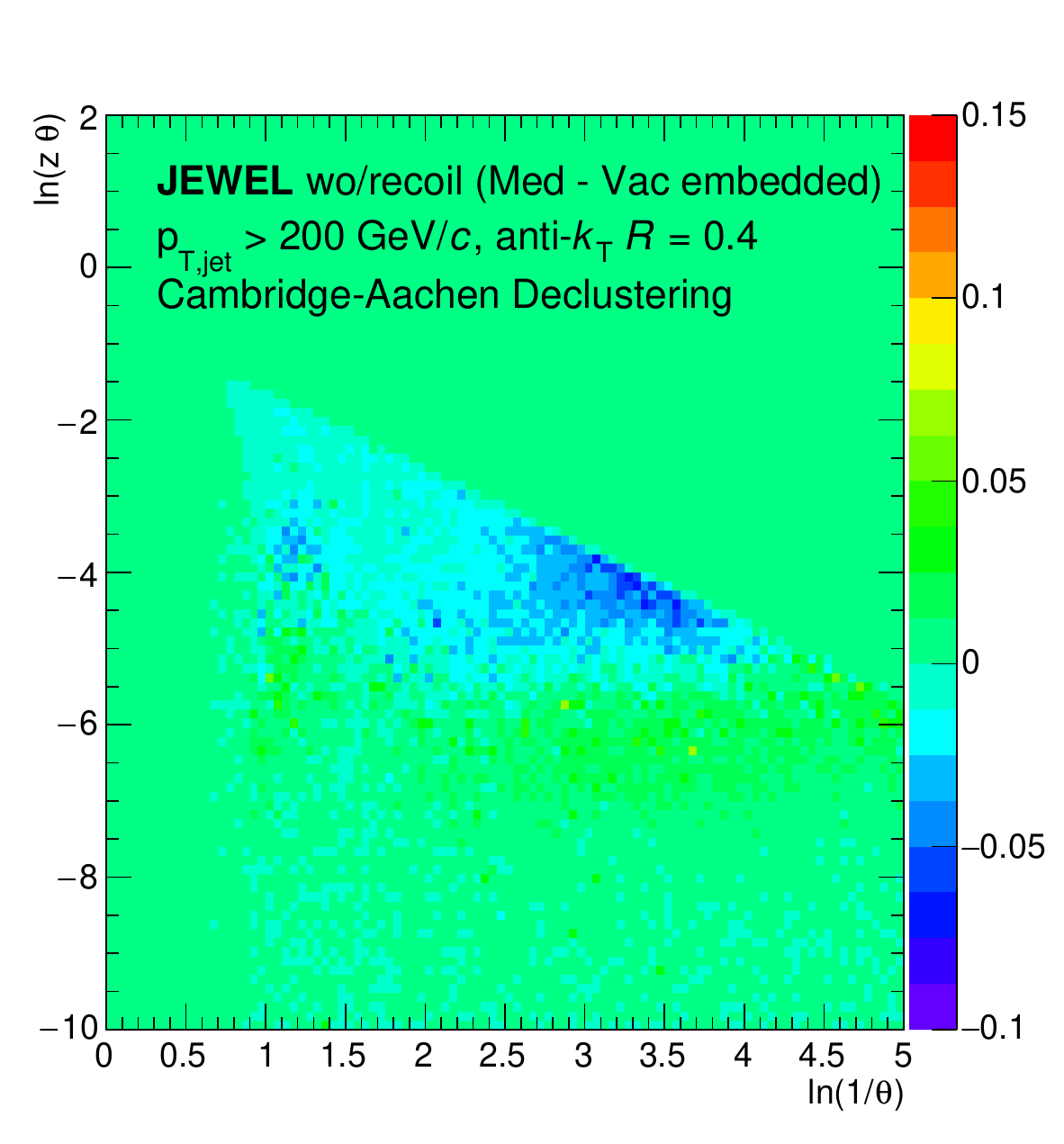}
\includegraphics[width=0.32\textwidth]{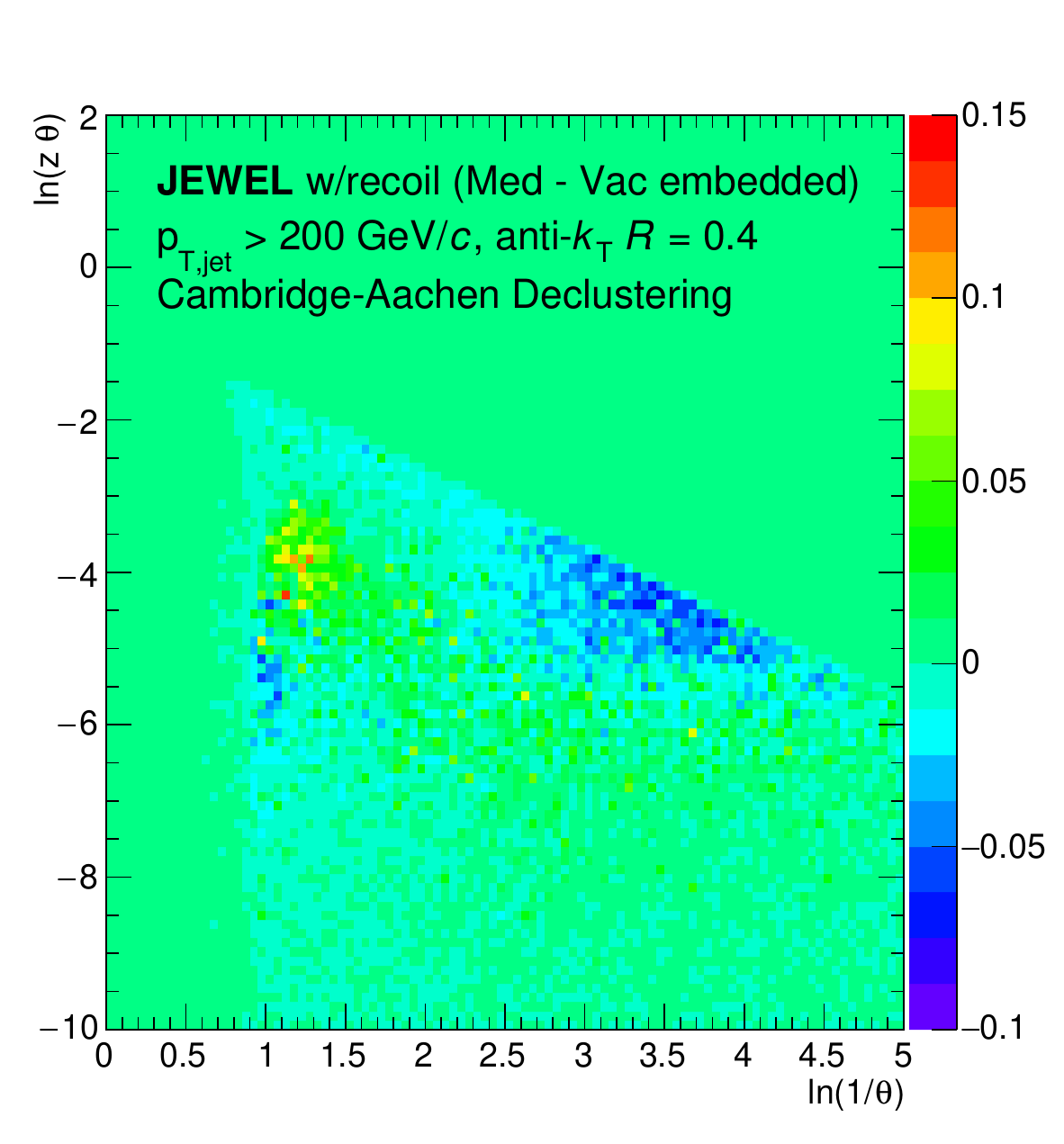}
\caption{Impact of the uncorrelated background in the splitting map of the medium parton showers QPYTHIA and JEWEL. 
Upper row: medium-modified MC models embedded in background. Lower row: difference between  embedded medium and vacuum showers.}
\label{fig:UncorrelatedBkgSignal}
\end{figure}

As expected, a generic enhancement of large-angle activity is a prominent feature for the in-medium showers as well.
Similar plots as discussed above are shown for embedded QPYTHIA and JEWEL showers in \autoref{fig:UncorrelatedBkgSignal} (upper row). 
Strikingly, all three plots share a similar dominant feature at large angles. This is confirmed by subtracting the generator-level events from the embedded ones. However, we observe that after embedding, the difference to the vacuum reference (also embedded) is still significant for most cases, see \autoref{fig:UncorrelatedBkgSignal} (lower row), meaning that the differences in the fragmentation pattern from different generators survive the presence of an underlying event, albeit with significant distortions.


\section{Jet substructure}
\label{sec:jetsubstructure}

In the last years, the study of jet substructure observables and associated techniques has expanded significantly in the context, in particular, of flavor-tagging (heavy-quark/light-quark/gluon discrimination), as well as in the identification of merged jets from Lorentz-boosted heavy particle decays in pp collisions at the LHC \cite{Ellis:2009su,Altheimer:2013yza}. Many different jet substructure variables have been systematically defined (generalized angularities: for example multiplicities, Les Houches Angularity, jet width or broadening, jet mass; eccentricity; groomed momentum fraction; N-subjettiness ratios; energy correlation functions; et cetera) with varying sensitivity to the momentum and angular properties of the jet constituents. On the theoretical side, many of these variables are IRC safe (at least in the QCD vacuum) and, thereby, amenable to high-precision pQCD calculations, in some cases up to NLO+NNLL accuracy (at least for the non-quenched reference case) \cite{Marzani:2017mva,Frye:2016aiz}. Implementation of such techniques for jets in heavy-ion collisions, and comparison with the corresponding pp results, provides new handles to study the medium-induced radiation pattern, as well as to quantitatively control the impact of non-perturbative (hadronization) and underlying-event contributions to the jet substructure. Further exploitation of such observables---e.g. for light-quark, gluon, and heavy-quark jet discrimination---can open a new window to better understand the color factor and mass dependence of
jet quenching phenomena.

As mentioned before, jet substructure techniques usually involve a step which reorganizes the constituents of a jet into a hierarchical tree where the nodes represent subsequent splitting processes.
This structure serves for further analysis using additional techniques such as jet grooming and tagging algorithms.
Grooming techniques usually reorganize the tree by discarding radiation that fail to pass given criteria, corresponding typically to soft and large-angle radiation. Taggers, on the other hand, aim at identifying the first splitting that passes a given criterion.
In this way a jet is decomposed into two sub-jets. Here, we will specifically focus on the first SoftDrop splitting as a tagger. Other examples of grooming methods, including trimming, pruning and filtering, are extensively reviewed, e.g., in \cite{Larkoski:2017jix,Marzani:2019hun}.
There has been a lot of progress recently utilizing these techniques for a wide range of substructure observables \cite{Butterworth:2008iy,Ellis:2009me,Krohn:2009th,Dasgupta:2013ihk,Larkoski:2014wba}, for a recent review see e.g. \cite{Larkoski:2017jix}. At least within the C/A algorithm, the subjets identified using grooming are in close correspondence to 
the first splitting of the parton evolution in the vacuum~\cite{Altarelli:1977zs,Larkoski:2015lea}.

While medium-modified jet fragmentation functions and other jet shape observables have been studied experimentally since many years, only recently have these substructure techniques been applied in the context of heavy-ion collisions.
Jet grooming was recently introduced as a tool to study the medium modification of leading partonic components in a parton shower~\cite{Sirunyan:2017bsd}, for related theoretical interpretations see \cite{Chien:2016led,Mehtar-Tani:2016aco,Milhano:2017nzm,Chang:2017gkt}. 

Given the proliferation of existing techniques, we will only refer to these as grooming techniques and concretely study one within the scope of this report, namely the Soft Drop procedure.
The Soft Drop algorithm reclusters the anti-$k_{\mathrm{T}}$ jet constituents using C/A to create an angular-ordered clustering tree. On this tree a pairwise declustering is performed. In each step of the declustering the softer branch is removed until a branching pair that satisfies
\beq
\label{eq:groompar}
\frac{\mathrm{min}(p_{\mathrm{T},i},p_{\mathrm{T},j})}{p_{\mathrm{T},i}+p_{\mathrm{T},j}} > z_{\text{cut}}\left( \frac{\Delta R_{ij}}{R_{0}} \right)^{\beta},
\eeq
is found, where the subscripts ``$i$'' and ``$j$'' indicate the subjets at that step of the declustering, $\Delta R_{ij}$ is the distance between the two subjets, $R_{0}$ is the cone size of the anti-$\kT$ jet, and $z_{\text{cut}}$ and $\beta$ are adjustable parameters. By varying $z_{\text{cut}}$ and $\beta$, specific regions of the emission phase space, see \autoref{fig:PS1}, can be isolated. For $\beta = 0$, this procedure is identical to the modified mass-drop tagger \cite{Dasgupta:2013ihk}, while $\beta \neq 0$ was introduced in \cite{Larkoski:2014wba}. It allows to design specific grooming settings sensitive to distinct regions of the kinematical phase space represented in the Lund plane. Equivalently, the parameters can be adjusted to suppress or enhance the effect of medium modifications.

\begin{figure}[t!]
\centering
\includegraphics[width=0.34\textwidth]{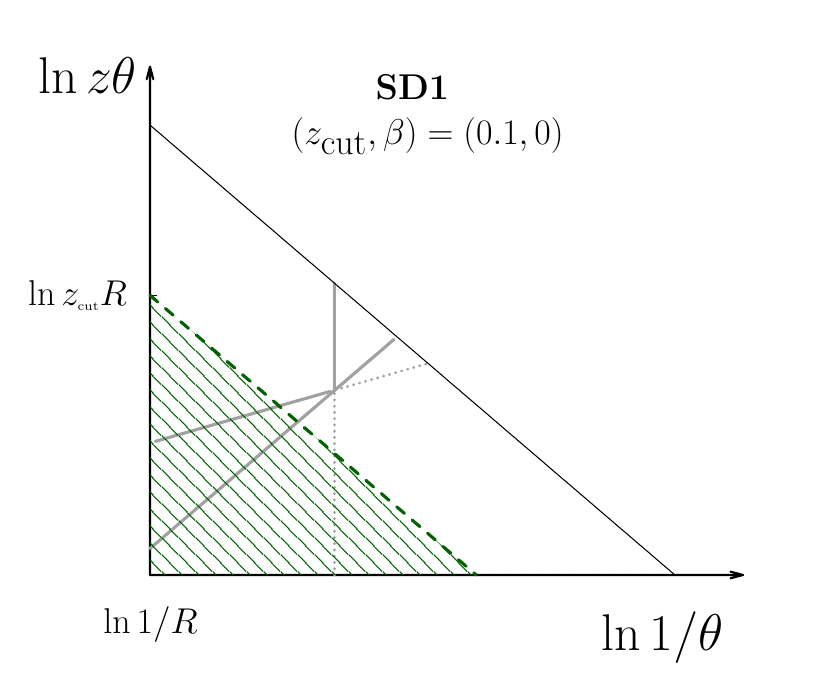}%
\includegraphics[width=0.34\textwidth]{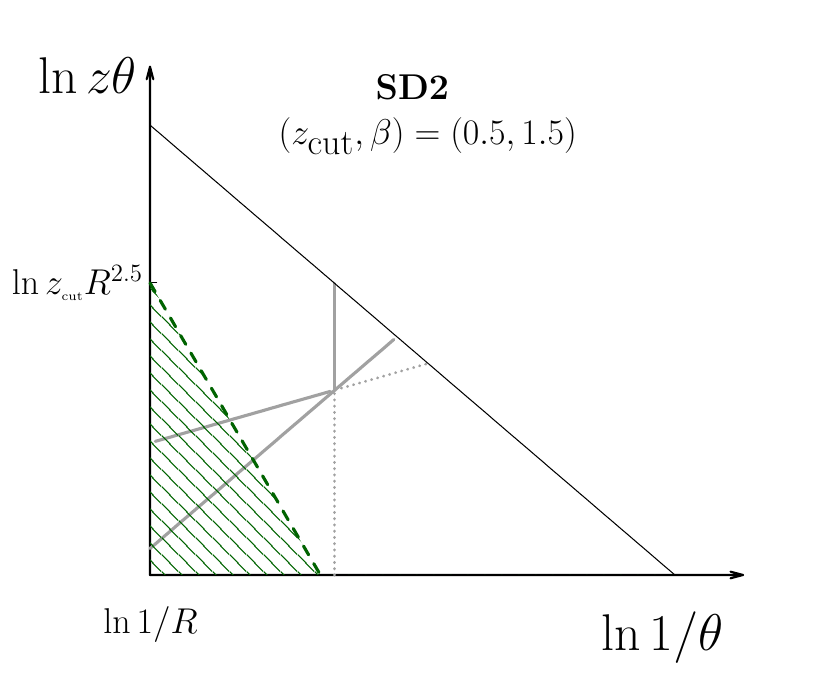}%
\includegraphics[width=0.34\textwidth]{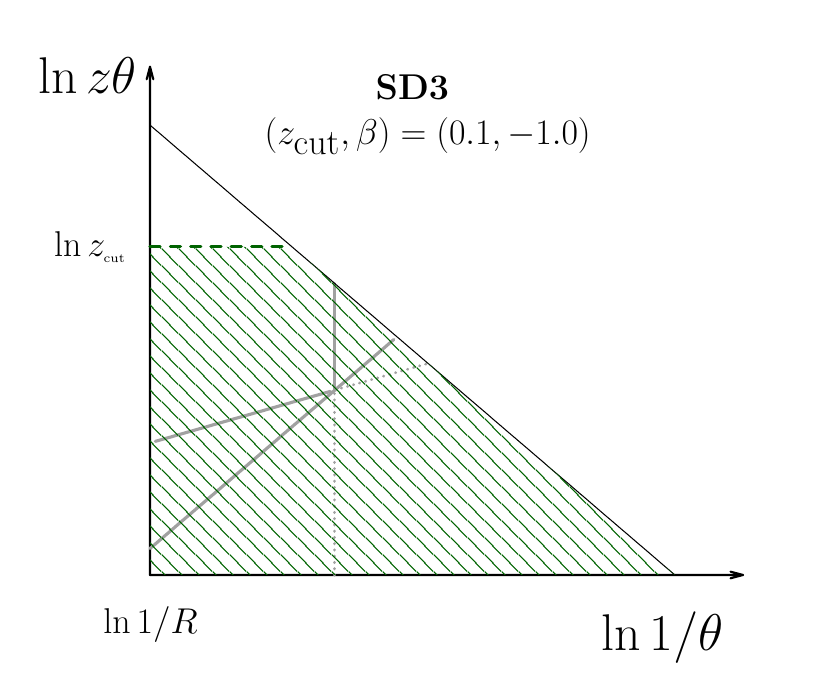}%
\caption{The three grooming settings studied in this report, see text for details. Shaded areas correspond to configurations that are groomed away.}
\label{fig:TheorySD}
\end{figure}

In this report, we compare the three grooming settings:
\begin{description}

\item[{\bf SD1:}] $z_{\text{cut}}=0.1$ and $\beta=0$: removes branches based only on the energy fraction;

\item[{\bf SD2:}]  $z_{\text{cut}}=0.5$ and $\beta=1.5$: has a stronger grooming at large angle;

\item[{\bf SD3:}]  $z_{\text{cut}}=0.1$ and $\beta=-1.0$: selects only hard radiation;

\end{description}
\autoref{fig:TheorySD} depicts how these settings remove parts of the phase space in the Lund plane. This will in turn affect the demands on statistics, especially for the SD3 setting. While the first setting is the more widely used in various studies of the SD procedure, the two latter are designed to suppress regions of phase space with a lot of medium activity, as identified in the diagrams in \autoref{fig:PS2}. One could, of course, devise other grooming strategies, or even combine various conditions, in order to ``carve'' out kinematical regimes of particular interest. We avoid such prescriptions here in order not to bias our jet sample excessively. On the other hand, it could be interesting to combine grooming strategies with specific reclustering algorithms, a point we briefly study in \autoref{sec:hadronization}.

\subsection{Groomed substructure observables and sensitivity to jet quenching}
\label{sec:groomedobservables}

After identifying the first splitting that satisfies Eq.~(\ref{eq:groompar}), we have access to the full kinematics of that branching step. The groomed jet energy ($\pT = E$) is now defined as $p_{{\rm \tiny T} g} \equiv p_{\mathrm{T},1}+p_{\mathrm{T},2}$, where the subscripts now refer to the identified subjets. We can then define the groomed momentum fraction, $z_g = \min \left(p_{{\rm \tiny T},1},p_{{\rm \tiny T},2}\right)/p_{{\rm \tiny T}g}$ and the angle $\Delta R_{12}$ between the subjets. In our numerical studies, we will focus on these two quantities but also introduce the groomed mass to energy ratio $M_g/\pT$, where $M_g$ is defined as in Eq.~(\ref{eq:DipoleMass}) with all relevant quantities being groomed. These observables shed light on how the branchings occur in course of the parton shower and are sensitive to medium effects as long as the branching originates from inside the medium, roughly $ t_{{\rm f}g}\equiv 2 p_{{\rm \tiny T}g}/M_g^2 < L$, see discussion above. For the chosen medium parameters, the samples analyzed with settings SD1 and SD2 will contain an admixture of in-medium and out-of-medium splittings, see \autoref{fig:TheorySD}, while SD3 picks exclusively out hard splittings originating from inside the medium. 

\begin{figure}[t]
\centering
\includegraphics[width=0.33\textwidth]{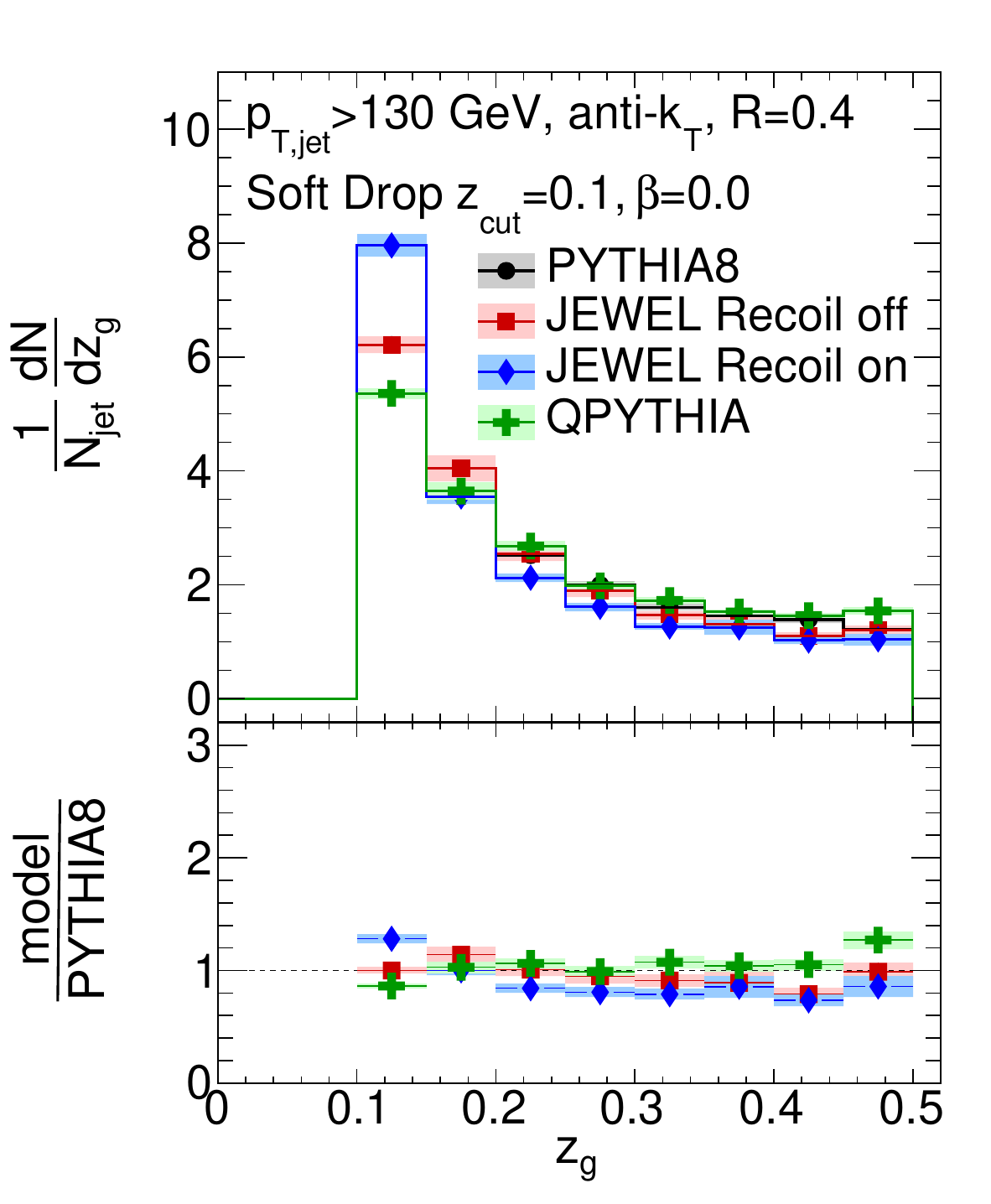}%
\includegraphics[width=0.33\textwidth]{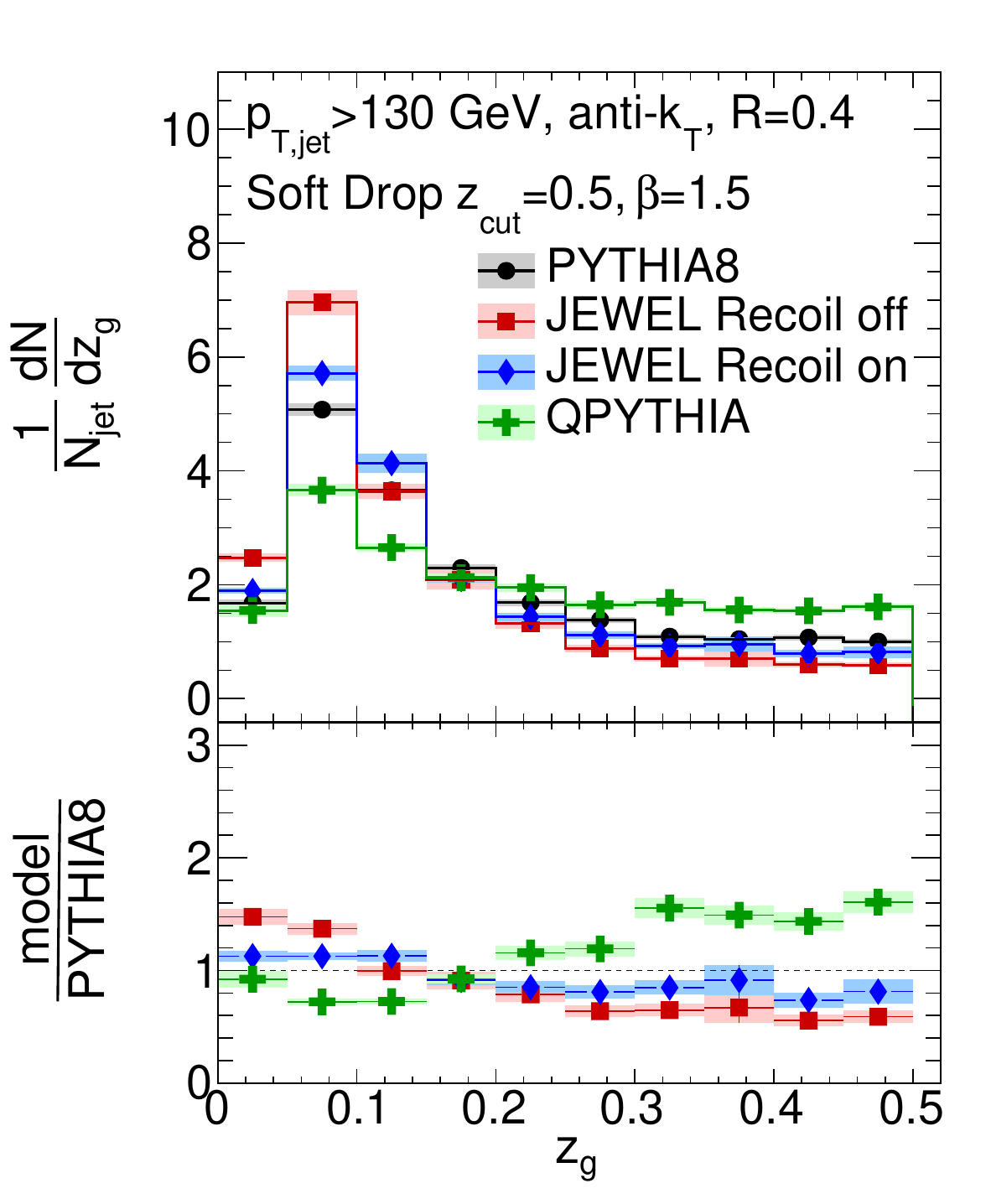}%
\includegraphics[width=0.33\textwidth]{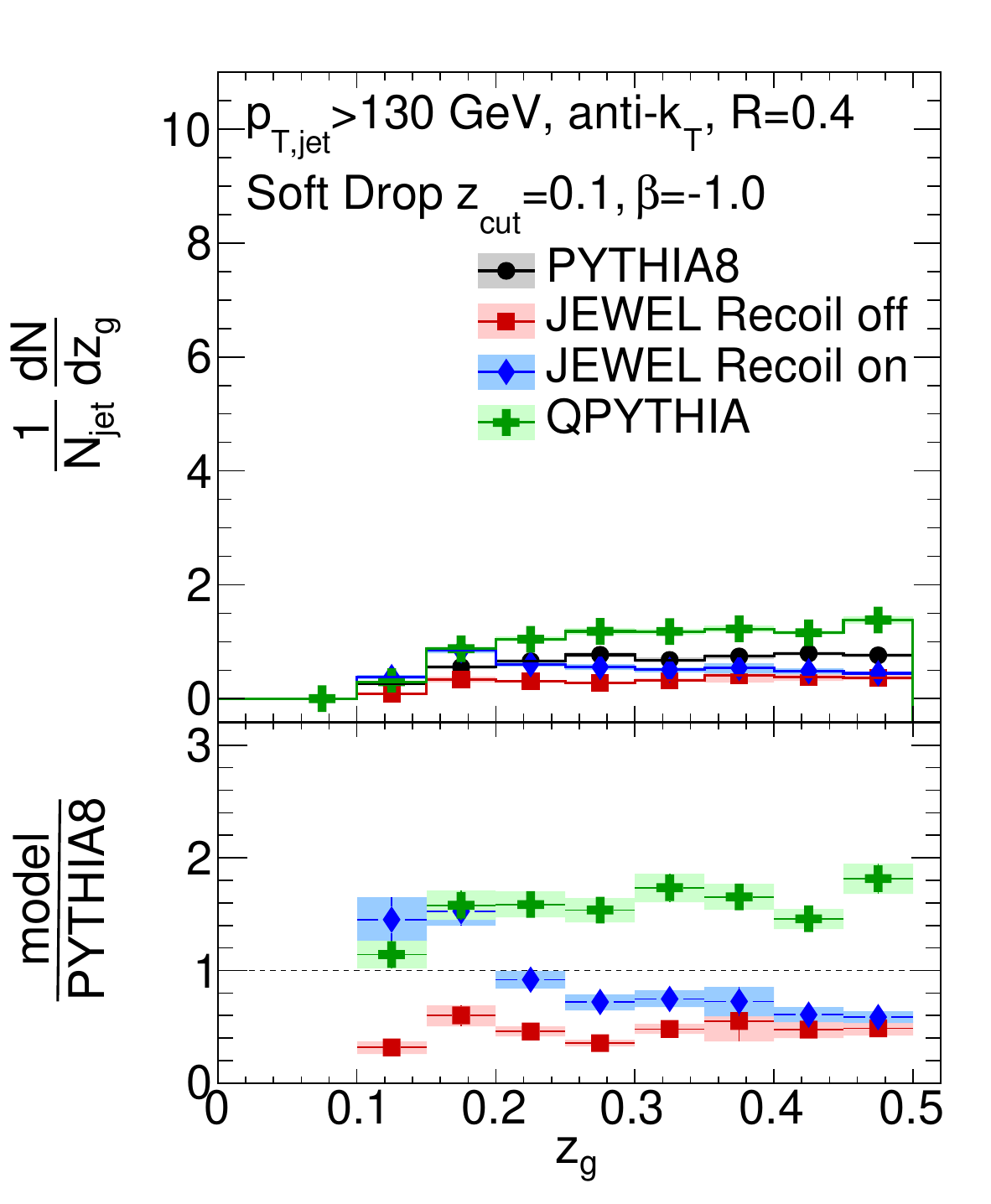}%
\caption{Groomed shared momentum fraction, $z_{\mathrm{g}}$, for three different grooming settings in simulations with and without jet quenching. Statistical errors have been included. The upper panels show the $z_{\mathrm{g}}$ distribution normalized by the total number of ungroomed jets while the lower panels show the ratio of JEWEL and QPYTHIA with respect to PYTHIA8.}
\label{fig:SDGenZG}
\end{figure}

\begin{figure}[th!]
\centering
\includegraphics[width=0.33\textwidth]{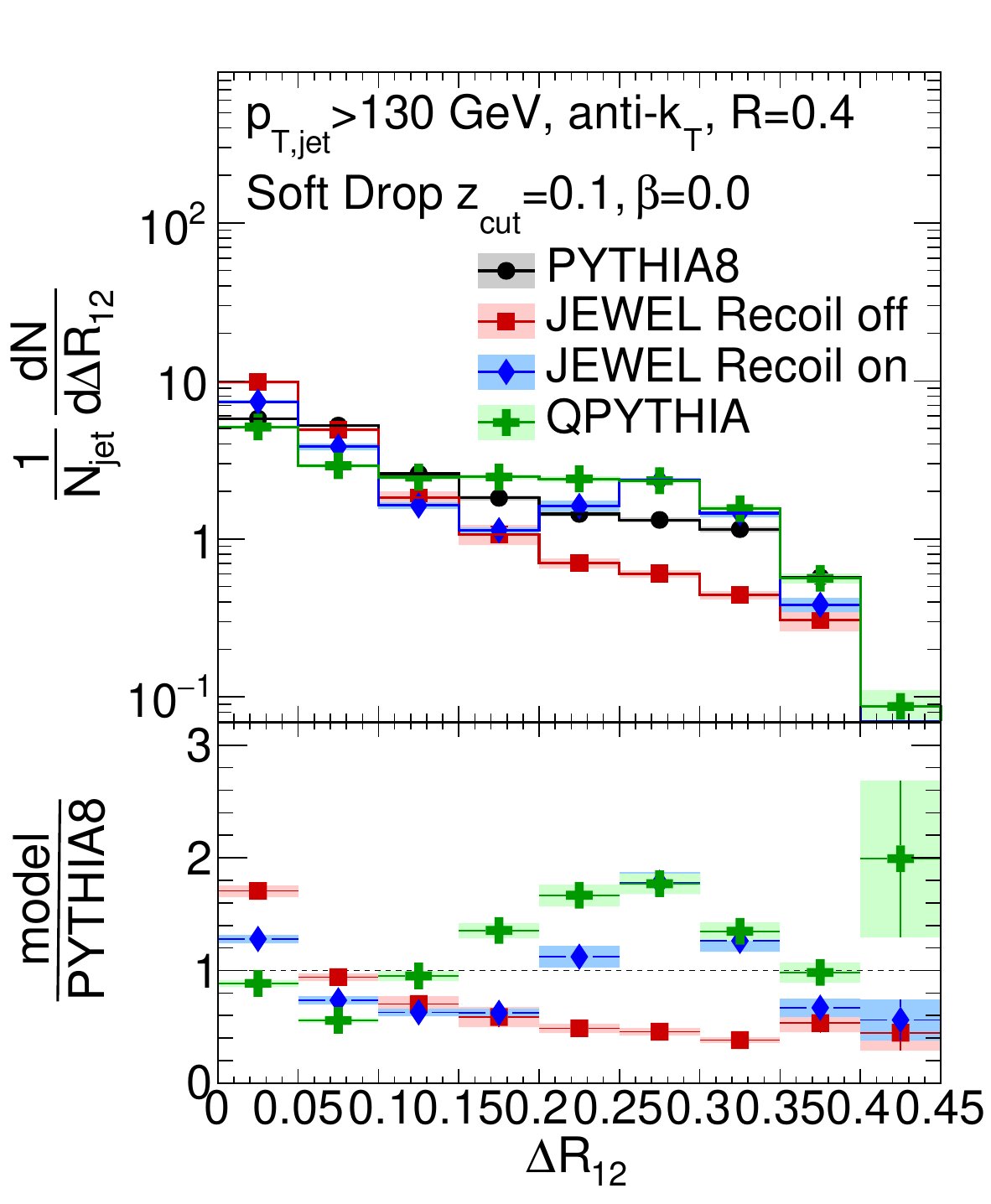}%
\includegraphics[width=0.33\textwidth]{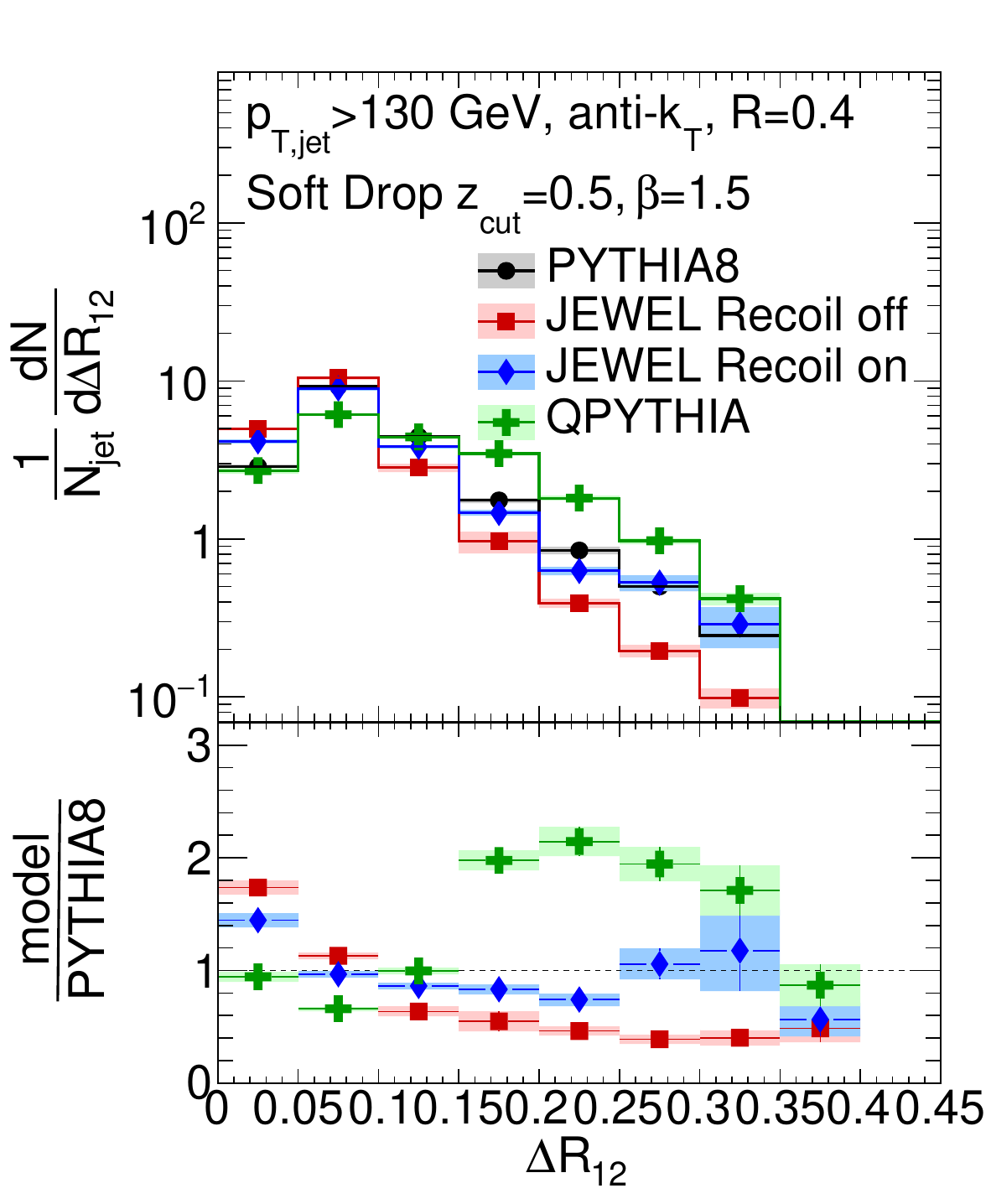}%
\includegraphics[width=0.33\textwidth]{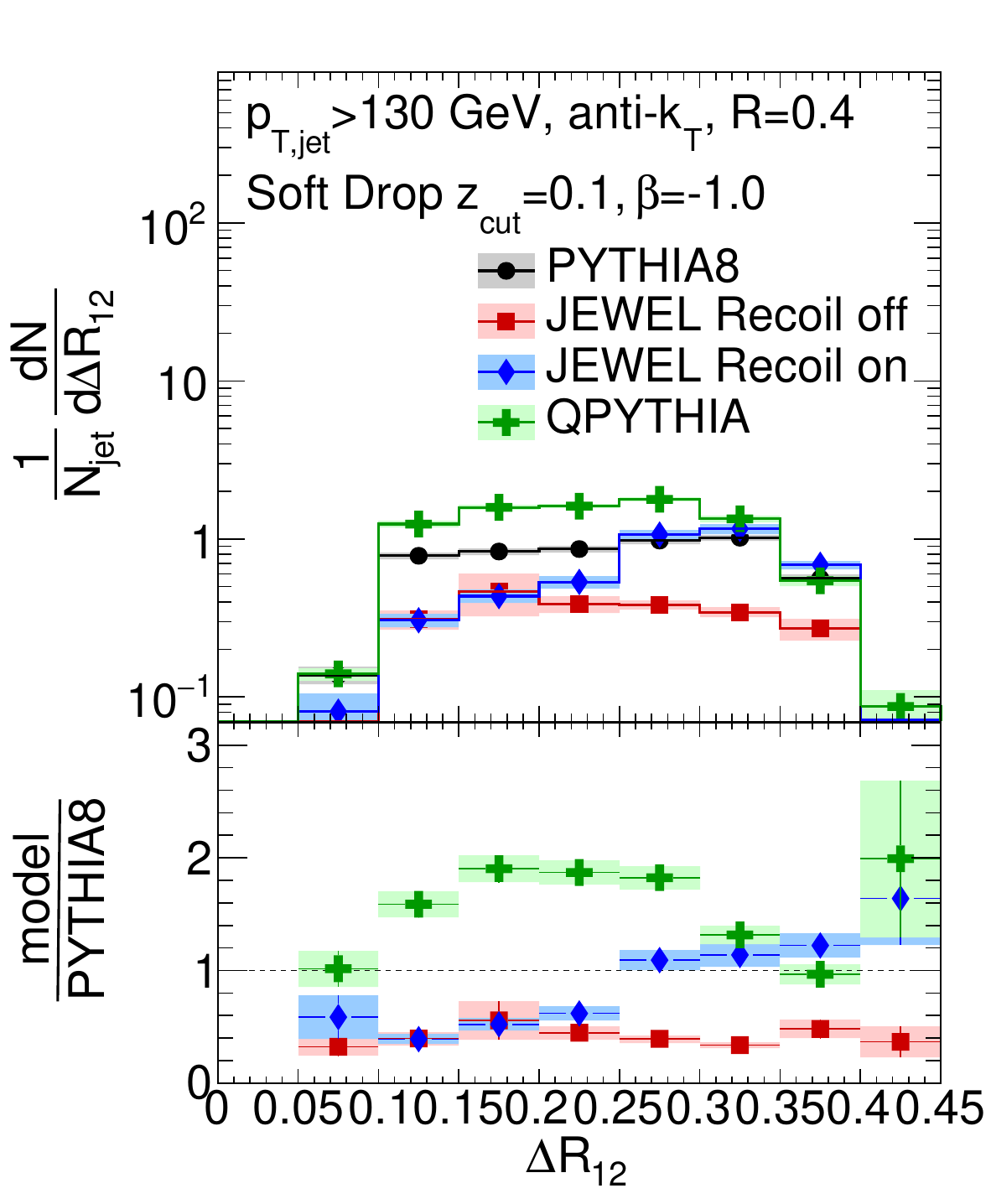}%
\caption{Distance between the two groomed subjets, $\Delta R_{12}$, for three different grooming settings in simulations with and without jet quenching. Statistical errors have been included. The uppers panels show the $\Delta R_{\mathrm{12}}$ distribution normalized by the total number of ungroomed jets while the lower panels show the ratio of JEWEL and QPYTHIA with respect to PYTHIA8.}
\label{fig:SDGenDR12}
\end{figure}

\begin{figure}[th]
\centering
\includegraphics[width=0.33\textwidth]{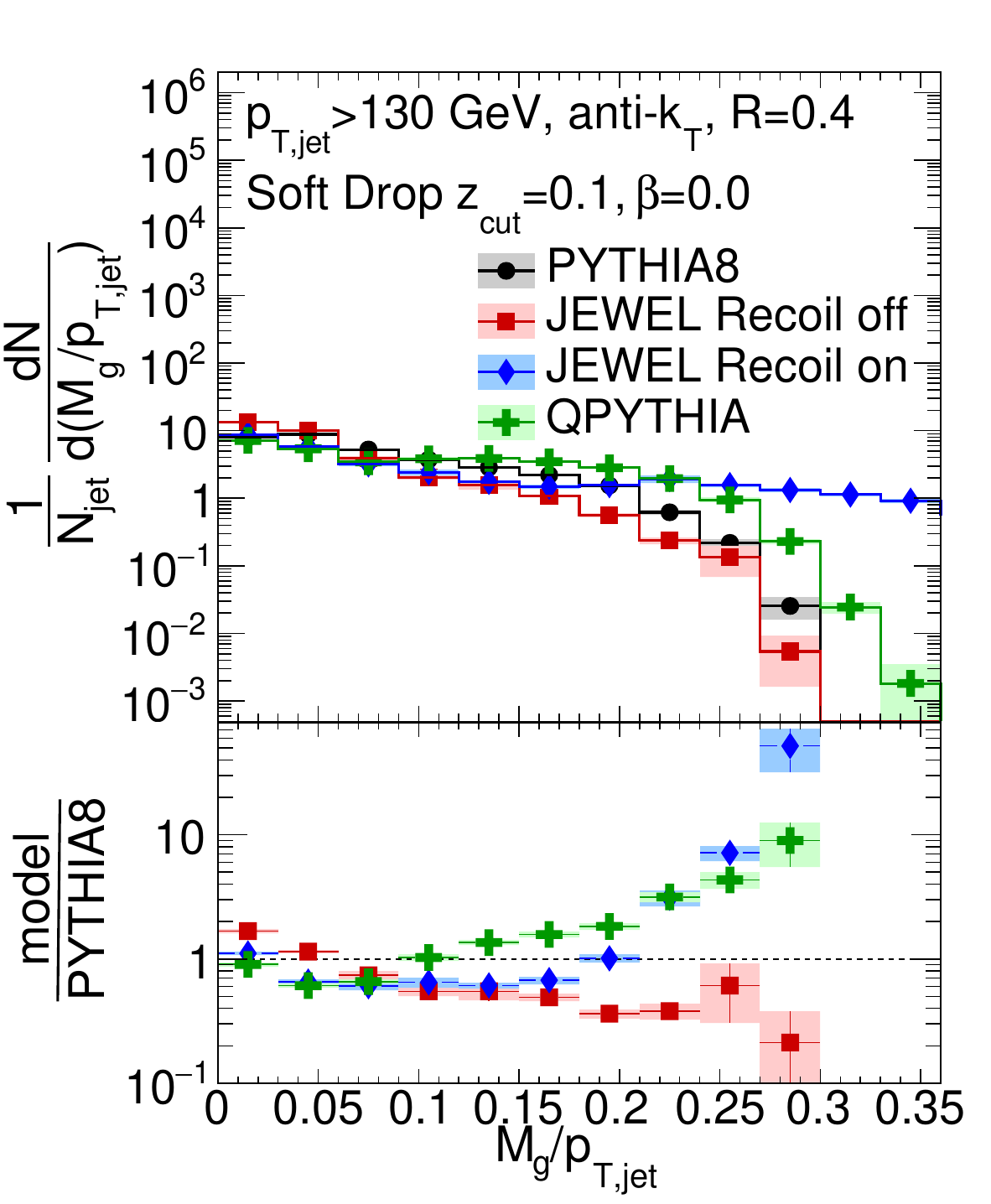}%
\includegraphics[width=0.33\textwidth]{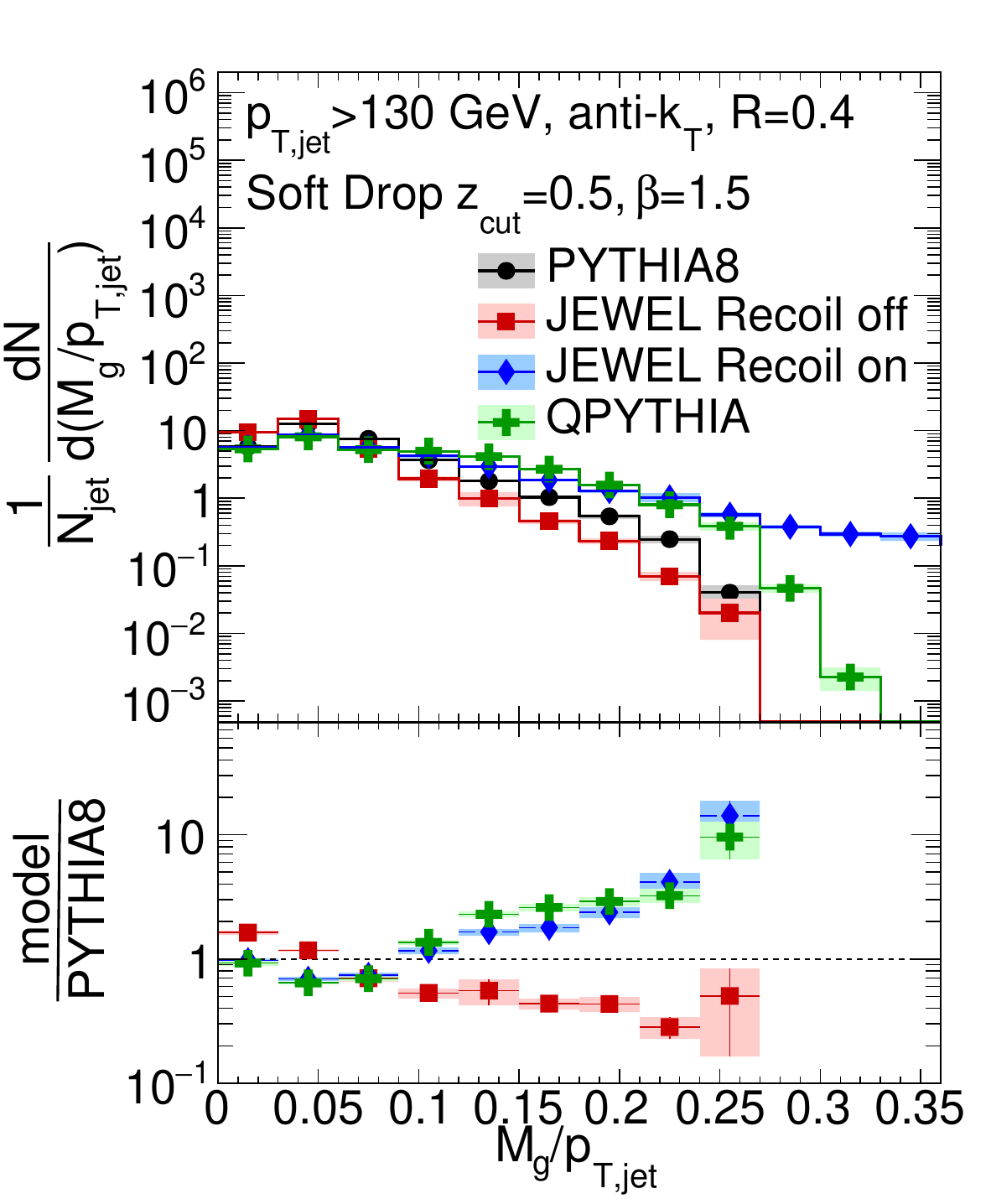}%
\includegraphics[width=0.33\textwidth]{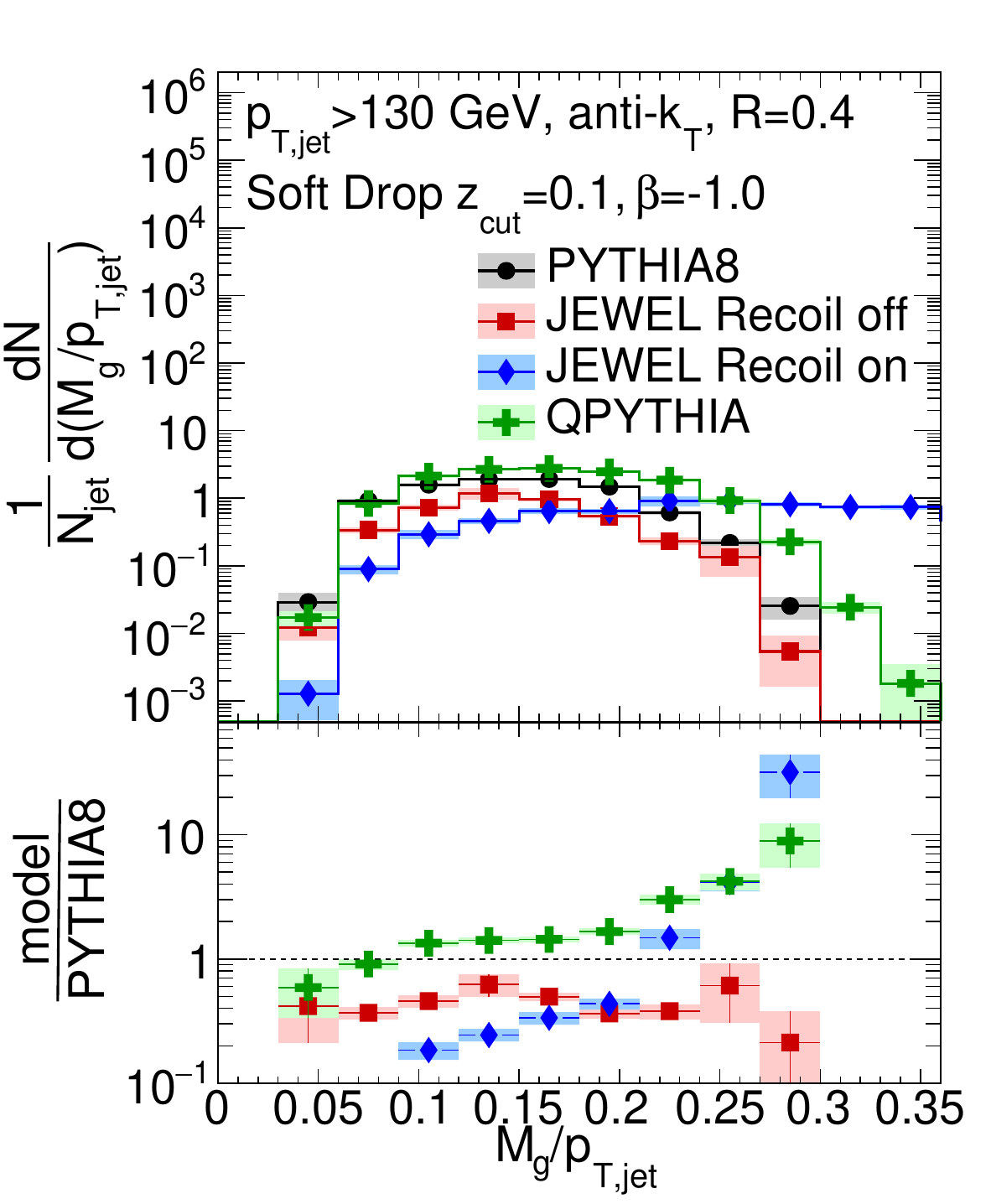}%
\caption{Groomed jet mass, $M_{\mathrm{g}}/p_{\mathrm{T,jet}}$, for three different grooming settings in simulations with and without jet quenching. Statistical errors have been included. The uppers panels show the $M_{\mathrm{g}}/p_{\mathrm{T,jet}}$ distribution normalized by the total number of ungroomed jets while the lower panels show the ratio of JEWEL and QPYTHIA with respect to PYTHIA8.}
\label{fig:SDGenMg}
\end{figure}

As in the previous section, the jet quenching Monte Carlo event generators we use in our study are QPYTHIA and JEWEL (with recoil effects turned on and off) and are shown in \autoref{fig:SDGenZG}, \ref{fig:SDGenDR12} and \ref{fig:SDGenMg}. Jets were reconstructed using anti-$\kT$ $R=0.4$ and for $\pT > 130$ GeV/c. 
The results in this section are obtained at generator level, without embedding. In particular, we have not introduced any detector resolution effects, such as a minimal angular cut-off $\Delta R_{\rm min}$.
Note, that the distributions are normalized by the total number of anti-$k_{\text{T}}$ (ungroomed) jets. The distributions are therefore not self-normalized and contain information how grooming affects the overall suppression of the jet yield. 

Figure~\ref{fig:SDGenZG} shows the momentum fraction $z_g$ distribution for different event generators. The vacuum baseline is represented by the PYTHIA8 data points and compared to results from the QPYTHIA and JEWEL jet quenching event generators.
In this figure, the perhaps most striking feature is the generally opposite trend of the two models. This can also be traced back to the discussion around \autoref{fig:PS2}. The modified parton shower in QPYTHIA makes the jets broader with respect to jets in vacuum and therefore many more jets survive the grooming. JEWEL however collimates the jets and therefore fewer jets are surviving the grooming with this setting.

We also note, that while for $\beta \geq 0$, see \autoref{fig:SDGenZG} (left and center), the number of jets for the different generators remains roughly constant while for the negative grooming setting $\beta < 0$, \autoref{fig:SDGenZG} (right), a large deviation from unity can be observed. Interestingly, QPYTHIA subjets are strongly enhanced in this regime while JEWEL ``Recoils off'' subjets are strongly suppressed, both by a factor $\sim1.5-2$. This is naturally in agreement with the features already observed in the Lund plane, see \autoref{fig:PS2}. For example, for QPYTHIA we note a strong enhancement at high-$\kT$ independently of the momentum fraction $z$, see \autoref{fig:PS2} (lower left panel), which reflects in the enhancement in \autoref{fig:SDGenZG} (right). Note also that the magnitude of effects are the biggest for the most aggressive setting that naively corresponds to early in-medium splittings.

Comparing the JEWEL results with and without recoil demonstrates that, for the chosen analysis settings, this observable is not very sensitive to recoil effects except for the small-$z_g$ region. In order to compare to the data presented in \cite{Sirunyan:2017bsd} for the $\beta=0$ setting, see also \cite{Milhano:2017nzm} for a study using JEWEL, where a significant deviation from vacuum baseline was observed. We again point out that no minimal angular cut-off was employed in our studies. Such a cut-off suppresses collinear vacuum radiation and, hence, amplifies the effects related to the medium.

Next we turn to studying the angular separation $\Delta R_{12}$ distribution of the groomed subjets. In the context of jet quenching, one particularly interesting question is to gauge whether substructures are quenched differently as a function of  their angular separation. The angular distance between the groomed sub-jets is plotted in \autoref{fig:SDGenDR12} for the three grooming settings. Once again, we see big differences between the MC models; JEWEL ``Recoils off'' being very collimated and QPYTHIA very broad. 
The JEWEL ``Recoils on'' setting interpolates between the two extremes and, most strikingly, exhibits an enhancement at intermediate angles, consistent with earlier studies of jet shape and fragmentation function \cite{KunnawalkamElayavalli:2017hxo}.

Once more, it is interesting to point out that the modifications are arguably the strongest for the most conservative SD setting, see \autoref{fig:SDGenDR12} (right). In particular, the JEWEL ``Recoils off'' samples are consistently suppressed for all angles. This could point to the importance of energy-loss that is not very sensitive to angle in JEWEL. The enhancement seen at small $\Delta R_{12}$ for $\beta \geq 0$, see \autoref{fig:SDGenDR12} (left, center), could also indicate a similar mechanism related to migration of narrow jets from higher $\pT$.

Finally, we study the groomed jet mass normalized by the ungroomed transverse momentum, $M_{\mathrm{g}}/p_{\mathrm{T,jet}}$, in \autoref{fig:SDGenMg}. 
This observable combines several of the features already seen before and seems particularly constraining of large-mass jet substructures. In this case, the QPYTHIA and JEWEL ``Recoils on'' samples give rise to similar distributions with a strong enhancement at large $M_g/\pT$. The enhancement is the largest for the latter model, putting strong constraints on the assumptions related to the free streaming of recoil fragments in JEWEL. In contrast, JEWEL ``Recoils off'' is more resilient and exhibits a mild suppression with respect to vacuum results at high-masses. This could again be interpreted as an effect of energy-loss.

To summarize, these generator level studies of the kinematics of the subjet samples obtained using Soft Drop illustrates the wide range of sensitivity to different kinematical regimes, and therefore different effects. Further studies, including embedding and involving more medium models, are planned for the future and could help further constrain large classes of medium effects. These improvements are also crucial for a realistic comparison to experimental data. Recently, results on the groomed mass in heavy-ion collisions at the LHC were released by the CMS collaboration \cite{Sirunyan:2018gct} and several jet substructure measurements in both pp and heavy-ion collisions have been reported by the ALICE collaboration \cite{Acharya:2019djg}.

\subsubsection{Sensitivity to hadronization effects}
\label{sec:hadronization}

The last stage of the jet fragmentation is the non-perturbative process of hadronization. This is a dynamical process that converts colored partons into color-singlet hadrons. In jet quenching event generators it is typically assumed that hadronization occurs outside of the medium, even if some modifications can persist from modifications of color flow at earlier stages, see e.g. \cite{Aurenche:2011rd,Beraudo:2011bh,Beraudo:2012bq}. A proof for this assumption does not exist and therefore hadronization uncertainties should be expected to be sizable. 

\begin{figure}[th]
\centering
\includegraphics[width=0.33\textwidth]{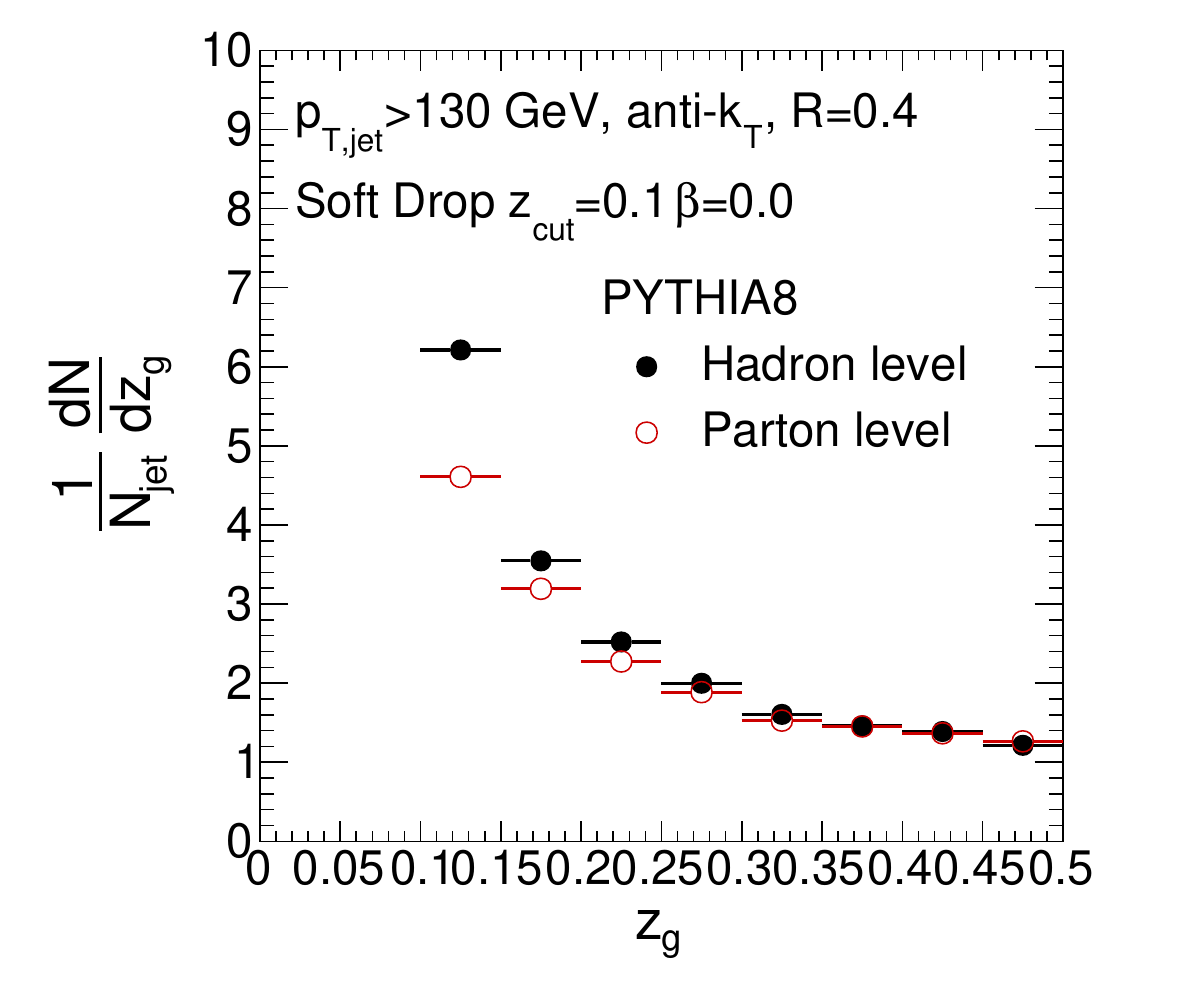}%
\includegraphics[width=0.33\textwidth]{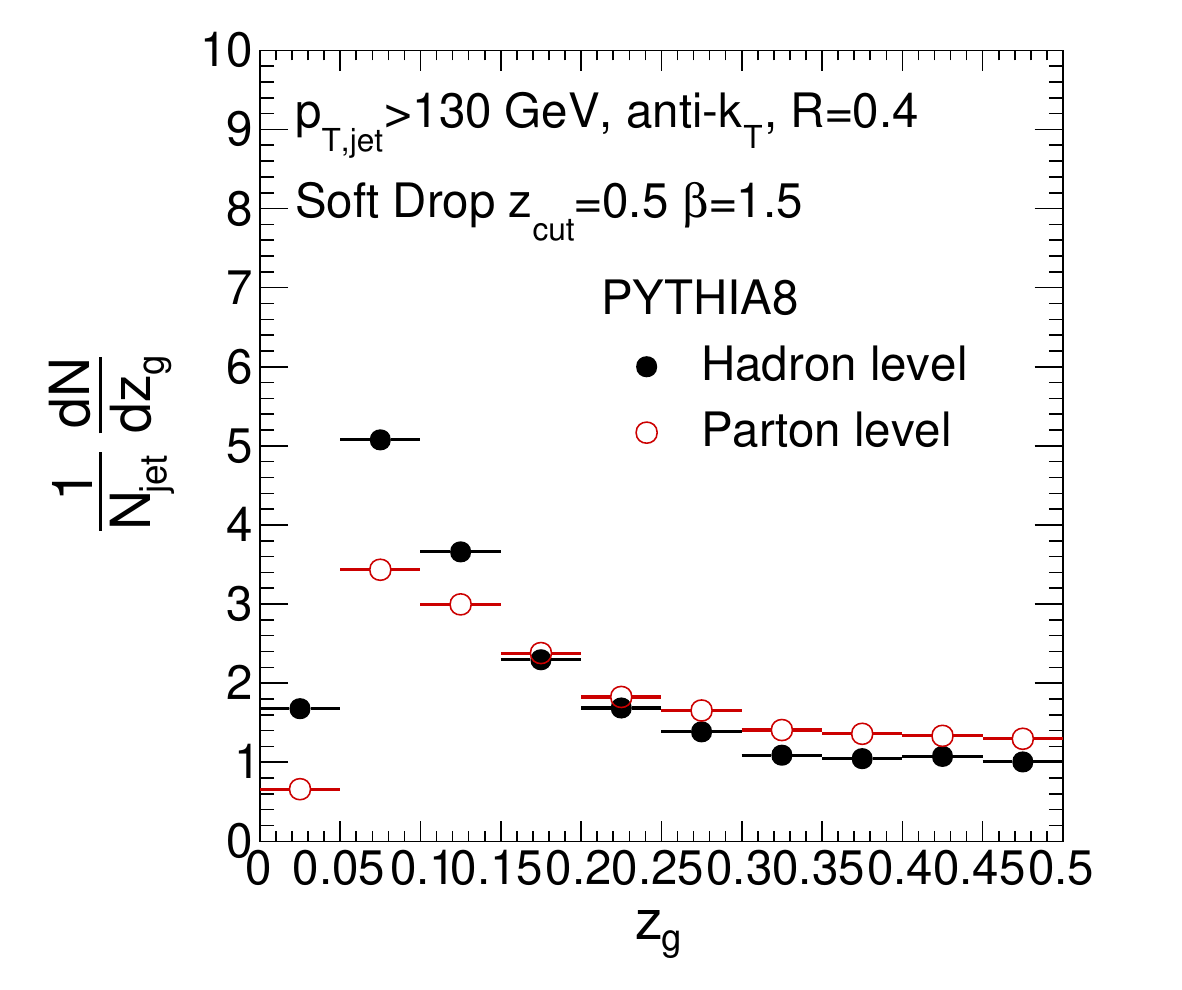}%
\includegraphics[width=0.33\textwidth]{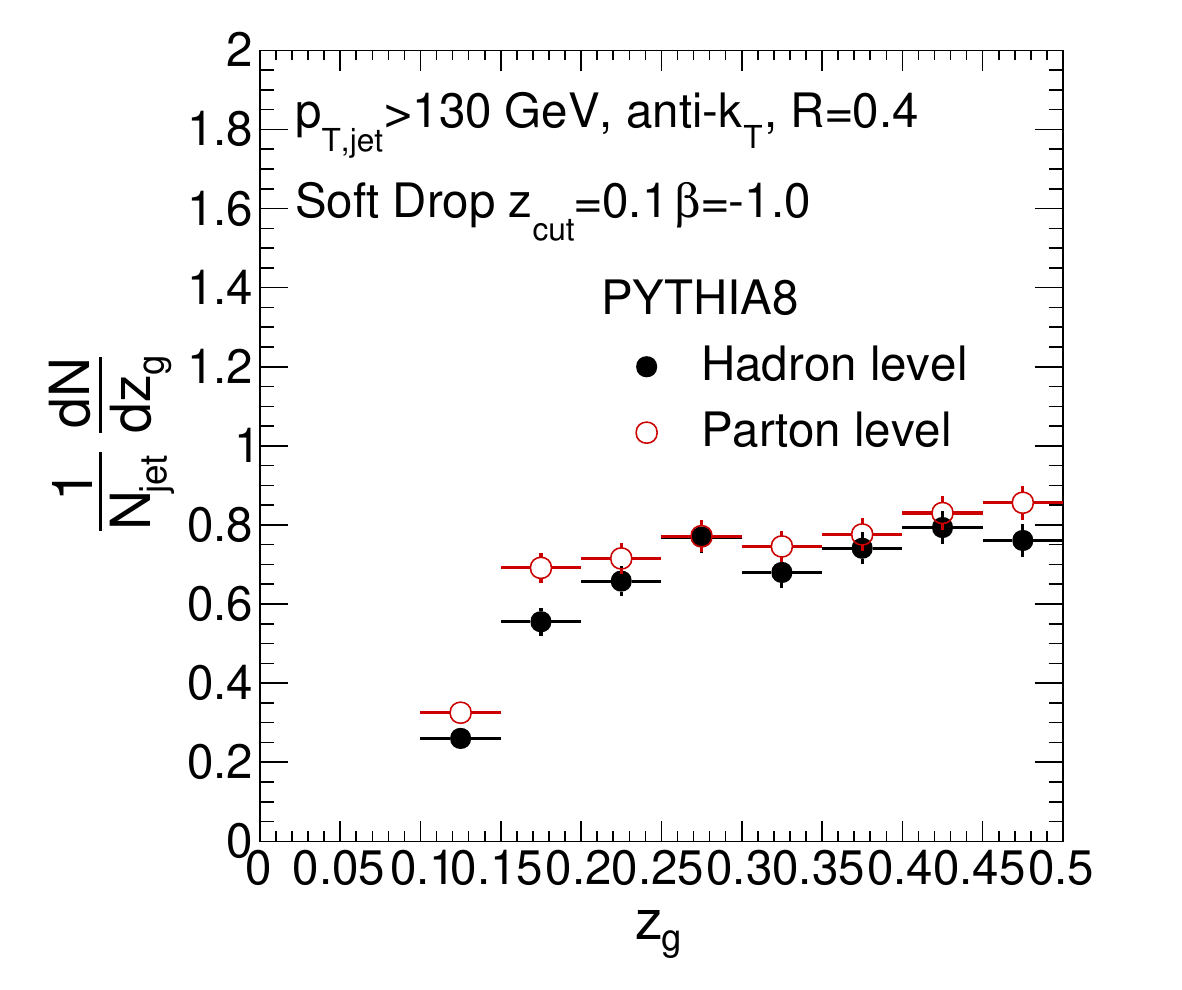}%
\caption{Groomed shared momentum fraction, $z_{\mathrm{g}}$, for three different grooming settings in simulations with and without hadronization with the PYTHIA8 event generator.}
\label{fig:SDGenZGHadVsPart}
\end{figure}

Even for jets in the QCD vacuum, it is well known that the SD procedure has some sensitivity to hadronization effects, for $\beta = 0$ see \cite{Dasgupta:2015yua}. From perturbative arguments, hadronization corrections to the jet $p_{T}$ grow like $R^{-1}$ \cite{Dasgupta:2007wa} and so are potentially important for subjet observables.
However, since hadronization is a process that happens locally in phase space, jets are less sensitive to the hadronization uncertainties than observables based on hadrons. In this paragraph we investigate how sensitive groomed subjet observables are to the hadronization process. For this purpose we compare the \zg~distribution in PYTHIA8 with and without hadronization for the three SD settings described above, as shown in \autoref{fig:SDGenZGHadVsPart}. It can be observed that the low-\zg\, region is particularly sensitive to hadronization effects. For grooming with negative $\beta$ the hadron- and parton-level results are most similar, see Fig,~\ref{fig:SDGenZGHadVsPart} (right), because with these grooming settings the soft splittings are rejected. Dedicated studies of these effects in conjunction with medium-modified hadronization are left for the future.

\subsubsection{Issues with changing reclustering algorithm}
\label{sec:reclusteringalgo}

\begin{figure}[ht]
\centering
\includegraphics[width=0.4\textwidth]{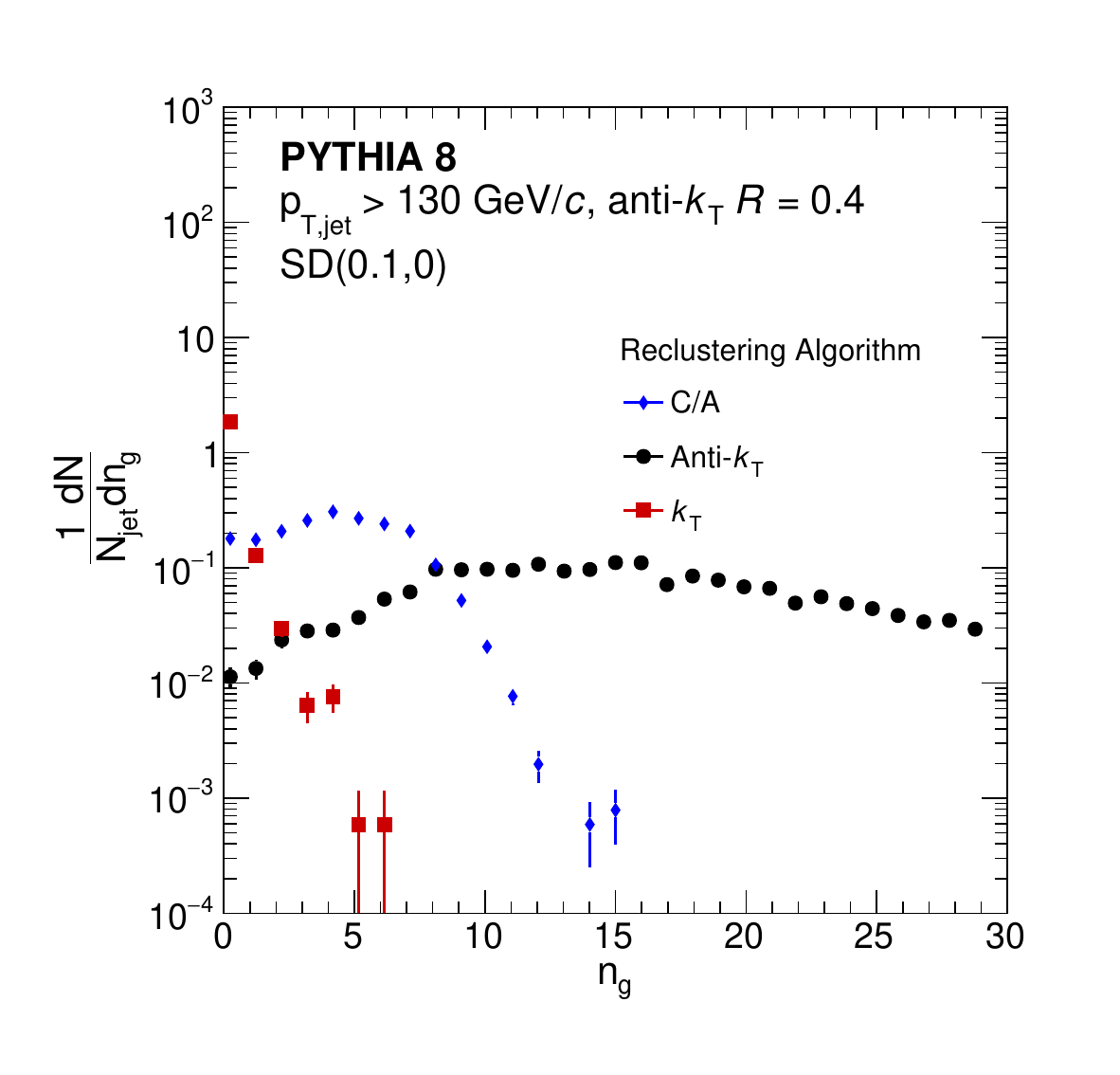}%
\includegraphics[width=0.4\textwidth]{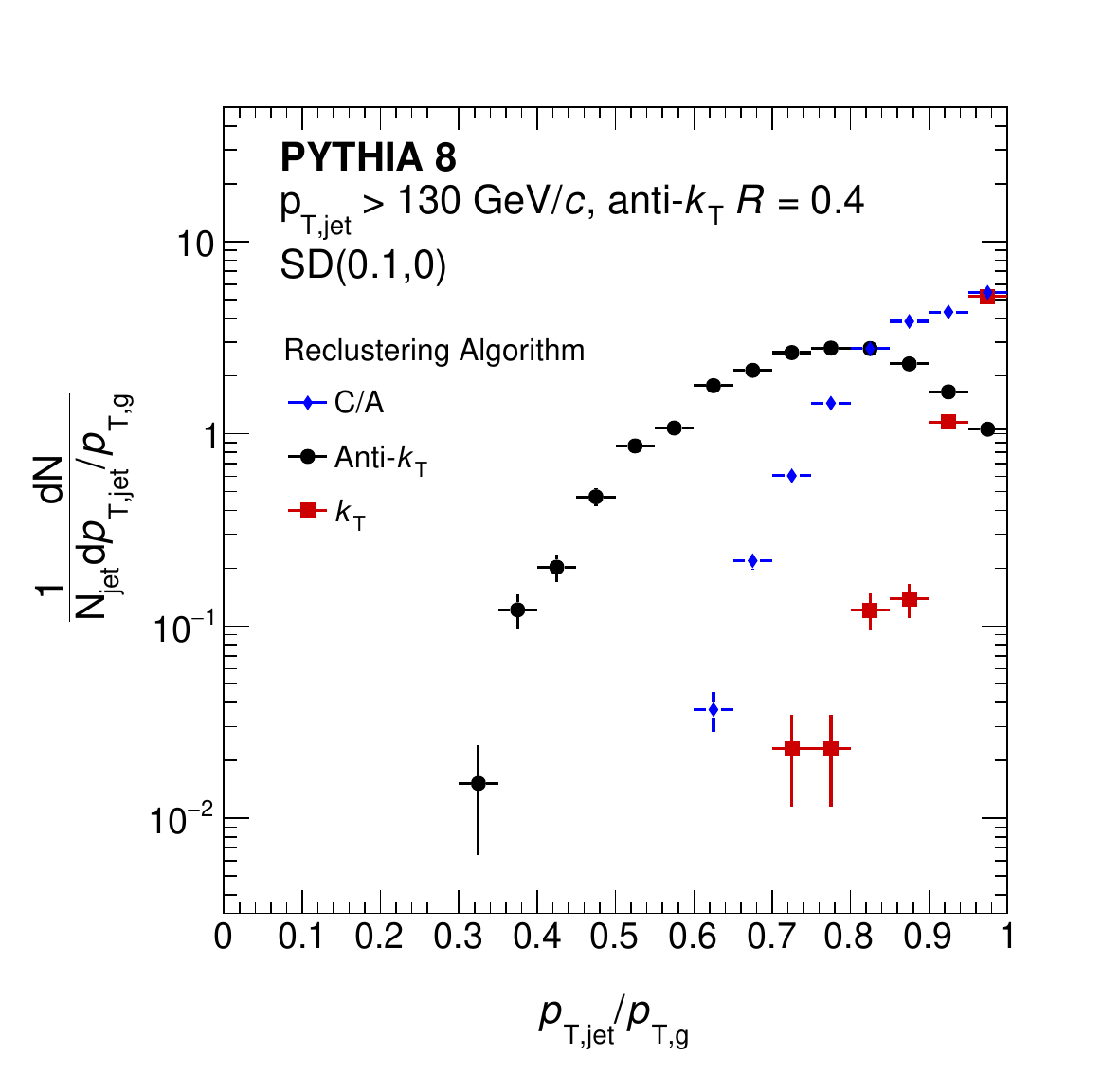}%
\caption{Effects of grooming on trees that are built up using different reclustering algorithms. Left plot: number of grooming steps. Right plot: ratio of jet $\pT$ before and after grooming.
}
\label{fig:PS2Vac_2}
\end{figure}

The change of algorithm also strongly affects what happens to the jet after grooming. In \autoref{fig:PS2Vac_2} (left), we show the distribution of the number of grooming steps for the three reconstruction algorithms discussed above. In particular, we note that the application of $\kT$ and anti-$\kT$ algorithms result in completely different grooming scenarios. While the jets reconstructed with the $\kT$ algorithm are mainly unaffected by grooming, in the anti-$\kT$ case of the order of 10--20 branches are groomed away. This also strongly affects the $\pT$ of the groomed jet, as seen in \autoref{fig:PS2Vac_2} (right), where the groomed jets in the anti-$\kT$ sample on average lose $\sim 20$\% of their energy. The C/A algorithm falls in between the two extremes, and is the only algorithm that maps the phase space with an approximately constant density, see \autoref{fig:PS2Vac} (left). Of the order of 5 branches are removed by the grooming procedure on average, which slightly reduces the jet $\pT$ by $\lesssim 10$\%.

\begin{figure}[th]
\centering
\includegraphics[width=0.32\textwidth]{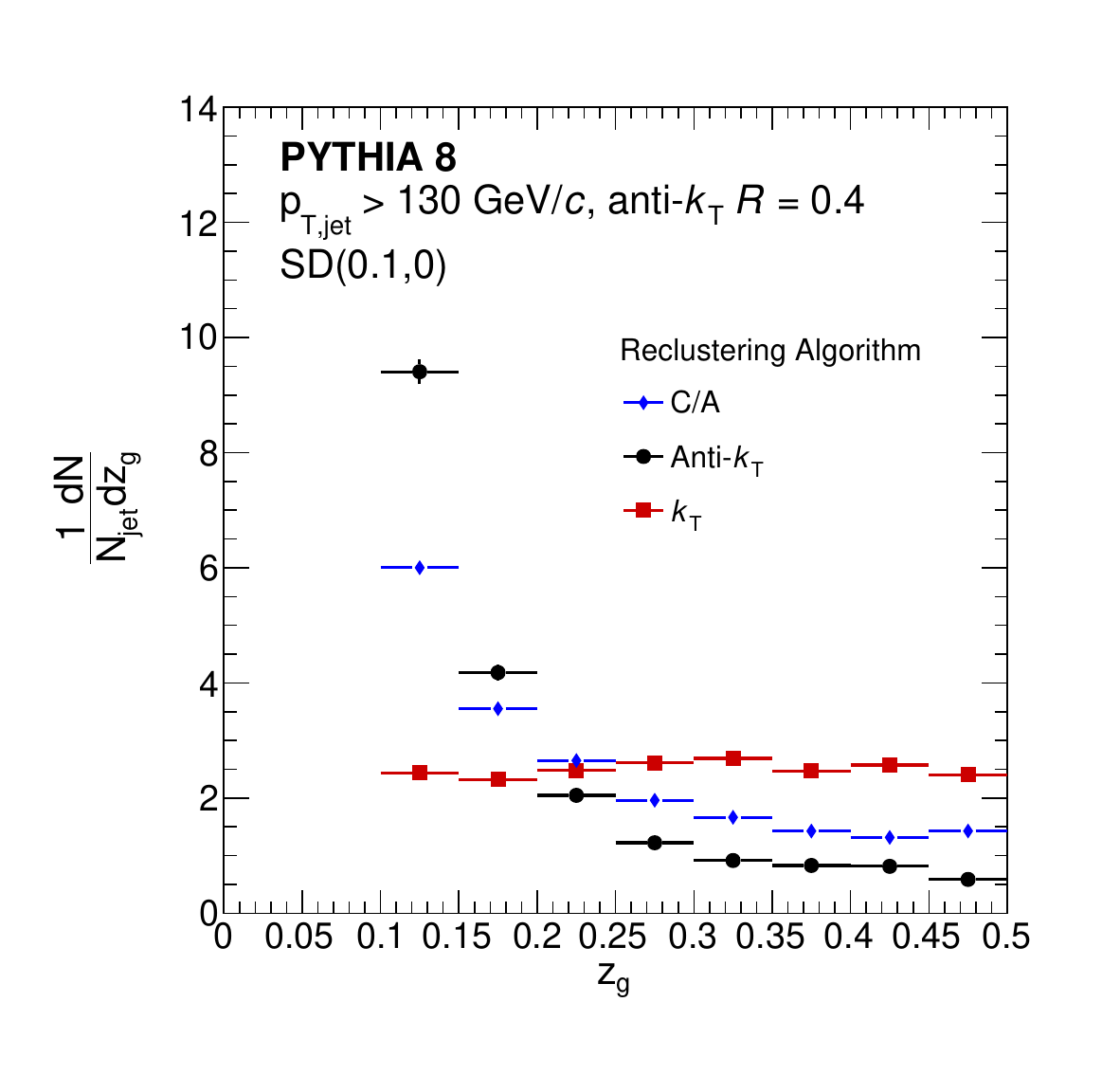}
\includegraphics[width=0.32\textwidth]{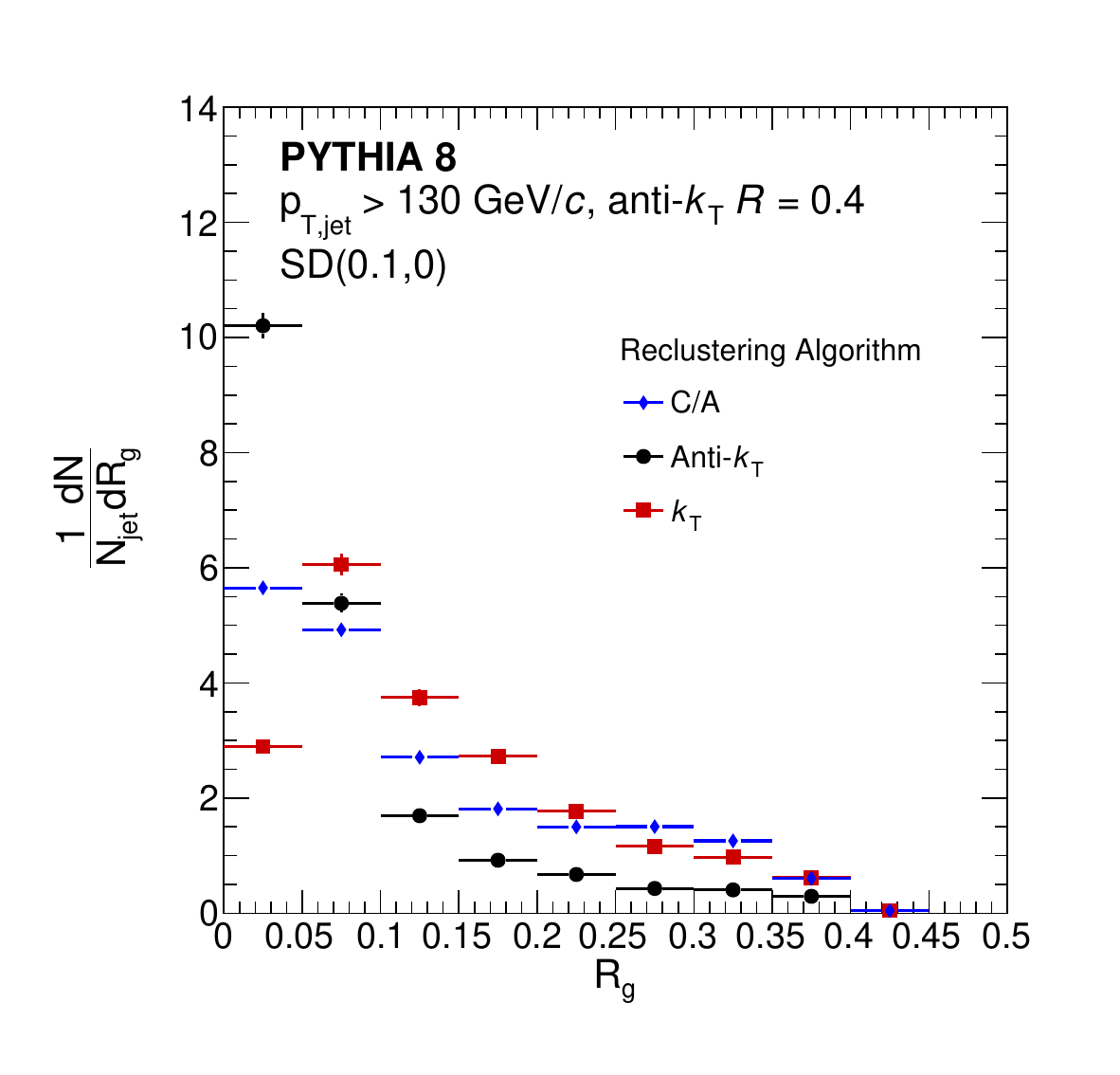}
\includegraphics[width=0.32\textwidth]{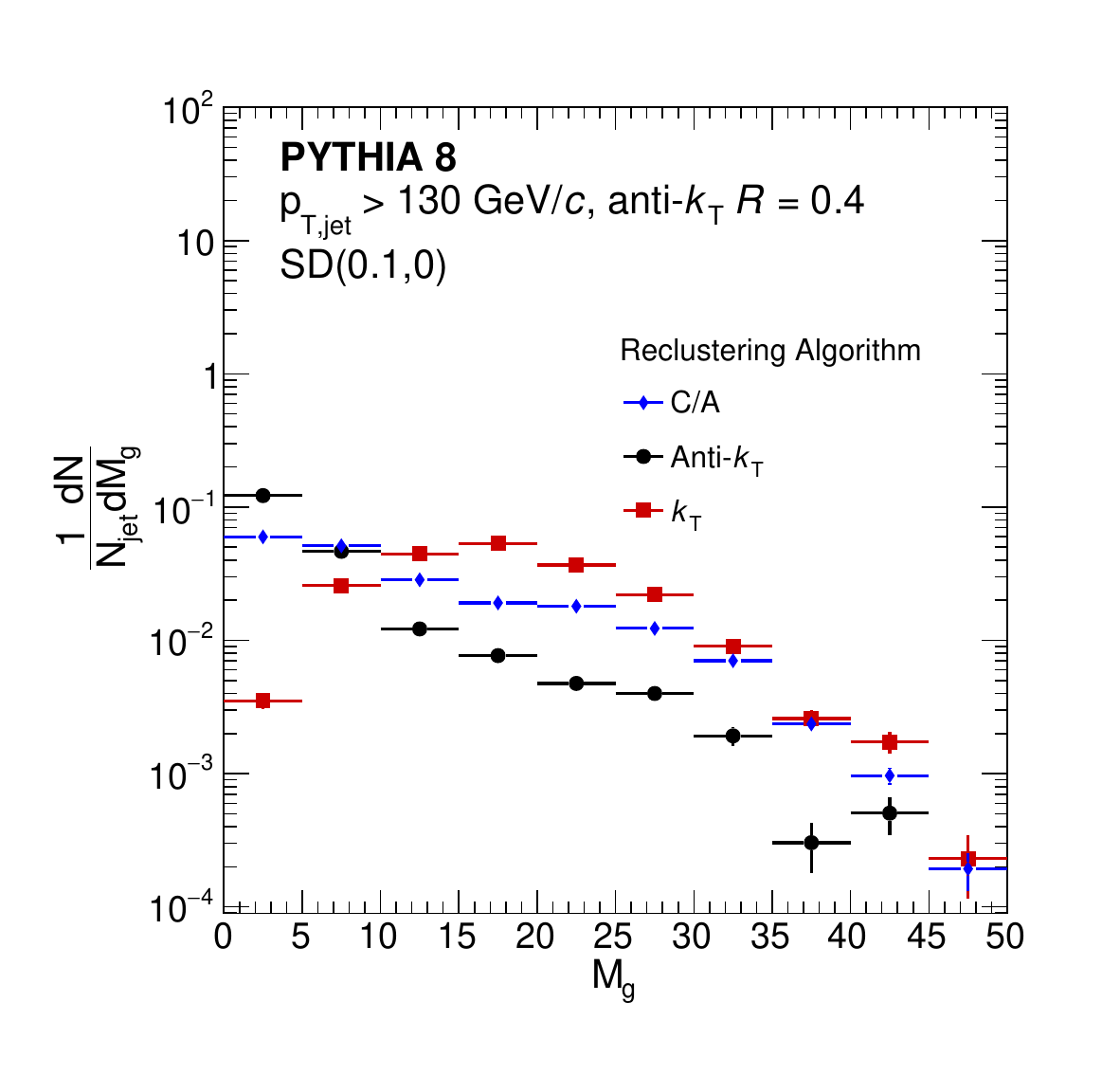}%
\caption{Subset of grooming variables, symmetry parameter ($z_{g}$), groomed mass ($M_{g}$) and groomed radius ($\Delta R_{12}$) for three different jet reclustering algorithms.}
\label{fig:SDClusteringComp}
\end{figure}

Finally, we studied the behavior of the three observables subject to different reclustering algorithms applied, see \autoref{fig:SDClusteringComp}. In this particular case, we limit ourselves only to looking at the PYTHIA8 (vacuum) samples.
In case of a grooming prescription that requires a semi-hard splitting, for instance like in the SD1 setting, the number of groomed branches will be large for anti-$k_{\rm \tiny T}$ reclustering ($\lesssim 30$) and very small for $k_{\rm \tiny T}$, for which the grooming conditions will be satisfied at the first iteration in most of the cases. Consistently, the groomed momentum fraction \zg\, probes very asymmetric splittings in the case of anti-$k_{\rm \tiny T}$ reclustering as can be seen in \autoref{fig:SDClusteringComp} (left). In contrast, $k_{\rm \tiny T}$-reclustered \zg\, picks up only symmetric splittings, resulting in an almost featureless distribution. Similar conclusions can be made for the $\Delta R_{12}$ distribution, \autoref{fig:SDClusteringComp} (center), and $M_g$, \autoref{fig:SDClusteringComp} (right), as well. Such artifacts arise due to the correlation  between either different regions on the primary Lund plane or between different Lund planes (e.g. primary and secondary) \cite{Dreyer:2018nbf},
and will therefore generally not be pursued further.

\subsection{Enhancing jet quenching observables using grooming}
\label{sec:dissecting}

Many jet quenching observables, such as the nuclear modification factor $R_{AA}$ and the momentum imbalance $x_{J\gamma}$ in photon-jet events, are considered benchmark measurements that quantify the amount of in-medium energy-loss and broadening. For reviews, see e.g. \cite{dEnterria:2009xfs,Majumder:2010qh}. However, their constraining power to discriminate between models has also been questioned. In some cases, the influence of background fluctuations can further obscure their constraining power.

In this section we present studies of conventional jet quenching observables that are enhanced by using grooming techniques.  As a first step, we apply SD grooming on the inclusive jet sample, extracting from each jet the grooming variables $z_g$ and $\Delta R_{12}$. 
This allows to further sub-divide the sample according to a measure involving these variables. 
For the purposes of this report, we have simply binned the fully inclusive sample according to the angle separating the two hardest subjets of a particular jet. This is motivated by the studies using splitting maps and the results obtained for the substructure observables previously. Another motivation is to differentiate between the modifications of the ``soft'' and the ``hard'' structure of the jet. The former is more dominant for inclusive observables and for non-restrictive SD settings, e.g. SD1 and SD2 in \autoref{fig:TheorySD} (left and central panels), while the latter would be more pronounced for conservative SD parameter choices, such as SD3 in \autoref{fig:TheorySD} (right panel).

Due to the exploratory scope of the workshop, we have not attempted to launch a systematic effort. Here, we only report on the two following studies at LHC energies:
\begin{itemize} 

\item the nuclear modification factor $R_{AA}$ binned in the angular separation $\Delta R_{12}$ as determined with SD2.

\item the $x_{J\gamma}$ distribution binned in the angular separation $\Delta R_{12}$ as determined with SD1. 

\end{itemize}
For both grooming settings, comparing small- and large-angle substructure configurations also gives an additional handle on the formation time of that particular splitting, see \autoref{fig:PS0} (right).

In both cases, jets were reclustered and groomed, and only jets that had a candidate subjet pair that fulfilled the Soft Drop condition were further analyzed.
More importantly, all results in this Section have been computed by embedding the MC jet samples into a centrality-dependent heavy-ion background \cite{deBarros:2012ws}, for details see \autoref{sec:uncorrelatedbackground}. As before, the background was subtracted using CS \cite{Berta:2014eza}, see \autoref{app:background} for further details. Therefore, these results reflect more realistically the magnitude of effects that should be expected to arise in heavy-ion collisions at the LHC.

The well-known nuclear modification factor $R_{AA}$ compares the yields of equivalent hard processes in heavy-ions and proton-proton collisions, and is given schematically as
\beq
\label{eq:RAA}
R_{AA} = \frac{\dd N_{AA}/\dd \pT^2 \dd y}{\langle T_{AA} \rangle \dd \sigma_{pp}/\dd \pT^2 \dd y} \,,
\eeq
where $\langle T_{AA} \rangle$ is the nuclear overlap function in a given centrality range,
is a standard benchmark for estimating/tuning medium parameters in theoretical calculations and Monte Carlo jet quenching models. By dividing the sample of inclusive high-$\pT$ jets into small- and large-angle configurations, we obtain more differential information regarding the accompanying modifications of the intra-jet structure. Similar studies, albeit using another method to dissect the jet sample into two-prong structures, were presented in \cite{Zhang:2015trf,Apolinario:2017qay}.
Note, however, that the suggested binning procedure could be sensitive to different physical mechanisms separately in the proton-proton and heavy-ion events. Disentangling this would demand further studies.

\begin{figure}[th]
\centering
\includegraphics[width=0.32\textwidth]{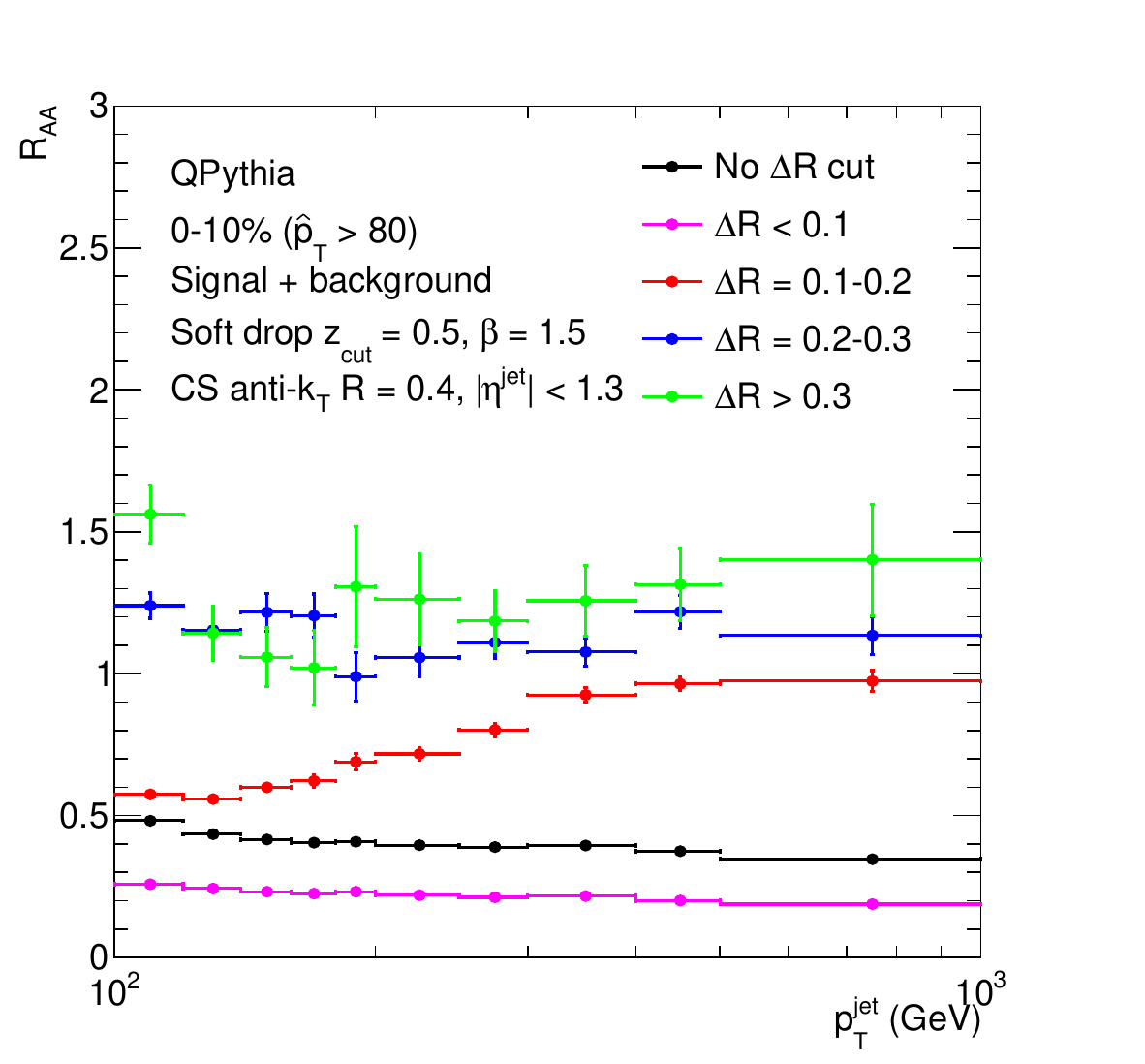}
\includegraphics[width=0.32\textwidth]{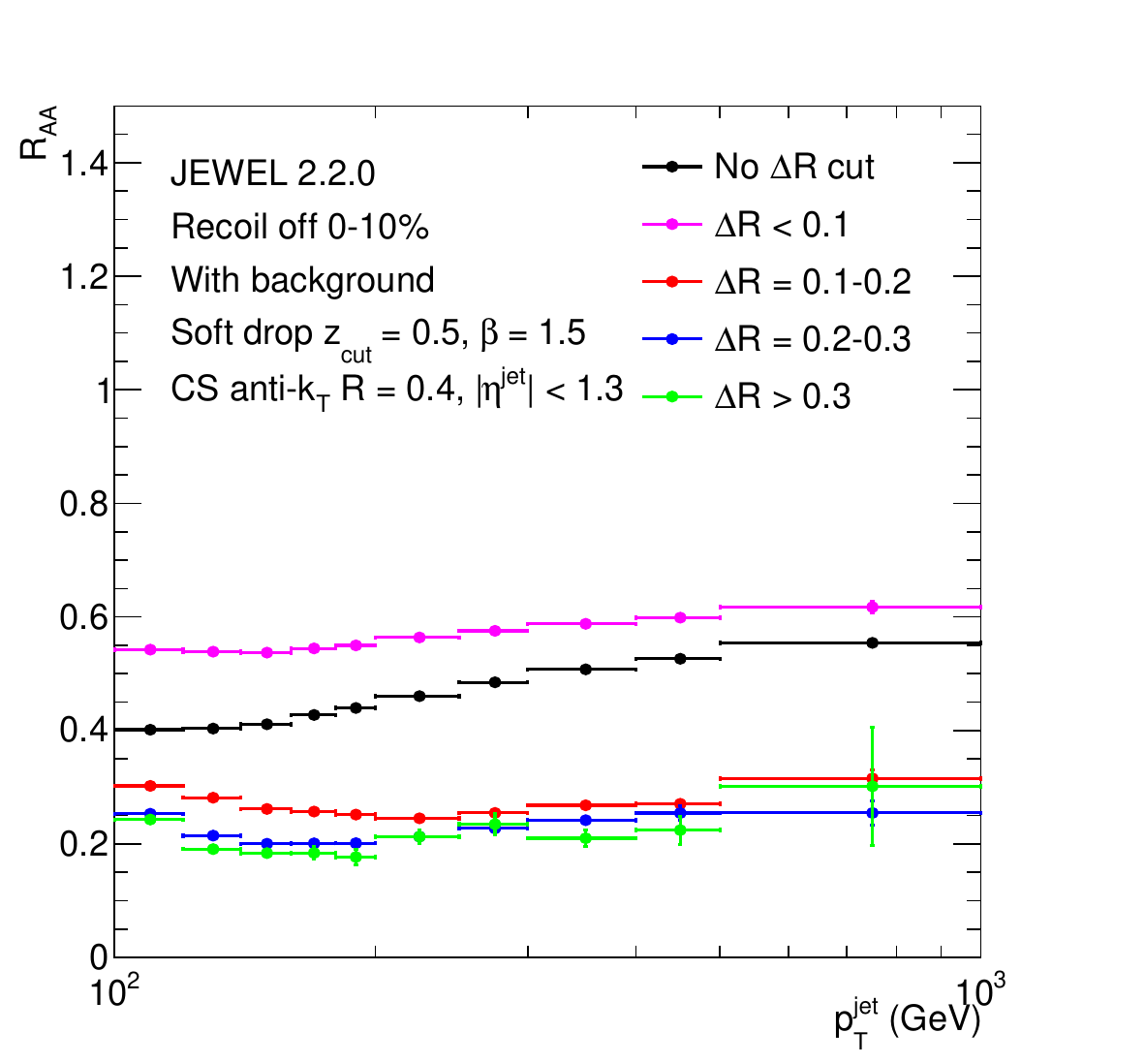}
\includegraphics[width=0.32\textwidth]{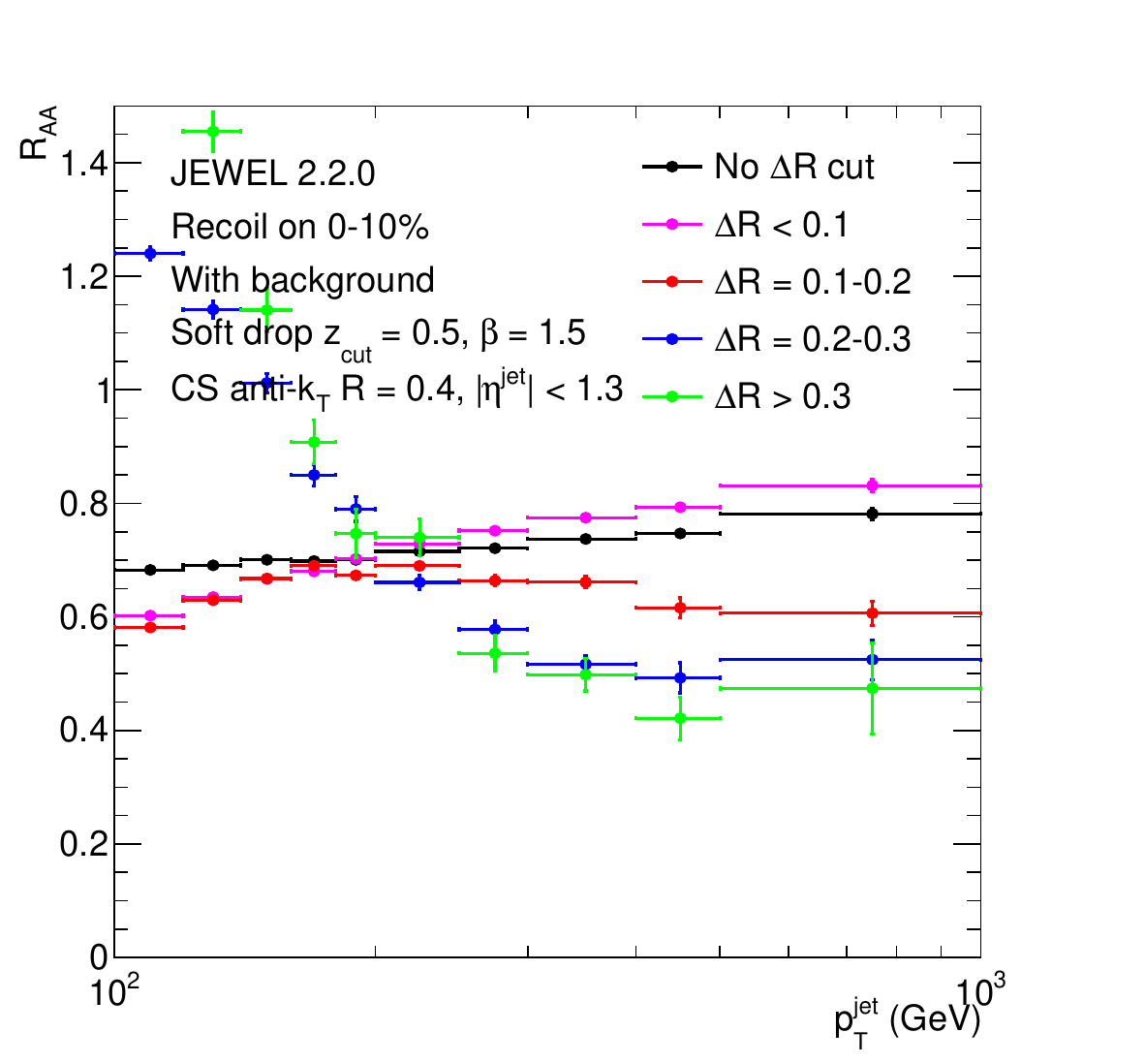}
\caption{Nuclear modification factor, \autoref{eq:RAA}, for subsamples of groomed jets binned as a function of $\Delta R$ of the leading sub-jets identified using SD1 for 0--10\% most central PbPb collisions.}
\label{fig:GroomedRAA}
\end{figure}

The jet samples generated from QPYTHIA, JEWEL ``Recoil off'' and JEWEL ``Recoil on'' that is used in the calculation of $R_{AA}$ in \autoref{fig:GroomedRAA}, have been binned according to the angular separation of the subjets identified using SD3 grooming, $\Delta R_{12}$. While all three models show a similar transverse momentum dependence of $R_{AA}$ for the fully inclusive sample (see black points in Fig.~\ref{fig:GroomedRAA}), large differences are seen for the more differential  results.\footnote{The overall magnitude of the inclusive $R_{AA}$ does not play an important role for the point we are trying to make here.} 

In QPYTHIA, the core of the jet is quenched stronger than the periphery, as expected from previous studies above, see \autoref{fig:GroomedRAA} (left). This fact basically related to the enhanced splitting of collinear modes. For the JEWEL ``Recoils off'' sample, see \autoref{fig:GroomedRAA} (center), the effect is completely opposite: the jet core is quenched much less than large-angle configurations. This also comes as no surprise in light  of other substructure observables that were analyzed above, see e.g. \autoref{sec:groomedobservables}, and reflects stronger energy-loss effects for large-angle substructure fluctuations thereby leading to more quenched partons \cite{Milhano:2015mng}. Finally including recoil effects, the JEWEL ``Recoils on'' sample, see \autoref{fig:GroomedRAA} (right), reveal a strong $\pT$-dependence of large-angle jets, leading to a big enhancement of $R_{AA}$ at relatively low transverse momenta. This implies an enhanced constraining power to details of medium recoil modeling in this observable.

Other benchmark observables in heavy-ion collisions include the $Z$-jet or photon-jet momentum asymmetry. Here, we will only focus on the latter, defined as the ratio of jet to photon momentum, 
\beq
x_{J\gamma} = \frac{p_\text{{\tiny T},jet}}{p_{\text{\tiny T},\gamma}} \,.
\eeq
In contrast to the nuclear modification factor \eqref{eq:RAA}, this observable does not immediately involve a comparison to a proton-proton baseline. The direct access to the photon energy in the measurement also would help constrain the effect of energy-loss or migration of jets between $\pT$-bins.

\begin{figure}[th]
\centering
\includegraphics[width=0.5\textwidth]{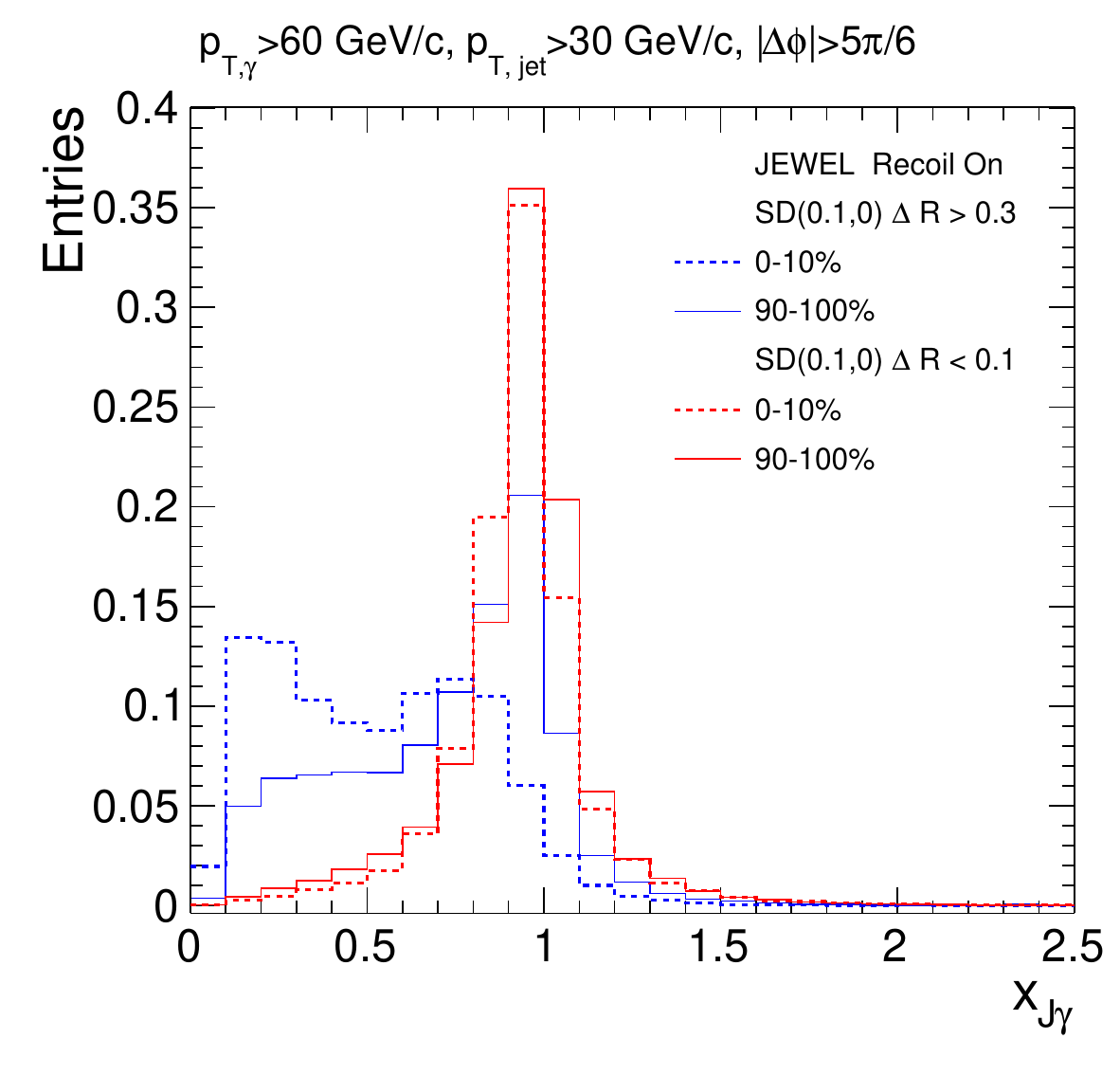}%
\caption{The $x_{J\gamma}$ distribution for subsamples of groomed jets that have been binned as a function of the angle found between the leading sub-jets using SD. }
\label{fig:GroomedGammaJet}
\end{figure}

In \autoref{fig:GroomedGammaJet} we present the resulting $x_{J\gamma}$ distribution for the JEWEL ``Recoils on'' samples in two centrality bins (corresponding to 0--10\% and 90--100\% centrality). This sample has been binned in subjet angular separation, as described above, this time using SD2 grooming. The same features that have been pointed out multiple times, also show up here as a function of collision centrality. Notably, the small-angle sample shows very little dependence on centrality, and is closely peaked around 1. The large-angle sample, on the other hand, which also corresponds to jets formed earlier in the medium, is strongly distorted. While the distribution for  90-100\% centrality is broader for the large-angle sample, we observe that the peak structure, which is clearly visible in \autoref{fig:GroomedGammaJet}), is completely removed when going to the most central collisions. This clean observable, augmented by sophisticated substructure techniques, could therefore prove very constraining with regard to the modeling of recoil effects inside the jet cone.

These proof-of-principle studies illustrate the enhanced sensitivity to more than one variable that can be obtained by differentiating the inclusive jet sample using a well-controlled procedure. 
In this section, we have analyzed medium-modified jet samples embedded in a heavy-ion background and utilized the angular separation between the leading subjets, as extracted in the Soft Drop procedure, in order to bin the jets into small- and large-angle configurations. While this should not come as a surprise in light of the previous results on the groomed substructure observables $z_g$, $\Delta R_{12}$ and $M_g/\pT$, we observe very different modification patters between the employed models and potentially large effects. Measurements of jets recoiling from $Z$-bosons or photons could prove as especially valuable in tracking how jets are modified in all variables, including the overall $\pT$ shift.
The results presented in this section are only exploratory and more systematic studies are left for the future.


\section{Outlook}
\label{sec:outlook}

The investigation of QCD jet observables in heavy-ion collisions is a community-wise effort, involving both experimentalists and theorists. 
Significant progress, both from the point of view of the development of experimental techniques as well as from theoretically motivated parametric estimates grounded on scale analysis and modeling within Monte Carlo parton showers, has led to a detailed qualitative understanding of how jets are modified in the medium created in the aftermath of heavy-ion collisions.
It is therefore worthwhile considering 
strategies that would be useful to further enhance jet observables as unique and valuable probes of the quark-gluon plasma.
As a first attempt at such a proposal, 
\emph{
in this report we have for the first time demonstrated how the
mapping  jet dynamics onto the kinematical Lund plane allows for a comprehensive comparison of jet quenching models.
This representation, which also could be useful for presenting experimental data, allows to pin down kinematical regimes where modifications arise and devise observables that are particularly constraining between existing and future models.}
It is as important to develop common synergies within the wide field of jet physics at colliders, based on modern perturbative QCD (fixed-order and resummation) techniques and on improved tools for high-energy experiments.

The organization of the 5th Heavy Ion Jet Workshop and CERN Theory Institute {\sl ``Novel tools and observables for jet physics in heavy-ion collisions''} provided an opportunity to initiate a first, community-wise effort in this direction.
While the concrete calculations and model studies presented in this report reflect the discussions that took place at that moment, the main messages are however relevant for the field at large. Let us summarize them in two points.
\begin{itemize}

\item We have introduced an operational way to map the full content of a jet splitting process, making use of the Lund diagram. Using kinematical arguments, we can make sense of enriched and depleted regions of phase space as results of medium interactions and recoil. An important caveat is that this idealized picture gets strongly distorted due to the presence of uncorrelated background but we have shown, through various exercises, that this aspect mostly affects the low-$\pT$ observables.

\item We have outlined a strategy, based on a selection of variables after jet reclustering/grooming, to single out jet samples enriched in configurations possessing specific properties as an aid to single out physics mechanisms, in particular \textsl{hard} (e.g. medium-induced bremsstrahlung, modifications of intra-jet structure due to energy loss) from \textsl{soft} (e.g. particle yield, sensitivity to recoil) medium effects and, similarly, \textsl{large-angle} from \textsl{small-angle} components.

\end{itemize}
We hope the topics we have reported here will trigger new and exciting future studies of jet, and in particular jet substructure, observables in heavy-ion collisions.

\section*{Acknowledgements} 
We thank all the participants of the 5th Heavy Ion Workshop and the TH institute {\sl ``Novel tools and observables for jet physics in heavy-ion collisions''}, that made it a memorable and productive meeting.
We especially thank the CERN TH department for hosting and especially Michelangelo Mangano and Angela Ricci for providing organizational support before and during the meeting.
M.S. was supported by Grant Agency of the Czech Republic (18-12859Y).
We acknowledge support from  Funda{\c c}{\~ a}o para a Ci{\^ e}ncia e Tecnologia (Portugal) under contracts CERN/FIS-NUC/0049/2015 and CERN/FIS-PAR/0022/2017 (LA, JGM, KZ), SFRH/BPD/103196/2014 (LA),  Investigador FCT - Development Grant IF/00563/2012 (JGM), and Investigador FCT - Starting Grant IF/00634/2015 (KZ).
DVP acknowledges funding from the Department of Energy under award DE-SC0018117.
HAA acknowledges support from the Science and Technology Facilities Council (https://stfc.ukri.org/).
CB was supported in part by Swedish Research Council, contracts number 2012-02283 and 2017-0034.
JM acknowledges support from United States Department of Energy, Office of Nuclear Physics, United States of America.
FD was supported by a Marie Sk\l{}odowska-Curie Individual Fellowship of the European Commission's Horizon 2020 Programme under contract number 660952 HPpQCD and Ministerio de Ciencia e Innovaci{\' o}n of Spain under projects FPA2014-58293-C2-1-P, FPA2017-83814-P and Unidad de Excelencia Mar{\' i}a de Maetzu under project MDM-2016-0692, by Xunta de Galicia (Conseller{\' i}a de Educaci{\' o}n) within the Strategic Unit AGRUP2015/11, and by FEDER.
YJL acknowledges support from MIT MISTI France Seed Fund, US Department of Energy Grant DE-SC0011088 and US Department of Energy Early Career Award DE-SC0013905.
KK acknowledges the support of Narodowe Centrum Nauki with grant DEC-2017/27/B/ST2/01985.

\appendix

\section{Monte Carlo parton showers}
\label{app:models}

This section briefly outlines the main physics ingredients of the MC in-medium parton shower generators used in course of the workshop. For detailed descriptions, we refer the interested reader to the original references.

\subsection{QPYTHIA}
\label{app:qpythia}

This appendix provides some details of the implementation of medium effects on the final-state parton shower as implemented in QPYTHIA \cite{Armesto:2009fj}. As the name suggests, the program builds on PYTHIA6 \cite{Sjostrand:2007gs,Sjostrand:2008vc}. The final-state shower is a mass-ordered (or virtuality-ordered) shower, where the Sudakov form factor is defined as
\beq
\label{eq:QPYTHIASudakov}
\Delta(t_1,t_0) = \exp\left[\int_{t_0}^{t_1} \frac{\dd t}{t} \int_{z_-}^{z^+} \dd z \, \frac{\alpha_s(t)}{2\pi}P(z)\right] \,,
\eeq
where the limits $z_\pm = z_\pm(t)$ implement the perturbative constraints and the evolution variable $t = M^2$ is the (squared )virtuality or invariant mass, see Eq.~\eqref{eq:DipoleMass}. The quantity in Eq.~\eqref{eq:QPYTHIASudakov} represents the probability of no splitting between the mass-scales $t_0$ and $t_1$ and can be used to determine the variables $(z,t)$ of the subsequent splitting in the shower by a standard dicing procedure. Although the shower is ordered in mass, angular ordering is enforced by a veto procedure.

In vacuum, the function $P(z)$ corresponds to the relevant Altarelli-Parisi splitting functions. However, in the medium one takes advantage of the fact that the medium-induced radiative spectrum comes simply in addition to the existing vacuum one \cite{Wang:2001ifa,Polosa:2006hb}, to substitute
\beq
P(z) \to P^\tot(z) = P(z) + \Delta P(z)
\eeq
in Eq.~(\ref{eq:QPYTHIASudakov}), where 
\beq
\Delta P(z) = \frac{2\pi t}{\alpha_s} \frac{\dd I^\med}{ \dd z \dd t} \,,
\eeq
where $\dd I^\med /(\dd z \dd t)$ is identified with the (double-differential) BDMPS spectrum \cite{Baier:1996sk}. In the current implementation of QPYTHIA it is computed in the multiple-soft scattering approximation, that neglects hard medium interactions, and in the soft limit $z \ll 1$. In addition, the splitting function $g \to q\bar q$ is not modified by this prescription since it is subleading.

\subsection{JEWEL}
\label{app:jewel}

JEWEL is also based on PYTHIA6, and handles exclusively the final-state parton shower routine. The main steps of the modified shower routine can be summarized in three points:
\begin{itemize}

\item Within the program, every interaction with the medium is treated similarly to the hard, partonic scattering itself and described by a $2 \to 2$ perturbative matrix element, suitably regularized in the IR due to the dressing of medium quasi-particles. Hence, one invokes so-called ``partonic parton densities'' to allow hard medium kicks to resolve additional (virtual) jet constituents in the course of the interaction. Scattering with the medium can also give rise to additional radiation.

\item The emission with the shortest formation time is realized first. This allows for a smooth interpolation between so-called vacuum emissions and the ones that are affected by medium interactions, often referred to as ``medium-induced''.

\item The Landau-Pomeranchuk-Migdal effect in QCD \cite{Wang:1994fx} is implemented by keeping track of the amount of re-scattering during the formation time of radiation.
\end{itemize}
Comparing to the analytical limits of single-gluon radiation spectrum, this approach allows to treat the kinematics of emission and the interactions on a more precise level.

In addition to the showering, JEWEL also permits to track the momenta and color flow of the recoiling medium constituents that happen to interact with the jet. Note that the medium parton is counted as a final-state particle directly after the scattering, and is not allowed to interact further. We refer to this mode as ``Recoils on''. The mode ``Recoils off'' refers to the case when these partons are not included in the event record and discarded. In the former case, it is imperative to subtract the thermal momentum of the partons before their interaction with the jet, since it forms part of the uncorrelated  thermal background in heavy-ion events. This is most reliably done with the so-called ``4MomSub'' procedure which consists of subtracting the sum momenta of medium constituents entering a given jet, for further details see \cite{KunnawalkamElayavalli:2017hxo}. 


\section{Background subtraction methods}
\label{app:background}

\begin{figure}[t!]
\centering
\includegraphics[width=0.5\textwidth]{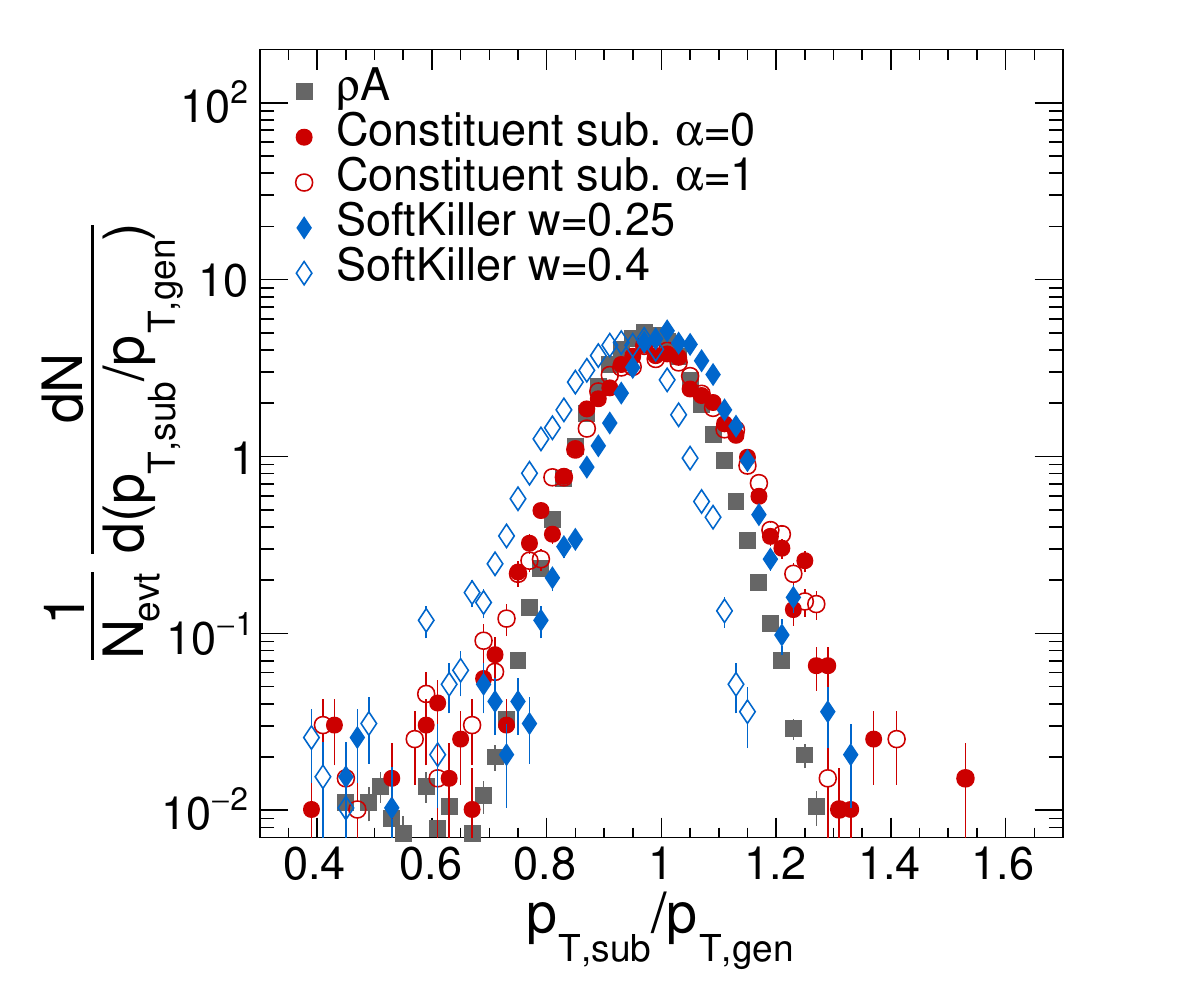}%
\includegraphics[width=0.5\textwidth]{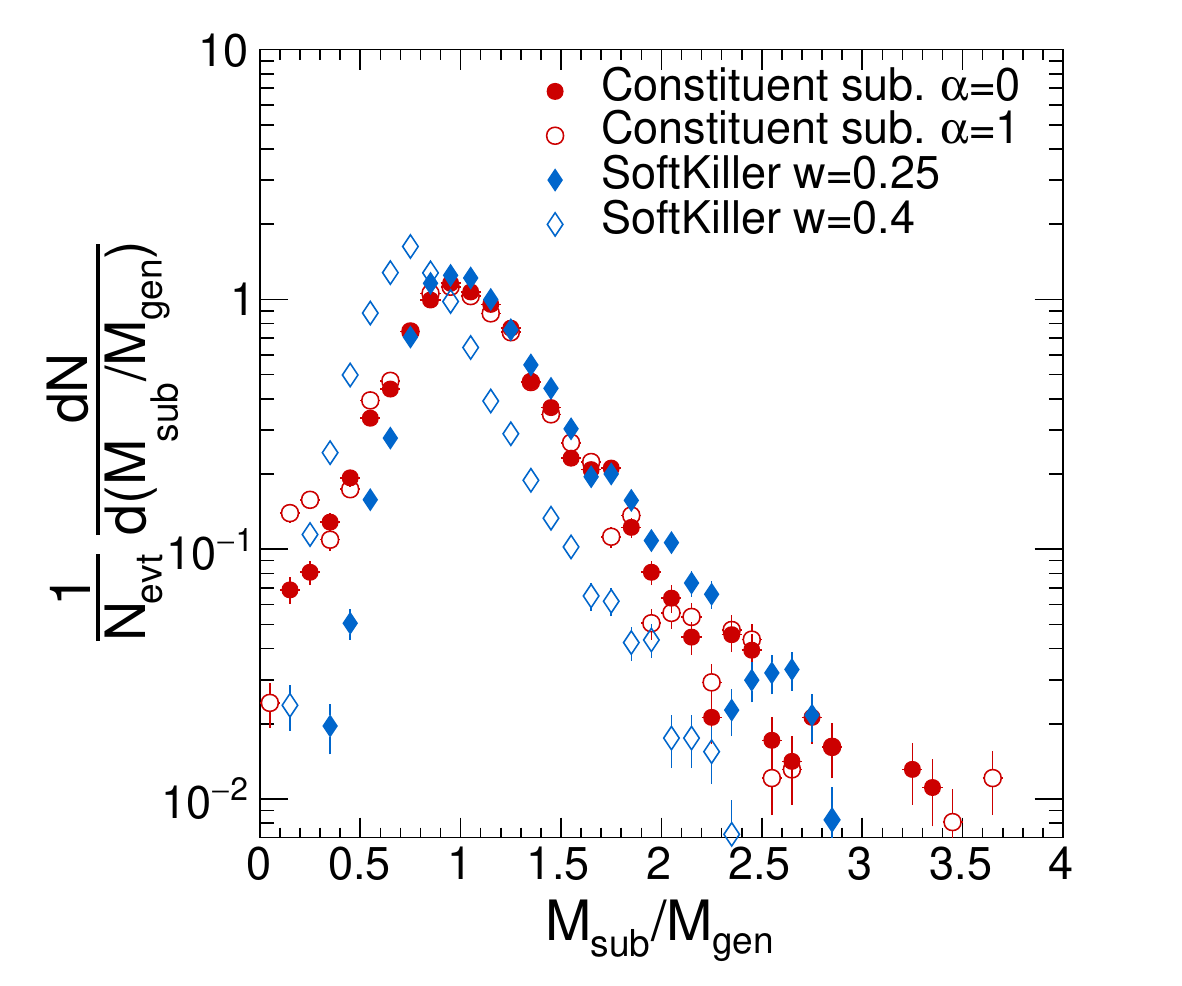}%
\caption{Jet energy (left) and mass response (right) for various jet subtraction techniques. PYTHIA jets are embedded into a thermal background corresponding to the characteristics observed in central PbPb collisions at the LHC.}
\label{fig:ResponseMethods}
\end{figure}

This sections provides some details about methods for background subtraction. We were mainly focussing on applications using area-based subtraction, constituent subtraction (CS) \cite{Berta:2014eza} and Soft Killer (SK) \cite{Cacciari:2014gra}. Other methods include \cite{Cacciari:2007fd}. The subtraction methods are evaluated by embedding jets generated with PYTHIA into a background resembling the properties of the underlying event observed at the LHC. We analyze anti-$\kT$ $R=0.4$ jets in pp collisions at $\sqrt{s_{NN}} = $ 5.02 TeV. The underlying event is generated assuming independent particle production according to a thermal Boltzmann distribution \cite{deBarros:2012ws}. Massless particles are generated randomly at central rapidities $|\eta|<3$. A Boltzmann distribution, with $\langle \pT \rangle$ = 1.2 GeV is used. This is equivalent to an average momentum density of $\langle \rho \rangle =250$ GeV and a total multiplicity of $\sim 7000$ particles, which corresponds roughly to the most central events in the CMS detector. \autoref{fig:ResponseMethods} shows the jet response for the transverse momentum (left) and mass (right) for the two background subtraction methods mentioned before. For both methods a free tunable parameter exists and we show the jet response for two different choices. The constituent subtraction is not very sensitive to the parameter $\alpha$ that regulates the $p_{\mathrm{T}}$ weighting of the distance parameter used in the algorithm. The response of Soft Killer is very sensitive to the width ($w$) used. By default the values is 0.4 but we observe that in the heavy ion environment a better performance is achieved using a smaller value since otherwise the algorithm subtracts part of the true jet signal.

\bibliographystyle{elsarticle-num}
\bibliography{THinstituteRefs}

\end{document}